\documentclass[final, times, 12pt]{elsarticle}

\usepackage{amssymb}
\usepackage{amsthm}

\journal{Computer Physics Communications}

\usepackage{graphicx}
\usepackage[utf8]{inputenc}
\usepackage{amsmath}
\usepackage[labelformat=simple]{subcaption}
\usepackage[top=2.5cm,bottom=2.5cm,left=2cm,right=2cm]{geometry}
\usepackage[english]{babel}

\usepackage[toc,page]{appendix} 
\usepackage{tabularx} 
\usepackage[table,xcdraw]{xcolor}

\usepackage{pgfplots}
\pgfplotsset{compat=newest}
\usepgfplotslibrary{groupplots}

\definecolor{myblue}{rgb}{0.050980,0.34118,0.69804}
\definecolor{myred}{rgb}{1,0.21961,0.28627}
\definecolor{mygrey}{rgb}{0.43137,0.43137,0.43137}

\usepackage{enumerate}
\usepackage[mathscr]{euscript}
\usepackage[T1]{fontenc}
\usepackage{caption}
\usepackage{multirow}

\usepackage{color}
\usepackage{amsfonts}
\usepackage{amstext}
\usepackage{alltt}
\usepackage{verbatim}
\usepackage{array}
\usepackage{booktabs}
\usepackage[colorlinks=true,linkcolor=blue,citecolor=blue,urlcolor=blue]{hyperref}
\usepackage[linesnumbered,ruled,resetcount]{algorithm2e}
\usepackage{nicefrac}
\usepackage{mathtools}
\usepackage{pdfpages}
\usepackage{bbm}
\usepackage{chngcntr}  
\usepackage{etoolbox}
\usepackage{empheq}
\usepackage{enumitem}
\usepackage{float}

\makeatletter
\let\old@algocf@pre@ruled\@algocf@pre@ruled
\renewcommand{\@algocf@pre@ruled}{%
  \Hy@raisedlink{\hyper@anchorstart{algocf.\thealgocf}\hyper@anchorend}%
  \old@algocf@pre@ruled}
\makeatother

\begin{document}

\renewcommand{\sectionautorefname}{Section}
\renewcommand{\subsectionautorefname}{Section}
\renewcommand{\subsubsectionautorefname}{Section}
\renewcommand{\algorithmautorefname}{Algorithm}

\begin{frontmatter}



\title{An optimal scaling to computationally tractable dimensionless models: Study of latex particles morphology formation}

\author[bcam]{Simone Rusconi\corref{cor}}
\author[lama]{Denys Dutykh}
\author[bcam,iker,stoilow]{Arghir Zarnescu}
\author[iker,quifis]{Dmitri Sokolovski}
\author[bcam,iker]{Elena Akhmatskaya}

\cortext[cor]{Contact address: \texttt{rusconis89@gmail.com}}

\address[bcam]{BCAM - Basque Center for Applied Mathematics, Alameda de Mazarredo 14, 48009 Bilbao, Spain}
\address[lama]{Univ. Grenoble Alpes, Univ. Savoie Mont Blanc, CNRS, LAMA, 73000 Chamb\'ery, France}
\address[iker]{IKERBASQUE, Basque Foundation for Science, Maria Diaz de Haro 3, 48013 Bilbao, Spain}
\address[stoilow]{``Simion Stoilow'' Institute of the Romanian Academy, 21 Calea Grivi\c{t}ei, 010702 Bucharest, Romania}
\address[quifis]{Departamento de Qu\'imica-F\'isica, Universidad del Pa\'is Vasco, UPV/EHU, Leioa, Spain}

\begin{abstract}
In modelling of chemical, physical or biological systems it may occur that the coefficients, multiplying various terms in the equation of interest, 
differ greatly in magnitude, if a particular system of units is used. Such is, for instance, the case of the Population Balance Equations (PBE) proposed to model the Latex Particles Morphology formation. The obvious way out of this difficulty is the use of dimensionless scaled quantities, although often the scaling procedure is not unique. In this paper, we introduce a conceptually new general approach, called Optimal Scaling (OS).
 The method is tested on the known examples from classical and quantum mechanics, and applied to the Latex Particles Morphology model, where it allows us to reduce the variation of the relevant coefficients from $49$ to just $4$ orders of magnitudes. The PBE are then solved
by a novel Generalised Method Of Characteristics, and the OS is shown to help reduce numerical error, and avoid unphysical behaviour of the solution. 
Although inspired by a particular application, the proposed scaling algorithm is expected find application in a wide range of chemical, physical and 
biological problems. 

\end{abstract}

\begin{keyword}
nondimensionalization \sep dimensionless models \sep Schr\"{o}dinger equation \sep multi-phase polymers \sep Population Balance Equations \sep Generalised Method Of Characteristics

\PACS 81.05.Zx \sep 82.20.Wt \sep 87.10.Ed 


\end{keyword}

\end{frontmatter}






\section{Introduction}
\label{sec:intro}
	
Obtaining a unitless or dimensionless formulation of governing equations plays an important role in mathematical and numerical analysis of any practical problem. First of all, it enables to specify important scaling parameters, which fully determine the system's behaviour. Henceforth, solving one dimensionless problem is equivalent to solving a whole class of dimensional problems sharing the same scaling parameters. Moreover, dimensionless equations allow one to estimate the relative magnitude of various terms and, thus, eventually to simplify the problem using asymptotic methods \cite{Nayfeh_PM2000}. Then, the floating-point arithmetic is designed in such a way that the rounding errors are minimal if computer manipulates numbers of the same magnitude \cite{Kahan:1979:PFS:1057520.1057522}. Also, the floating-point numbers have the highest density in the interval $(0,1)$ and their density decreases when moving further away from this interval. So, it is always more advantageous to manipulate numerically the quantities with magnitude $\approx 10^0$, in order to avoid severe round-off errors and to improve the conditioning of the problem in hands.

For these reasons, nondimensionalization, or more ambiguously ``scaling'', reformulates dimensional models in terms of dimensionless quantities only. As stated above, one of the purposes is in decreasing the computationally intractable orders of magnitude of physical quantities. In addition, the scaling procedure allows for an estimation of characteristic properties of a given system, such as intrinsic frequency, typical length, or time duration.

Once a system of equations in physical units is given, the following steps \cite{Holmes2009_ND} are usually made to scale the model. First, one must identify all independent and unknown variables. Then, each identified variable is replaced with its dimensionless counterpart, defined by dividing it over a dimensional characteristic constant to be determined. Thus, the system of equations is rewritten in terms of dimensionless quantities and coefficients. Finally, numerical values of the characteristic constants are found by imposing as many dimensionless coefficients as possible equal to 1. The computed constants determine the characteristic properties of a given system.	

The characteristic constants are found by solving a system of $ N = \min\{N_d,N_x\} $ equations with $N_x$ unknowns, with $N_d$ and $N_x$ 
being  the total numbers of dimensionless coefficients and characteristic constants, respectively. Since the complexity of realistic models often leads to $N_d > N_x$, potential difficulties may affect the scaling procedure as explained in \cite{Holmes2009_ND}. First, at most $N_x$ out of $N_d$ coefficients can be imposed to 1, with no control on the orders of magnitude of the remaining $N_d-N_x$ coefficients. Second, the choice of the coefficients to be imposed to 1 represents a problem-specific task, with no clue for a proper selection in the general case \cite{Scaling_Langtangen}. Finally, the characteristic properties of the system may not be correctly estimated, if the dimensionless coefficients and variables become very large or very small \cite{Scaling_Langtangen}. For these reasons, we propose an alternative approach, which provides a well defined routine for a rational choice of characteristic constants for an arbitrary equation.

The novel technique may be beneficial for models with a number of dimensionless coefficients larger than the number of characteristic constants to estimate. This is exactly the case of the Population Balance Equations (PBE) model \cite{Ramkrishna2000} for Latex Particles Morphology formation \cite{DDPM_2016,PhDThesis_Rusconi_PMCQS}. The multi-phase particle morphology is defined as a pattern formed by the phase-separated domains comprising the multi-phase particle. Assembling the composite (multi-phase) polymer particles is of great practical interest, due to their performance superiority over particles with uniform compositions. Multi-phase polymers are used in many important applications, such as, coatings, additives for constructing materials, cosmetics, diagnostic tests and drug delivery. Properties of a multi-phase polymer particle strongly depend on the particle morphology, and thus the control of particle morphology is a key factor for success in producing high quality polymer materials. The synthesis of new morphologies is time and resources consuming, as it largely relies on heuristic knowledge. Therefore, the predictive modelling of such processes is vital for practitioners.

The paper is structured as follows. In \autoref{sec:optimal_scaling_procedure}, we present the novel scaling procedure, which we call Optimal Scaling. Meaning to reduce computational efforts and numerical errors, we also derive an analytical formula for determining the characteristic constants within the proposed technique. The analytical solution is general and holds for a wide range of equations. In addition, \autoref{sec:optimal_scaling_procedure} analyses and tests the new methodology on three popular models presenting examples of classical and quantum dynamics. In particular, we consider the Projectile Problem \cite{Barenblatt1996_book}, the Schr\"{o}dinger Equation for an hydrogen atom in a magnetic field \cite{Fassbinder_1996_Hmagn} and the Landau-de Gennes model for liquid crystals \cite{Gartland_2018_liquid_cryst}. Then, \autoref{sec:modelling_DDPM} introduces the PBE model for dynamic development of Latex Particles Morphology, and derives the corresponding dimensionless model with the help of the Optimal Scaling procedure. \autoref{sec:num_treat_DDPM} proposes the Generalised Method Of Characteristics (GMOC) for the numerical treatment of the presented model and investigates the effect of different scaling approaches on the performance of GMOC applied to model the Latex Particles Morphology formation. Conclusions are provided in \autoref{sec:concl_discuss}.

\section{Optimal Scaling Procedure}		
\label{sec:optimal_scaling_procedure}

	Given $\tilde{f}: \mathbb{R}^{N_x} \times (0,\infty)^{N_p} \rightarrow \mathbb{R}$, let us consider the equation 
	
	\begin{equation}
	\tilde{f}
	\left(
	\tilde{x}_1, \dots, \tilde{x}_{N_x};
	\tilde{p}_1, \dots, \tilde{p}_{N_p}
	\right)
	= 0,
	\label{eqn:dimensional_equation}
	\end{equation}

	\noindent where $\tilde{x}_1, \dots, \tilde{x}_{N_x}$ correspond to the unknown variables and the independent quantities, while $\tilde{p}_1, \dots, \tilde{p}_{N_p}$ are the positive parameters assuming experimental values. All variables and parameters in \eqref{eqn:dimensional_equation} are expressed in their physical units. Our aim is to derive from the given equation the corresponding dimensionless model with computationally tractable orders of magnitude for parameters and involved variables. The following derivation can be generalised to the vector case, by considering $\tilde{f}$ as a single component of the given vector operator.
	
	The dimensionless counterparts of the variables  $\tilde{x}_1, \dots, \tilde{x}_{N_x}$ are defined as:		
	
	\begin{equation}
	x_1 \equiv \tilde{x}_1 / \theta_1,
	\dots,
	x_{N_x} \equiv \tilde{x}_{N_x} / \theta_{N_x},
	\label{eqn:change_of_var}
	\end{equation}
		
	\noindent where $ \theta \equiv \left\{ \theta_1, \dots, \theta_{N_x} \right\} \in (0,\infty)^{N_x}$ are strictly positive characteristic constants with the dimensions and orders of magnitude of the corresponding variables. The change of variables \eqref{eqn:change_of_var} allows rewriting \eqref{eqn:dimensional_equation} as a dimensionless equation 

	\begin{equation}
	f
	\left(
	x_1, \dots, x_{N_x};
	\lambda_1, \dots, \lambda_{N_d}
	\right)
	= 0,
	\quad 
	f: \mathbb{R}^{N_x} \times (0,\infty)^{N_d} \rightarrow \mathbb{R},
	\label{eqn:dimensionless_equation}
	\end{equation}
	
	\noindent where the dimensionless coefficients $\lambda \equiv \left\{ \lambda_1, \dots, \lambda_{N_d} \right\} \in (0,\infty)^{N_d}$ are functions of the characteristic constants $\theta$ and the physical parameters $\tilde{p} \equiv \{ \tilde{p}_1, \dots, \tilde{p}_{N_p} \}$, i.e. $ \lambda_i = \lambda_i ( \theta ; \tilde{p} ) $, $\forall i=1, \dots, N_d$. Given the physical parameters $\tilde{p}$, the characteristic constants $\theta$ should be proposed by the user in order to specify the coefficients $\lambda$, i.e. $\lambda_i=\lambda_i(\theta)$, $\forall i=1, \dots, N_d$. In other words, the characteristic constants $\theta$ correspond to the degrees of freedom in the scaling procedure.
	
	One of the most common approaches for the definition of the characteristic constants $\theta$ consists in imposing as many $\lambda$ as possible equal to 1, starting from the coefficients of the highest orders polynomial or derivative terms in \eqref{eqn:dimensionless_equation}, as explained in \cite{Holmes2009_ND} and discussed in \autoref{sec:intro}. In the case $N_d \le N_x$, it is possible to impose $\lambda_i(\theta)=1$, $\forall i=1, \dots, N_d$, and the constants $\theta$ are defined by solving a system of $N_d$ equations with $N_x$ unknowns. However, such a case rarely corresponds to realistic models, whose complexity leads to $N_d > N_x$. As a result, it is impossible to solve the system of equations $\lambda_i(\theta)=1$, $\forall i=1, \dots, N_d$. If $N_d > N_x$, the approach presented in \cite{Holmes2009_ND} ensures at most $N_x$ coefficients $\lambda$ equal to 1, while the remaining $N_d-N_x>0$ are defined by plugging the computed $\theta$, with no control on their orders of magnitude.
	
	For this reason, we suggest an alternative approach. More specifically, we propose to look for such constants $\theta$ (we call them \emph{optimal scaling factors} $\theta_{\mathrm{opt}} \in (0,\infty)^{N_x}$), for which the deviation $C$ of $\lambda(\theta) \equiv \left\{ \lambda_1(\theta), \dots, \lambda_{N_d}(\theta) \right\} \in (0,\infty)^{N_d}$ from $\mathbf{1} \equiv \left\{1, \dots, 1\right\} \in (0,\infty)^{N_d}$ is minimal:
	
	\begin{equation}
	\theta_{\mathrm{opt}}
	\equiv 
	\mathop{\mathrm{argmin}}_{\theta \in (0,\infty)^{N_x}}
	C(\theta),
	\label{eqn:theta_opt_def}
	\end{equation}

	\noindent where the functional $C(\theta) : (0,\infty)^{N_x} \to [0,\infty)$ measures the distance of the magnitude of coefficients $\lambda$ from $10^0$. The cost function $C(\theta)$ in \eqref{eqn:theta_opt_def} can evaluate the constants $\theta$ through the Euclidean distance $C_{\mathrm{eucl}}(\theta)$ between the vector $\left\{ \log_{10}(\lambda_i) \right\}_{i=1}^{N_d}$, holding the orders of magnitude of the coefficients $\lambda$, and the vector $\left\{ \Theta_i \right\}_{i=1}^{N_d}$, holding the desired orders of magnitude $\Theta_i = \log_{10}(10^0) = 0$, $\forall i = 1, \dots, N_d$:
	
	\begin{equation}
	C(\theta)
	=
	C_{\mathrm{eucl}}(\theta)
	\equiv	
	\sum_{i=1}^{N_d}
	\left[ \log_{10}(\lambda_i(\theta)) - \Theta_i \right]^2.
	\label{eqn:C_eucl_def}
	\end{equation}
	
	\noindent The choice of $C(\theta)$ in \eqref{eqn:theta_opt_def} is not unique, and definitions alternative to \eqref{eqn:C_eucl_def} can be explored. In particular, the functional $C(\theta)$ in \eqref{eqn:theta_opt_def} can be chosen as the maximum norm $C_{\max}(\theta)$:
	
	\begin{equation}
	C(\theta)
	=
	C_{\max}(\theta)
	\equiv	
	\max_{i=1, \dots, N_d}
	| \log_{10}(\lambda_i(\theta)) - \Theta_i |.
	\label{eqn:C_max}
	\end{equation}

	\noindent The functional \eqref{eqn:C_max} is less regular than \eqref{eqn:C_eucl_def}. However, numerical routines, such as the Simulated Annealing Algorithm
\cite{Kirk1983, Cerny1985, 10.2307/3214721, 10.2307/3690278, YaoLi1991, Zabinsky1993}, can be successfully applied for finding the minimum of non-smooth cost functions. In this study, we have used the implementation proposed in \cite{10.2307/3214721} due to its computational efficiency (up to $10^2$ times faster than algorithm \cite{Kirk1983,Cerny1985} for the tested systems).

	The proposed \emph{Optimal Scaling} procedure is summarised in \autoref{algo:optimal_scaling}, for the dimensional equation \eqref{eqn:dimensional_equation} with computationally intractable values. 

	\begin{algorithm}[!h]
	Given $\tilde{f}(\tilde{x};\tilde{p})=0$, where $\tilde{x} \equiv \{ \tilde{x}_1, \dots, \tilde{x}_{N_x} \}$ correspond to unknown variables and independent quantities, expressed in physical units, and $\tilde{p} \equiv \{ \tilde{p}_1, \dots, \tilde{p}_{N_p} \}$ are known dimensional positive parameters, introduce the strictly positive constants $\theta \equiv \{ \theta_1, \dots, \theta_{N_x} \}$ with the same dimensions as $\tilde{x}_1, \dots, \tilde{x}_{N_x}$\;
	Define the dimensionless variables $x \equiv \{ x_1 = \tilde{x}_1 / \theta_1 , \dots, x_{N_x} = \tilde{x}_{N_x} / \theta_{N_x} \}$\;
	Rewrite $\tilde{f}(\tilde{x};\tilde{p})=0$ in the unitless form $f(x;\lambda)=0$, with $\lambda(\theta) \equiv \{ \lambda_i(\theta;\tilde{p}) \}_{i=1}^{N_d} \in (0,\infty)^{N_d}$ being the dimensionless coefficients of the equation $f(x;\lambda)=0$\;
	Define the functional $C(\theta)$ to measure the distance of the magnitude of $\lambda(\theta)$ from $10^0$\;
	Compute $\theta_{\mathrm{opt}} = \mathop{\mathrm{argmin}}_{\theta \in (0,\infty)^{N_x}} C(\theta)$, using the appropriate minimisation procedure\;
	Return the dimensionless and computationally tractable equation $f(x;\lambda(\theta_{\mathrm{opt}}))=0$\
	\caption{The Optimal Scaling procedure to reduce the equation $\tilde{f}(\tilde{x}_1, \dots, \tilde{x}_{N_x};\tilde{p}_1, \dots, \tilde{p}_{N_p})=0$ to dimensionless and computationally tractable variables, when the positive parameters $\tilde{p}_1, \dots, \tilde{p}_{N_p}$ assume known experimental values.}  
    \label{algo:optimal_scaling}
    \end{algorithm}
    
	The Optimal Scaling procedure results in dimensionless equations with computationally tractable orders of magnitude for the variables of interest. In addition, it is independent from any resolution methodology, since it can be applied before running the chosen technique. Preferably, the scaling procedure should not imply a significant computational effort. In what follows, we show that if the Euclidean norm is chosen in \eqref{eqn:theta_opt_def} as a cost function, then the minimisation problem can be solved analytically.  

\subsection{Analytical Solution for Optimal Scaling}
\label{sec:anal_sol_min_Ceucl}

	We consider the minimisation problem \eqref{eqn:theta_opt_def} with the cost function $C(\theta) = C_{\mathrm{eucl}}(\theta)$ \eqref{eqn:C_eucl_def}. The Buckingham $\Pi$-theorem \cite{Barenblatt2003_book} ensures the following shape for the coefficients $\lambda$ defined in \eqref{eqn:dimensionless_equation}: 
	
	\begin{equation}
	\lambda_i(\theta) 
	=
	\kappa_i \, \theta_1^{\alpha^i_1} \, \dots \, \theta_{N_x}^{\alpha^i_{N_x}},
	\quad
	\kappa_i \in (0,\infty),
	\quad
	\alpha^i_j \in \mathbb{R},
	\quad
	\forall i=1, \dots, N_d,
	\quad
	\forall j=1, \dots, N_x.
	\label{eqn:HP_shape_lambda}	
	\end{equation}	 	
	
	\noindent From \eqref{eqn:HP_shape_lambda}, it follows	
	
	\begin{equation}
	\frac{\partial C_{\mathrm{eucl}}(\theta)}{\partial \theta_j}
	=
	\frac{2}{\ln(10) \, \theta_j}
	\sum_{i=1}^{N_d}
	\alpha^i_j
	\left[ \log_{10}(\lambda_i(\theta)) - \Theta_i \right],
	\quad
	\forall j=1, \dots, N_x.
	\end{equation}
	
	\noindent Imposing $\partial C_{\mathrm{eucl}}(\theta) / \partial \theta_j = 0$, $\forall j=1, \dots, N_x$, one obtains
	
	\begin{equation}
	\prod_{i=1}^{N_d}
	\, \theta_1^{\alpha^i_1 \alpha^i_j}
	\, \dots \,	
	\theta_{N_x}^{\alpha^i_{N_x} \alpha^i_j}
	= 
	\hat{\kappa}_j,
	\quad
	\hat{\kappa}_j \equiv
	\prod_{i=1}^{N_d} \kappa^{-\alpha^i_j}_i
	\, 10^{\sum_{i=1}^{N_d} \alpha^i_j \Theta_i},
	\quad
	\forall j=1, \dots, N_x.
	\label{eqn:grad_Ceucl_0_solved}
	\end{equation}	 
	
	\noindent Discarding the units of measure, the change of variables $\theta_j = 10^{\rho_j}$, $\forall j=1, \dots, N_x$, allows rewriting \eqref{eqn:grad_Ceucl_0_solved} as a symmetric linear system,
		
	\begin{equation}
	\left( \sum_{i=1}^{N_d} \alpha^i_1 \alpha^i_j \right)
	\, \rho_1
	\, + \,
	\dots
	\, + \,
	\left( \sum_{i=1}^{N_d} \alpha^i_{N_x} \alpha^i_j \right)
	\, \rho_{N_x}
	=
	\log_{10}(\hat{\kappa}_j),
	\quad
	\forall j=1, \dots, N_x,
	\label{eqn:lin_system_dCeucl_0}	
	\end{equation}
	
	\noindent whose solution provides $\theta_{\mathrm{eucl}} = \left\{ 10^{\rho_j} \right\}_{j=1}^{N_x}$.
	
	Here $\theta_{\mathrm{eucl}}$ is the point of minimum of the cost function $C_{\mathrm{eucl}}(\theta)$. As has been noticed previously, $C_{\mathrm{eucl}}$ is not a unique choice for the cost function. In the following sections, we shall demonstrate that the performance of the Optimal Scaling method does not depend on a choice of the cost function. Such an observation can be explained by the equivalence of norms on the finite dimensional space $(0,\infty)^{N_x}$ \cite{doi:10.1137/1.9781611971446}. Nevertheless, the availability of an analytical solution makes $C_{\mathrm{eucl}}$ the metric of choice for Optimal Scaling.
	
	The following subsections analyse and test the Optimal Scaling procedure on three case studies: the Projectile Problem \cite{Barenblatt1996_book}, the Schr\"{o}dinger Equation for an hydrogen atom in a magnetic field \cite{Fassbinder_1996_Hmagn} and the Landau-de Gennes model for liquid crystals \cite{Gartland_2018_liquid_cryst}.

\subsection{Case Study I: the Projectile Problem}
\label{sec:scaling_projectile}

	In the projectile problem described in \cite{Barenblatt1996_book}, a ball is thrown vertically from the surface of the Earth. The height $x(t)$ above ground reached at time $t$ is measured in meters [m], while the elapsed time $t$ in seconds [s]. As shown in \cite{Barenblatt1996_book}, the variable $x(t)$ satisfies the dimensional Ordinary Differential Equation (ODE) 
	
	\begin{equation}
	\frac{d^2x}{dt^2} = - \frac{g \, R^2}{(x+R)^2},
	\quad
	x(0) = 0,
	\quad
	\frac{dx}{dt}(0) = v_0,
	\label{eqn:dimensional_ODE_projectile}
	\end{equation}
	
	\noindent with $g=9.8$ m$/$s$^2$ the gravitational acceleration, $ R = 6.3781 \times 10^6$ m the radius of the Earth and $v_0 = 25$ m$/$s the initial velocity. The dimensionless counterpart of ODE \eqref{eqn:dimensional_ODE_projectile} is obtained by the change of variables
	
	\begin{equation}
	t = t_c \, \tau,
	\quad
	x = x_c \, \xi,
	\label{eqn:change_of_var_projectile}
	\end{equation}	  

	\noindent where $t_c$ [s] and $x_c$ [m] belong to the set of the strictly positive characteristic constants $\theta$ defined in \eqref{eqn:change_of_var}. The transformation \eqref{eqn:change_of_var_projectile} leads to
	
	\begin{equation}
	\frac{d^2 \xi}{d\tau^2} = - \frac{\lambda_1}{(1 + \lambda_2 \xi)^2},
	\quad
	\xi(0) = 0,
	\quad
	\frac{d\xi}{d\tau}(0) = \lambda_3,
	\label{eqn:dimensioless_ODE_projectile}
	\end{equation}
	
	\noindent with the dimensionless coefficients $\lambda_1(\theta), \lambda_2(\theta), \lambda_3(\theta) > 0$ defined as
	
	\begin{equation}
	\lambda_1(\theta) \equiv \frac{g \, t_c^2}{x_c},
	\quad
	\lambda_2(\theta) \equiv \frac{x_c}{R},
	\quad
	\lambda_3(\theta) \equiv \frac{t_c \, v_0}{x_c},
	\quad
	\theta \equiv \{t_c,x_c\}.
	\label{eqn:lambdas_def_projectile}
	\end{equation}
	
	\noindent The scaling \eqref{eqn:change_of_var_projectile} results in $N_d=3$ dimensionless coefficients $\lambda \equiv \{ \lambda_1,\lambda_2,\lambda_3 \}$, computed as functions of $N_x=2$ characteristic constants $\theta \equiv \{t_c,x_c\}$. Several methods can be applied to assign suitable numerical values to coefficients $\lambda$ and constants $\theta$. In particular, \autoref{tab:scaling_projectile} reports a comparison between the Optimal Scaling procedure, with two different choices of the cost function $C(\theta)$ (Methods (d)-(e)), and the approach presented in  \cite{Holmes2009_ND} imposing $N_x=2$ coefficients $\lambda$ equal to 1, $\lambda_i(\theta)=\lambda_j(\theta)=1$, $\forall i \neq j \in \{1,2,3\}$ (Methods (a)-(c)).
	
	As remarked above, there is no control on the magnitude of $N_d-N_x=1$ coefficient $\lambda$, if $N_x=2$ coefficients $\lambda$ are set to 1. In fact, \autoref{tab:scaling_projectile} shows that the coefficients $\lambda$ differ up to \emph{five} orders of magnitude, if Methods (a)-(c) are applied. However, the Optimal Scaling procedure is able to reduce such difference to up to \emph{two} orders of magnitude, once the Euclidian cost function is employed, see \autoref{tab:scaling_projectile}, Method (d). To give an insight to such a behaviour, it is possible to compare the solutions for coefficients $\lambda$, provided by Optimal Scaling (Method (d)), with the corresponding results obtained using Methods (a), (b) and (c). The Optimal Scaling solutions can be expressed as the powers $\alpha=1/6,1/6,2/3$ of corresponding solutions of Methods (a), (b) and (c). Since $\alpha \in [0,1)$, the values of coefficients $\lambda$ would be closer to 1 as long as Optimal Scaling with $C_{\mathrm{eucl}}(\theta)$ (Method (d)) is applied instead of Methods (a), (b) and (c).
	
	Finally, we remark that the cost functions $C_{\mathrm{eucl}}(\theta)$ and $C_{\max}(\theta)$ provide similar solutions
	as seen from the numerical values of $\theta_1=t_c$ and $\theta_2=x_c$ reported in \autoref{tab:scaling_projectile}, Methods (d)-(e). In addition, the scaling factors $\theta_1,\theta_2$, found by the Optimal Scaling procedure (Methods (d)-(e)), are close to the ones obtained by Method (a).
						
	\begin{table}[!h]
    \footnotesize
    \centering
	\begin{tabular}{ | p{2.6cm} | p{0.8cm} p{1.25cm} | p{0.8cm} p{1.25cm} | p{0.8cm} p{1.25cm} | p{0.8cm} p{1.5cm} | p{0.8cm} p{1.5cm} |}	
	
	\hline
	
	\multirow{2}{*}{\textbf{Method}} &
    \multicolumn{2}{c}{$\theta_1 = t_c$} &
    \multicolumn{2}{c}{$\theta_2 = x_c$} &	
    \multicolumn{2}{c}{$\lambda_1$} &
    \multicolumn{2}{c}{$\lambda_2$} &
    \multicolumn{2}{c|}{$\lambda_3$} \\	
    
    	& \scriptsize{\emph{Solution}} & \scriptsize{\emph{Value} [s]} 
	& \scriptsize{\emph{Solution}} & \scriptsize{\emph{Value} [m]}
	& \scriptsize{\emph{Solution}} & \scriptsize{\emph{Value}} 
	& \scriptsize{\emph{Solution}} & \scriptsize{\emph{Value}}
	& \scriptsize{\emph{Solution}} & \scriptsize{\emph{Value}} \\    
    
    	\hline
    	
    	\rowcolor[HTML]{EFEFEF}
   	
   	(a) $\lambda_{1,2}(\theta)=1$ &
   	$\sqrt{\frac{R}{g}}$ & $8.1 \times 10^2$ &
   	$R$ & $6.4  \times 10^6$ &
   	& $1.0 \times 10^0$ &
   	& $1.0 \times 10^0$ & 
   	$ \sqrt{ \frac{v_0^2}{g \, R} }$ & $3.2 \times 10^{-3}$ \\
   		
	(b) $\lambda_{2,3}(\theta)=1$ &
   	$\frac{R}{v_0}$ & $2.6 \times 10^5$ &
   	$R$ & $6.4  \times 10^6$ &
   	$ \frac{g \, R}{v_0^2} $ & $1.0 \times 10^5$ &
   	& $1.0 \times 10^0$ &
   	& $1.0 \times 10^0$ \\
   	
   	\rowcolor[HTML]{EFEFEF}   	
   	
	(c) $\lambda_{1,3}(\theta)=1$ &
   	$\frac{v_0}{g}$ & $2.6 \times 10^0$ &
   	$\frac{v_0^2}{g}$ & $6.4 \times 10^1$ &
   	& $1.0 \times 10^0$ &
   	$\frac{v_0^2}{g \, R}$ & $1.0 \times 10^{-5}$  &
   	& $1.0 \times 10^0$ \\  
   	
   	(d) $\min_\theta C_{\mathrm{eucl}}(\theta)$ &
   	$\sqrt{\frac{R}{g}}$ & $8.1 \times 10^2$ &	
   	$\sqrt[6]{\frac{R^5 \, v_0^2}{g}}$ & $9.4 \times 10^5$ &
   	$\sqrt[6]{\frac{g \, R}{v_0^2}}$ & $6.8 \times 10^0$ &
   	$\sqrt[6]{\frac{v_0^2}{g \, R}}$ & $1.5 \times 10^{-1}$  &
   	$\sqrt[3]{\frac{v_0^2}{g \, R}}$ & $2.2 \times 10^{-2}$ \\ 	 	
   	  	   	
	\rowcolor[HTML]{EFEFEF}   	
   	

	(e) $\min_\theta C_{\mathrm{max}}(\theta)$ &
   	- & $7.3 \times 10^2$ &
   	- & $3.7 \times 10^5$ &
   	- & $1.4 \times 10^1$ &
   	- & $5.8 \times 10^{-2}$ &
   	- & $5.0 \times 10^{-2}$  \\     	  	   	

	\hline   	
   	   	
	\end{tabular}
	\caption{The Projectile Problem. Analytical solutions and numerical values of constants $\theta \equiv \{\theta_1=t_c,\theta_2=x_c\}$, with corresponding dimensionless coefficieints $\lambda_1,\lambda_2,\lambda_3$ \eqref{eqn:lambdas_def_projectile}, for scaling the dimensional ODE \eqref{eqn:dimensional_ODE_projectile}. The following scaling methods are compared: Methods (a), (b), (c) impose $\lambda_i(\theta)=\lambda_j(\theta)=1$, $\forall i \neq j \in \{1,2,3\}$, and Methods (d), (e) find the argument of minimum of the cost functions $C_{\mathrm{eucl}}(\theta)$ \eqref{eqn:C_eucl_def} and $C_{\mathrm{max}}(\theta)$ \eqref{eqn:C_max}. The minimisation of $C_{\mathrm{eucl}}$ is performed by means of the analytical solution derived in \autoref{sec:anal_sol_min_Ceucl}. 
	The R function \emph{optim} \cite{stats_Rpackage}, with the method SANN for Simulated Annealing Algorithm \cite{10.2307/3214721}, is applied to minimise $C_{\mathrm{max}}$. The numerical routine is stopped when the maximum number $10^5$ of allowed evaluations of the cost function is exceeded. The simulation requires $\approx 0.3$ sec of computation on a 2.70GHz processor.}
	\label{tab:scaling_projectile}
    \end{table}	 
    	 	
\begin{figure}[!h]
	\centering
	\begin{subfigure}[h]{.495\linewidth}
	\centering
	\includegraphics[scale=0.44]{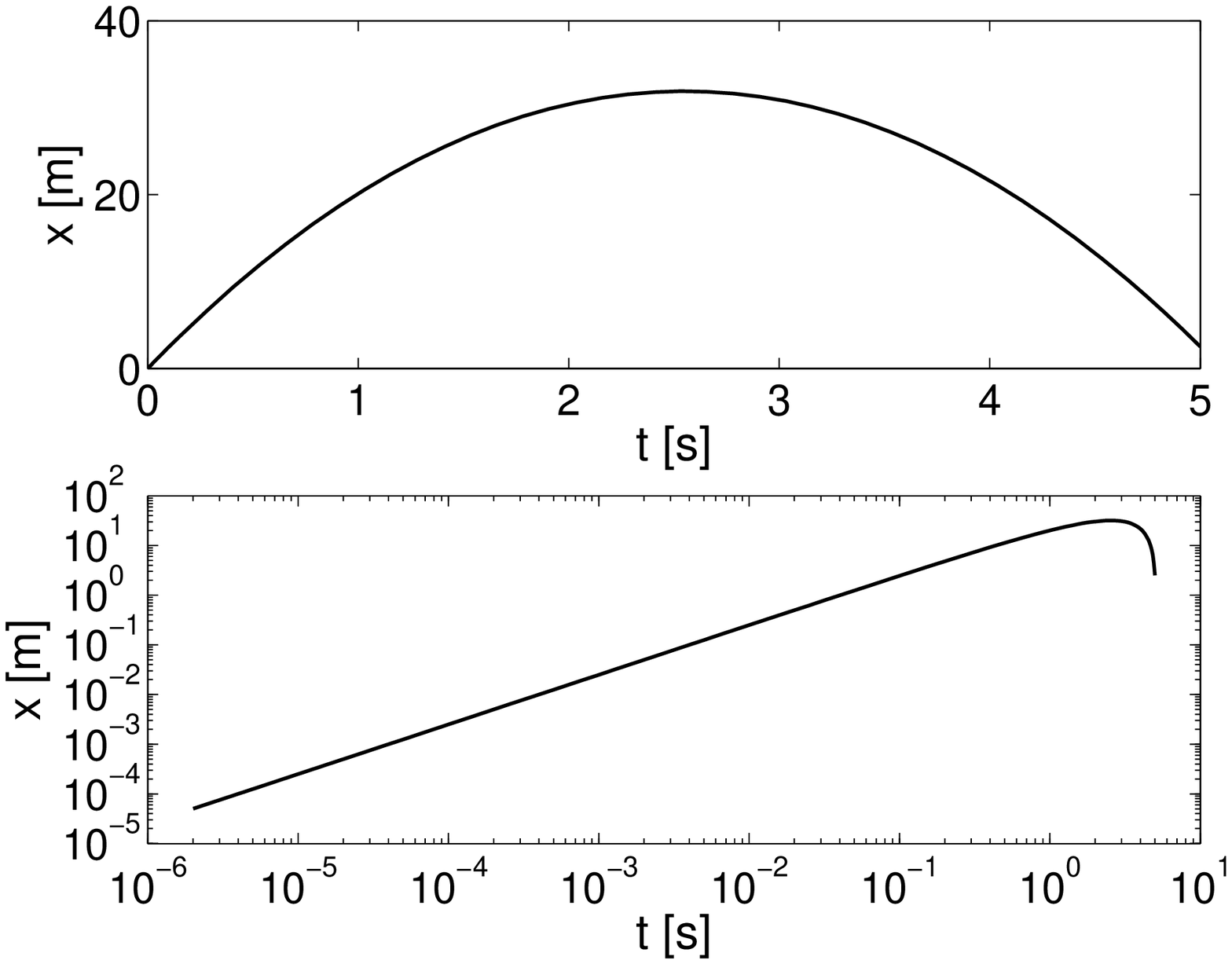}
	\caption{Solution $x(t)$ of dimensional ODE \eqref{eqn:dimensional_ODE_projectile}.}
	\label{fig:sol_dimensional_projectile}
	\end{subfigure}
	\begin{subfigure}[h]{.495\linewidth}
	\centering
	\includegraphics[scale=0.44]{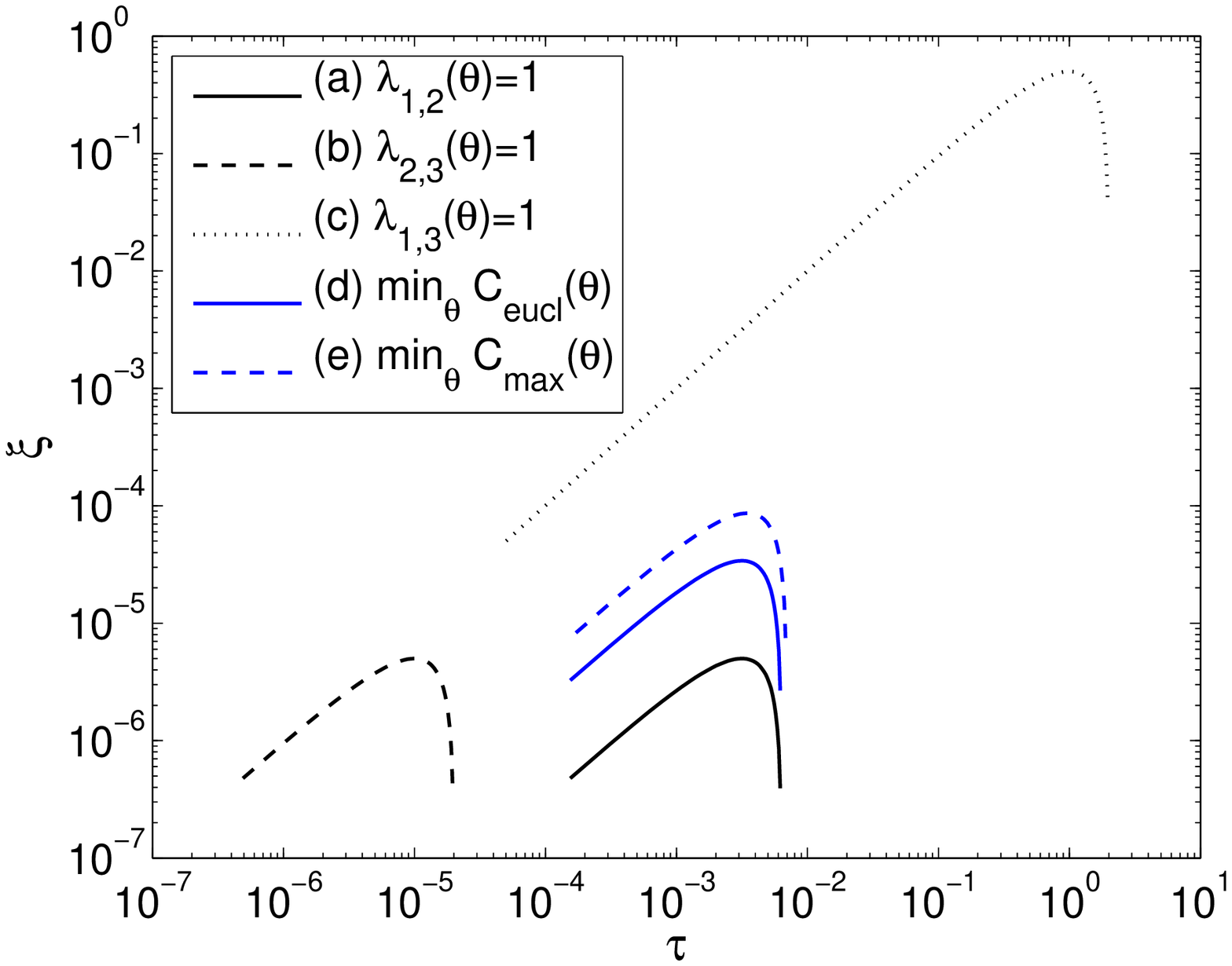}
	\caption{Solutions $\xi(\tau)$ of dimensionless ODE \eqref{eqn:dimensioless_ODE_projectile}.}
	\label{fig:sol_dimensionless_projectile}
	\end{subfigure}	
	\caption{The Projectile Problem. (i) Solution $x(t)$ of the dimensional ODE \eqref{eqn:dimensional_ODE_projectile} for $t \in [0,t_{\max}]$, with $t_{\max} = 5$ s (top). The same values of $x(t)$ are also shown in logarithmic scale (bottom). (ii) Solutions $\xi(\tau)$, $\tau \in [0,\tau_{\max}]$, $\tau_{\max} \equiv t_{\max} / t_c$, of the dimensionless ODE \eqref{eqn:dimensioless_ODE_projectile}, corresponding to Methods (a)-(e) reported in \autoref{tab:scaling_projectile}. The initial data $x(0)=0$ and $\xi(0)=0$ are not drawn in the logarithmic scale. The solutions of ODEs \eqref{eqn:dimensional_ODE_projectile} and \eqref{eqn:dimensioless_ODE_projectile} are computed by the MATLAB routine \emph{ode45} \cite{ode45_Matlab}, requiring fractions of a second of computation on a 2.70GHz processor.}
	\label{fig:solutions_ODE_projectile}
	\end{figure}

    	While \autoref{tab:scaling_projectile} reports different methodologies to transform the dimensional ODE \eqref{eqn:dimensional_ODE_projectile} into the corresponding dimensionless model \eqref{eqn:dimensioless_ODE_projectile}, the effect of the performed scalings is better visible when comparing the solutions of the arising ODEs, as shown in \autoref{fig:solutions_ODE_projectile}. In particular, \autoref{fig:sol_dimensionless_projectile} suggests that the Optimal Scaling procedure (blue lines, Methods (d)-(e)) maps the solution of the dimensional ODE \eqref{eqn:dimensional_ODE_projectile}, shown in \autoref{fig:sol_dimensional_projectile}, to values of $\xi$ and $\tau$ averaging the solutions obtained by Methods (a), (b) and (c) (\autoref{fig:sol_dimensionless_projectile}, black lines). In this sense, the Optimal Scaling procedure is able to mitigate the extreme values Methods (a), (b) and (c) may lead to. Such statement is in agreement with the behaviour of coefficients $\lambda$, reported in \autoref{tab:scaling_projectile}.
		
\begin{figure}[!h]
	\centering
	\begin{subfigure}[h]{.32\linewidth}
	\centering
	\includegraphics[scale=0.33]{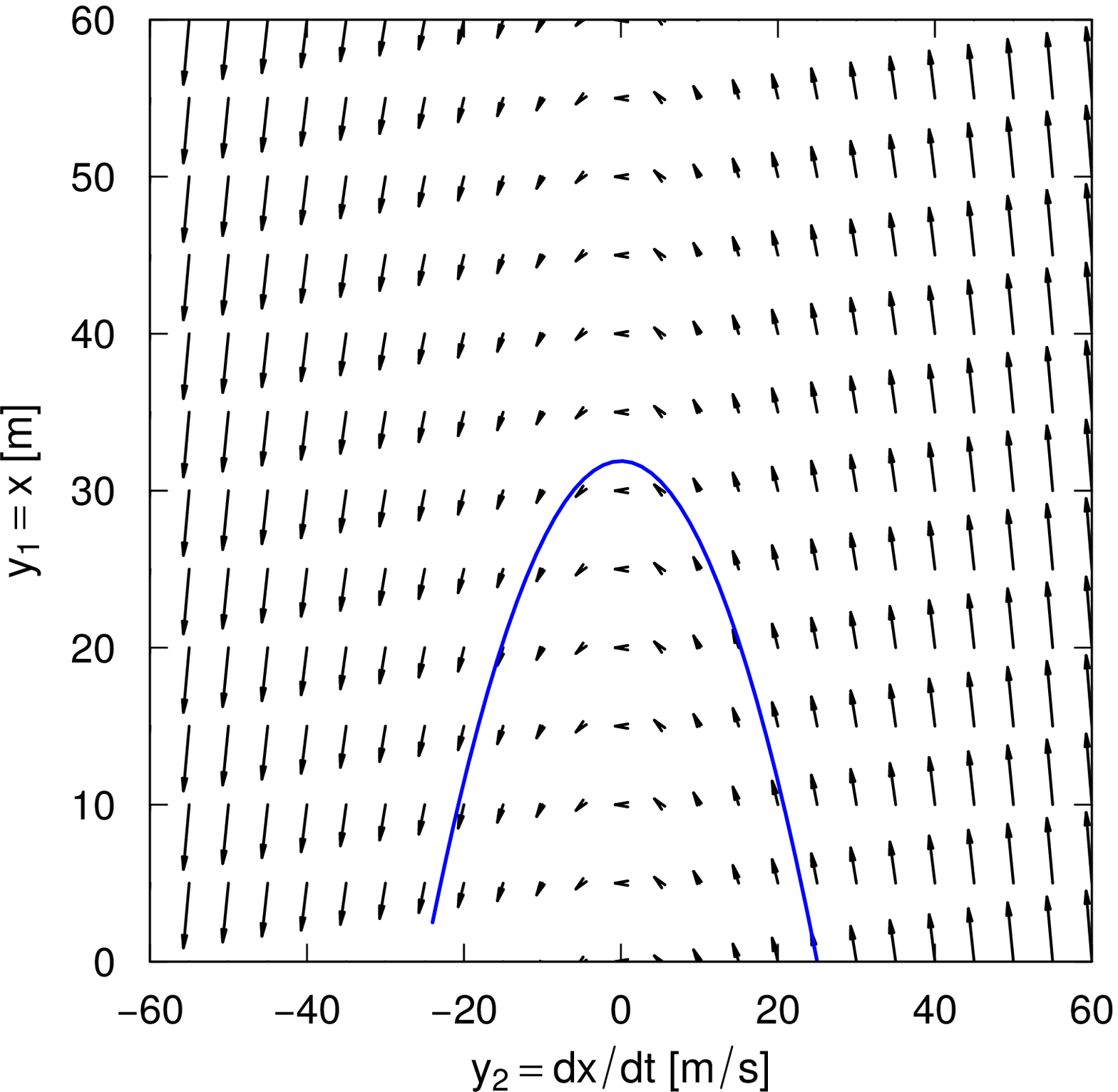}
	\captionsetup{width=0.85\linewidth}	
	\caption{Flow in phase space of dimensional ODE \eqref{eqn:dimensional_ODE_projectile}.}
	\label{fig:flows_projectile_dimensional}	
	\end{subfigure}
	\begin{subfigure}[h]{.32\linewidth}
	\centering
	\includegraphics[scale=0.33]{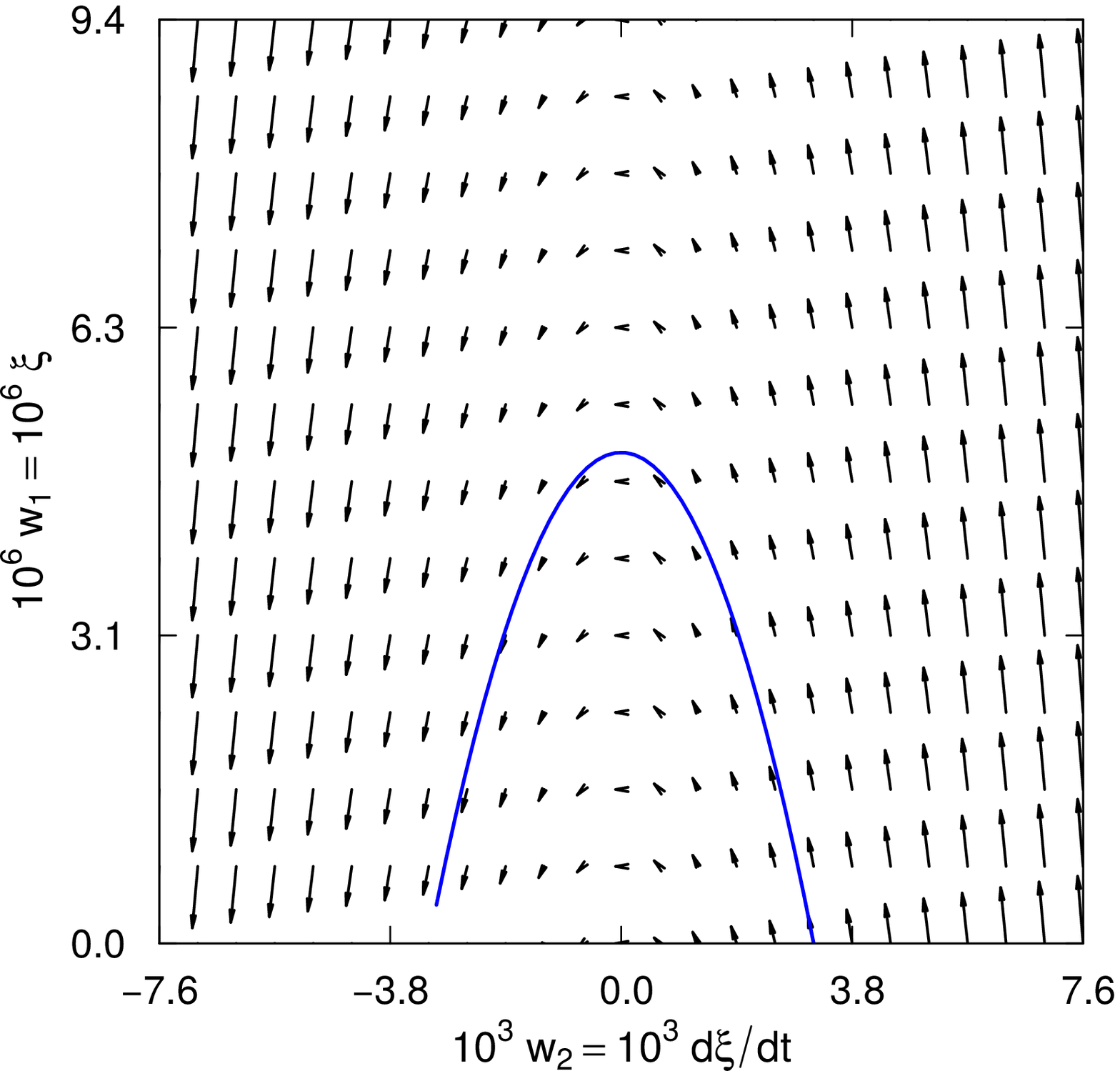}
	\captionsetup{width=0.85\linewidth}
	\caption{Flow in phase space of \eqref{eqn:dimensioless_ODE_projectile} for (a) $\lambda_{1,2}=1$.}
	\end{subfigure}	
	\begin{subfigure}[h]{.32\linewidth}
	\centering
	\includegraphics[scale=0.33]{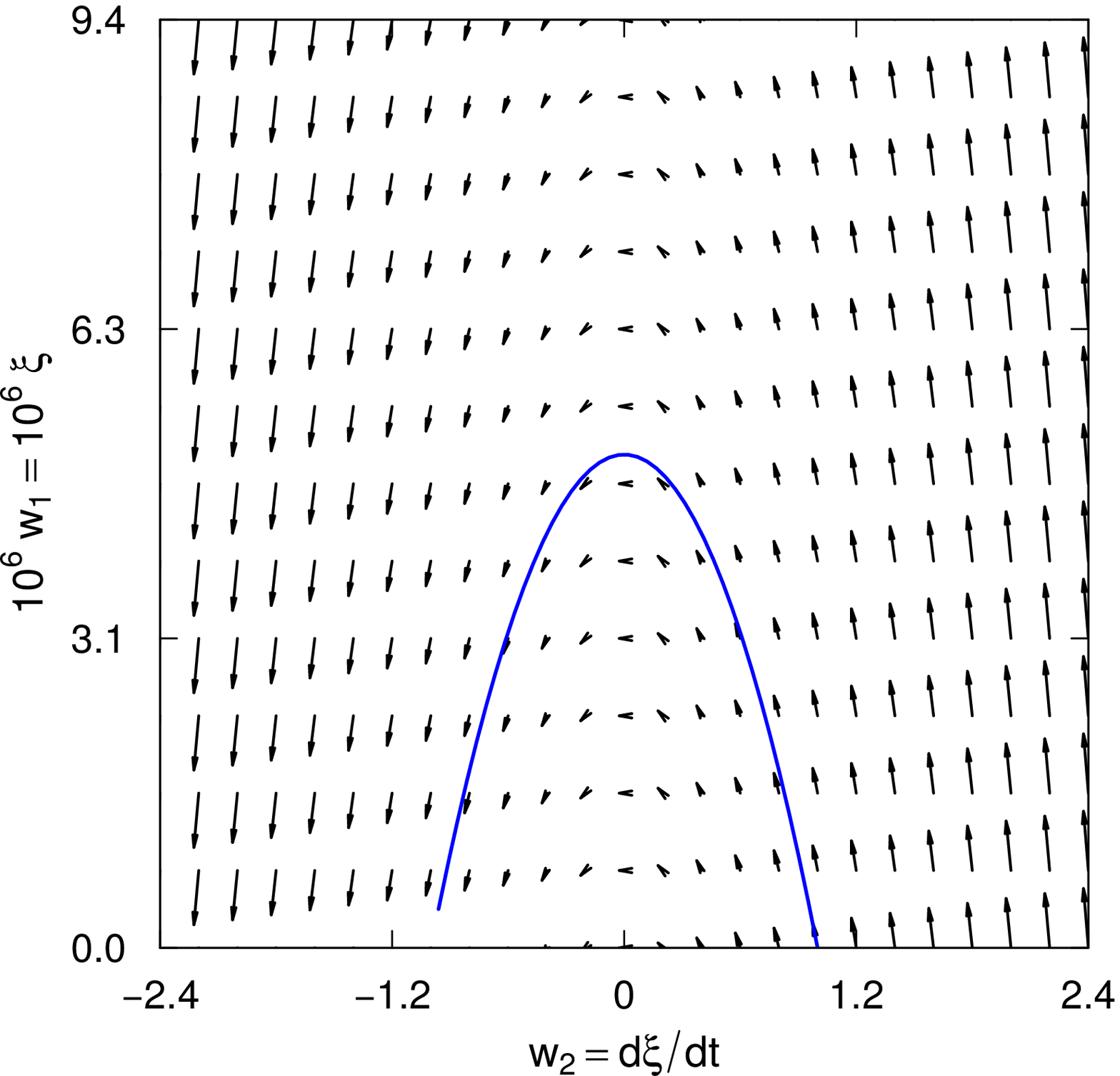}
	\captionsetup{width=0.85\linewidth}
	\caption{Flow in phase space of \eqref{eqn:dimensioless_ODE_projectile} for (b) $\lambda_{2,3}=1$.}
	\end{subfigure}		
	\begin{subfigure}[h]{.32\linewidth}
	\centering
	\includegraphics[scale=0.33]{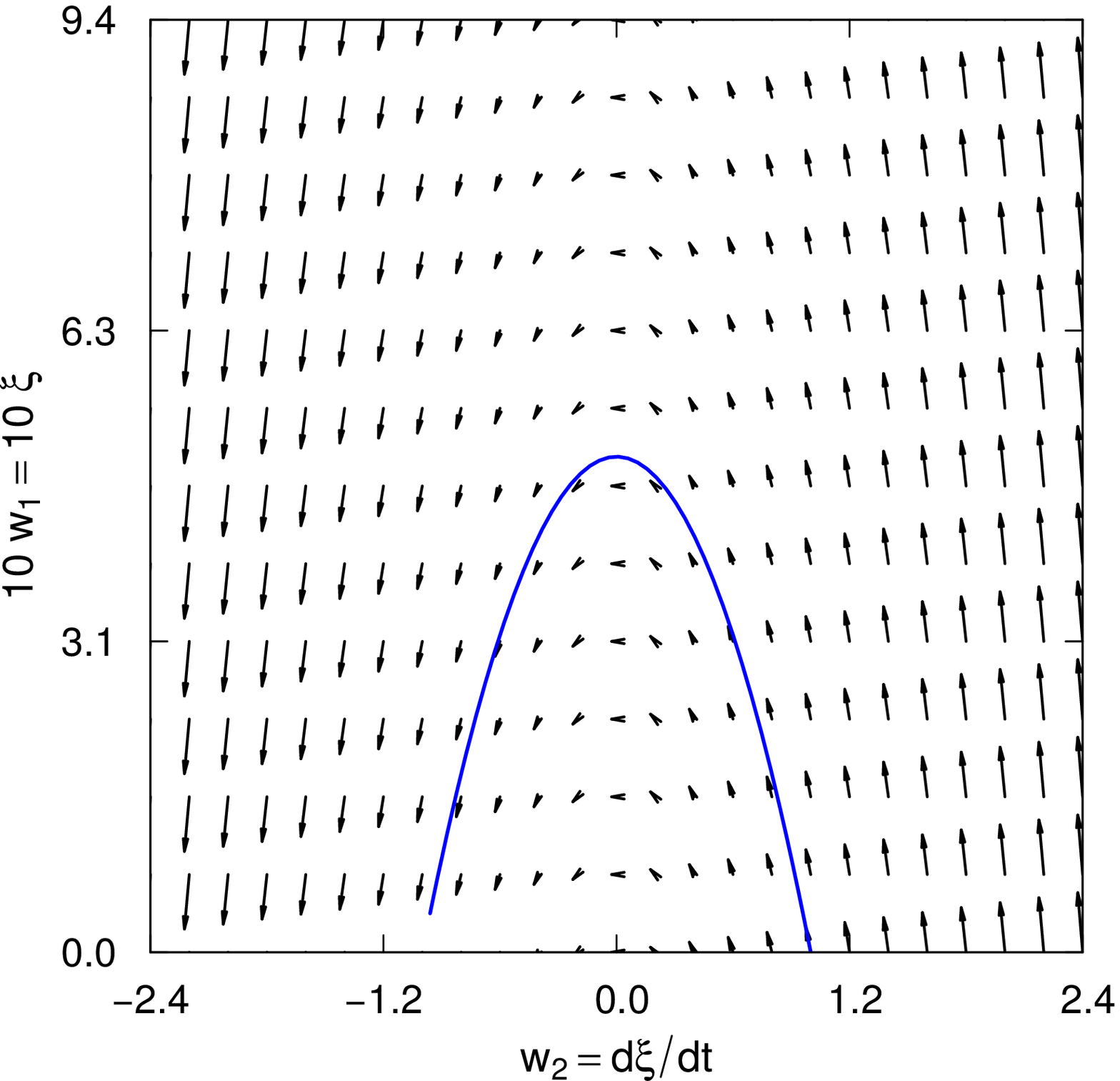}
	\captionsetup{width=0.85\linewidth}
	\caption{Flow in phase space of \eqref{eqn:dimensioless_ODE_projectile} for (c) $\lambda_{1,3}=1$.}
	\end{subfigure}
	\begin{subfigure}[h]{.32\linewidth}
	\centering
	\includegraphics[scale=0.33]{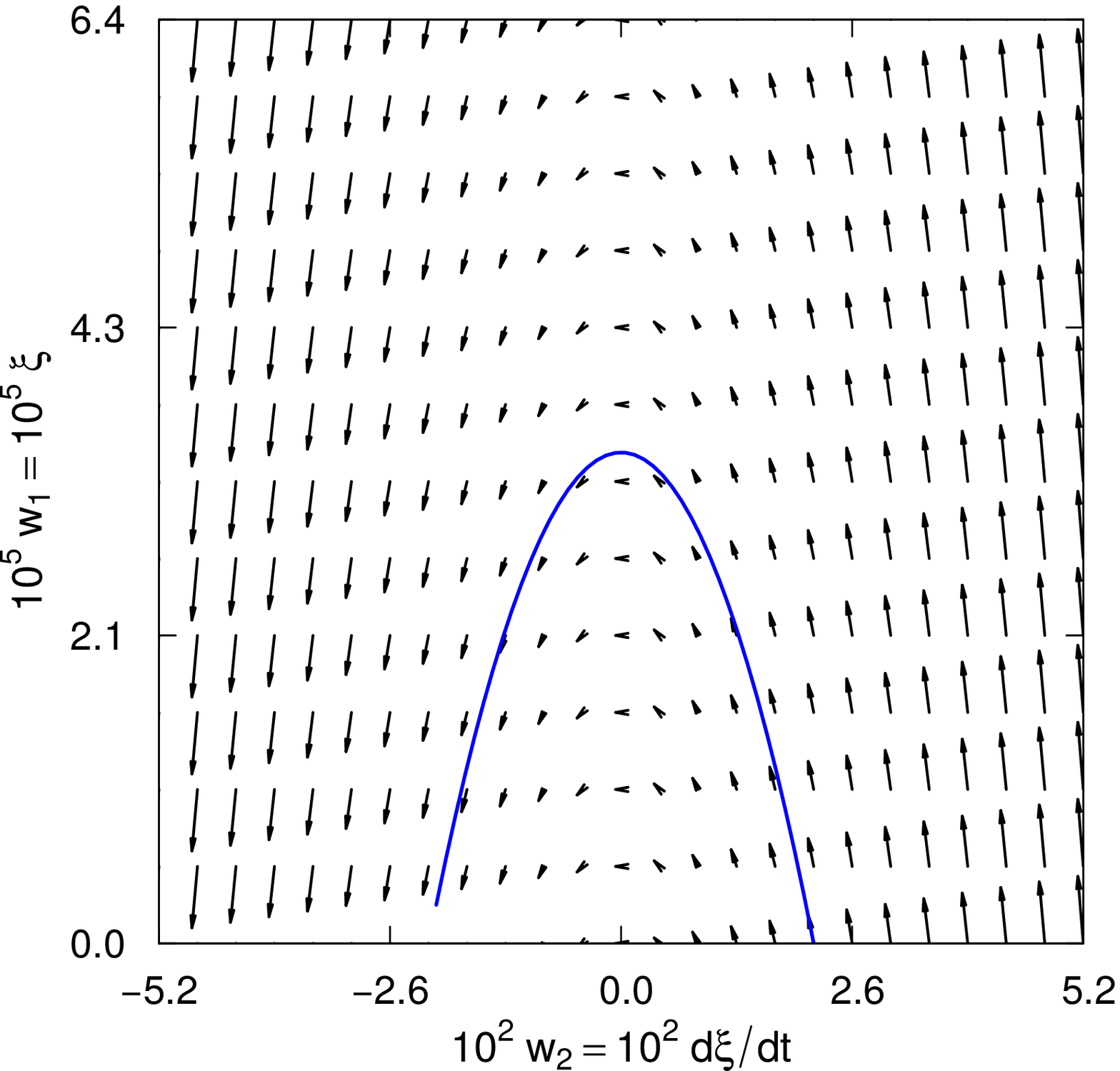}
	\captionsetup{width=0.85\linewidth}
	\caption{Flow in phase space of \eqref{eqn:dimensioless_ODE_projectile} for (d) $\min_\theta C_{\mathrm{eucl}}(\theta)$.}
	\end{subfigure}		
	\begin{subfigure}[h]{.32\linewidth}
	\centering
	\includegraphics[scale=0.33]{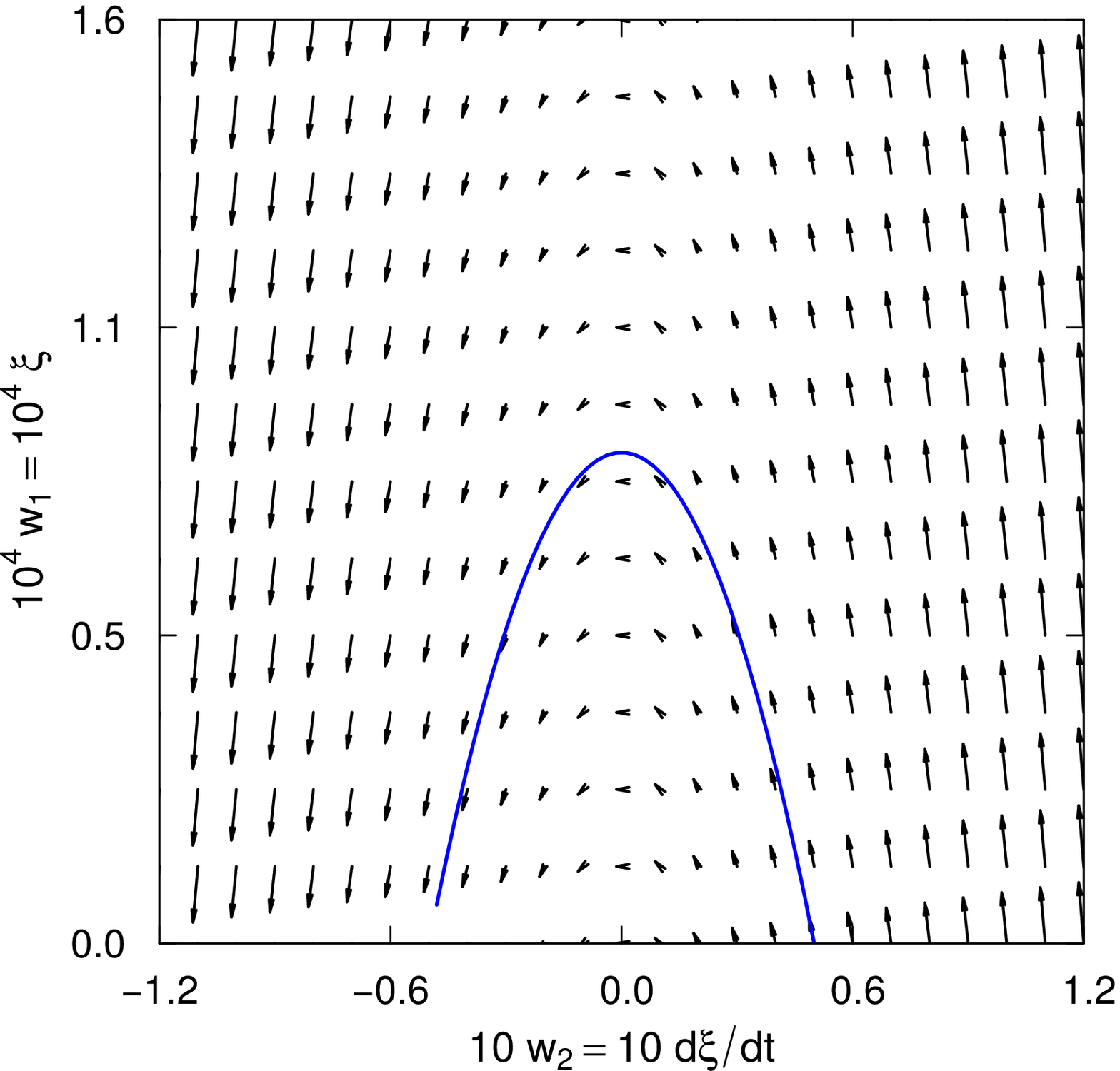}
	\captionsetup{width=0.85\linewidth}
	\caption{Flow in phase space of \eqref{eqn:dimensioless_ODE_projectile} for (e) $\min_\theta C_{\mathrm{max}}(\theta)$.}
	\end{subfigure}				
	\caption{The Projectile Problem. Flows in phase spaces of ODEs \eqref{eqn:dimensional_ODE_projectile} and \eqref{eqn:dimensioless_ODE_projectile}, corresponding to Methods (a)-(e) of \autoref{tab:scaling_projectile}. The arrows show the vectors \eqref{eqn:tangent_vectors_projectile} in the intervals \eqref{eqn:domains_phase_space_projectile}, with $x_{\max} = 60$ m and $v_{\max} = 60$ m/s. The blue lines provide the solutions of \eqref{eqn:dimensional_ODE_projectile} and \eqref{eqn:dimensioless_ODE_projectile} reported in \autoref{fig:solutions_ODE_projectile}.}
	\label{fig:flows_phase_spaces_projectile}
	\end{figure}

	The effect of the scaling procedures can be also analysed by inspecting the flow in the phase spaces of position and momentum of the ODEs \eqref{eqn:dimensional_ODE_projectile} and \eqref{eqn:dimensioless_ODE_projectile}. Given $y_1 = x$, $y_2 = dx/dt$, $w_1 = \xi$ and $w_2 = d\xi/d\tau$, 
	the flow is represented in \autoref{fig:flows_phase_spaces_projectile} by the tangent vectors with components $(y_1',y_2')$ and $(w_1',w_2')$:
	
	\begin{equation}
	\begin{cases}	
	y_1' & = y_2, \\
	y_2' & = - g \, R^2 / ( y_1 + R )^2,
	\end{cases}
	\quad
	\begin{cases}	
	w_1' & = w_2, \\
	w_2' & = - \lambda_1 / ( 1 + \lambda_2 \, w_1 )^2.
	\end{cases}
	\label{eqn:tangent_vectors_projectile}
	\end{equation}	  
	
	\noindent The vectors \eqref{eqn:tangent_vectors_projectile} are shown in \autoref{fig:flows_phase_spaces_projectile} for
		
	\begin{equation}
	y_1 \in [0,x_{\max}],
	\quad
	y_2 \in [-v_{\max},v_{\max}],
	\quad
	w_1 \in [0,x_{\max}/x_c],
	\quad
	w_2 \in [ - (t_c/x_c) v_{\max}, (t_c/x_c) v_{\max} ],
	\label{eqn:domains_phase_space_projectile}
	\end{equation}
	
	\noindent such that the intervals \eqref{eqn:domains_phase_space_projectile} are of the scale of the solutions of \eqref{eqn:dimensional_ODE_projectile} and \eqref{eqn:dimensioless_ODE_projectile} (\autoref{fig:flows_phase_spaces_projectile}, blue lines). In other words, $x_{\max}$ and $v_{\max}$ in \eqref{eqn:domains_phase_space_projectile} are chosen to be of the scale of the solution of the dimensional model \eqref{eqn:dimensional_ODE_projectile} (\autoref{fig:flows_projectile_dimensional}, blue line). Then the flows of scaled models \eqref{eqn:dimensioless_ODE_projectile} are shown in \autoref{fig:flows_phase_spaces_projectile} for the properly scaled corresponding intervals \eqref{eqn:domains_phase_space_projectile}. \autoref{fig:flows_phase_spaces_projectile} suggests that the flow is not deformed by the compared scaling procedures, preserving the qualitative behaviour of the flux in the different scales.
			
\subsection{Case Study II: the Schr\"{o}dinger Equation}
\label{sec:scaling_schrodinger}

	Aimed to demonstrate the versatility of Optimal Scaling procedure, we test the proposed methodology on the Schr\"{o}dinger Equation (SE) for an hydrogen atom in a magnetic field \cite{Fassbinder_1996_Hmagn}. Even if its solution is beyond the scope of this paper, SE provides an illustrative case study, as discussed below.
	
	Assume a constant magnetic field along a direction indicated by the dimensionless unit vector $\vec{n} \in \mathbb{R}^3$, i.e $\vec{B} = B \vec{n}$, with $B$ expressed in Tesla [T]. Despite its simplicity, this model is rich in effects, which include classical chaos \cite{Fassbinder_1996_Hmagn}, localisation of wave functions \cite{0953-4075-21-20-001}, irregular spectra \cite{0022-3700-6-9-002}, etc. Then, the SE for the atom's electron (spin omitted) is given by
	
	\begin{equation}
	i \hbar \partial_t \psi(\vec{r},t) 
	=
	- \frac{\hbar^2}{2\mu} \nabla^2 \psi
	- i \frac{ \hbar e B }{ 2 \mu }
	( \vec{n} \cdot \vec{r} \times \vec{\nabla} \psi )
	+ \frac{ e^2 B^2 }{ 8  \mu }
	[ r^2 - (\vec{r} \cdot \vec{n})^2 ] \psi
	- \frac{e^2}{4 \pi \epsilon r} \psi.
	\label{eqn:dimensional_SE}
	\end{equation}
	
	\noindent The position $\vec{r} \in \mathbb{R}^3$ is measured in meters [m] and the elapsed time $t$ in seconds [s], with $r \equiv \sqrt{\vec{r} \cdot \vec{r}}$. The wave function $\psi$, expressed in m$^{-3/2}$, should satisfy the normalisation condition
	
	\begin{equation}
	\int \! |\psi(\vec{r},t)|^2 \, d\vec{r} = 1,
	\label{eqn:Dimensional_Norm_Cond_SE}
	\end{equation}
	
	\noindent and the constants in \eqref{eqn:dimensional_SE} are defined in \autoref{tab:SE_nomen_def_exp_val}. 
		
	\begin{table}[!h]
	\centering
	\begin{tabular}[t]{lll}

	\hline
	\textbf{Nomenclature} & \textbf{Definition} & \textbf{Value} \\
	\hline

	\rowcolor[HTML]{EFEFEF}	
	$\hbar$ & Reduced Planck's constant & $1.05457172647 \times 10^{-34}$ J s \\
	
	$\mu$ & Electron's mass & $9.1093829140 \times 10^{-31}$ kg\\

	\rowcolor[HTML]{EFEFEF}	
	$e$ & Electron's charge & $1.60217656535 \times 10^{-19}$ C\\
	
	$B$ & Magnetic flux intensity & $45$ T  \\
	 	
	\rowcolor[HTML]{EFEFEF}	
	$( 4 \pi \epsilon )^{-1}$ & Coulomb force constant & $8.987551787368 \times 10^9$ J m C$^{-2}$ \\
	
	\hline			

	\end{tabular}
	\caption{The Schr\"{o}dinger Equation. Nomenclature and definition of parameters in the Schr\"{o}dinger Equation \eqref{eqn:dimensional_SE}, with corresponding values. The value of $B$ is chosen as a current (2015) world record for continuous field magnets (National High Magnetic Field Laboratory, USA). The symbols J, s, kg, C, T and m stand for Joule, second, kilogram, Coulomb, Tesla and meter respectively.}
	\label{tab:SE_nomen_def_exp_val}
	\end{table}
	
	\noindent The SE \eqref{eqn:dimensional_SE} can be written in terms of dimensionless variables by means of the transformation
	
	\begin{equation}
	\vec{r} = \alpha_0 \, \vec{\rho},
	\quad
	t = \beta_0 \, \tau,
	\quad
	\psi = \gamma_0 \, \phi,
	\label{eqn:change_of_var_SE}
	\end{equation}	  

	\noindent with $\alpha_0$ [m], $\beta_0$ [s] and $\gamma_0$ [m$^{-3/2}$] being the strictly positive characteristic constants $\theta$ defined in \eqref{eqn:change_of_var}. By plugging \eqref{eqn:change_of_var_SE} in \eqref{eqn:dimensional_SE} and \eqref{eqn:Dimensional_Norm_Cond_SE}, we have
	
	\begin{equation}
	i \partial_{\tau} \phi(\vec{\rho},\tau) 
	=
	- \frac{\lambda_1}{2} \nabla^2 \phi
	- i \lambda_2 ( \vec{n} \cdot \vec{\rho} \times \vec{\nabla} \phi )	
	+ \lambda_3 [ \rho^2 - ( \vec{\rho} \cdot \vec{n} )^2 ] \phi
	- \lambda_4 \frac{\phi}{\rho},
	\quad
	\lambda_5 \int \! |\phi(\vec{\rho},\tau)|^2 \, d\vec{\rho} = 1,
	\label{eqn:dimensioless_SE}
	\end{equation}
	
	\noindent with $\rho \equiv \sqrt{\vec{\rho} \cdot \vec{\rho}}$ and the dimensionless coefficients $\lambda(\theta) > 0$ defined as
	
	\begin{equation}
	\lambda_1(\theta) \equiv \frac{ \hbar \, \beta_0 }{ \mu \, \alpha_0^2 },
	\quad
	\lambda_2(\theta) \equiv \frac{ e \, B \, \beta_0 }{ 2 \, \mu },
	\quad
	\lambda_3(\theta) \equiv 
	\frac{ e^2 \, B^2 \, \alpha_0^2 \, \beta_0 }{ 8 \, \mu \, \hbar },
	\quad
	\lambda_4(\theta) \equiv 
	\frac{ e^2 \, \beta_0 }{ 4 \pi \epsilon \, \hbar \, \alpha_0 },
	\quad
	\lambda_5(\theta) \equiv \alpha_0^3 \, \gamma_0^2,
	\label{eqn:lambdas_def_SE}
	\end{equation}
	
	\noindent for $\theta \equiv \{\alpha_0,\beta_0,\gamma_0\}$. Now, setting $\lambda_1=\lambda_4=\lambda_5=1$ leads to the often used \emph{atomic units} and the scaling factors $\theta = \theta_{\mathrm{atom}}$:
	
	\begin{equation}
	\theta_{\mathrm{atom}} = \{\alpha_0,\beta_0,\gamma_0\},
	\quad
	\alpha_0 = \frac{ 4 \pi \epsilon \, \hbar^2 }{ \mu \, e^2 },
	\quad
	\beta_0 = \frac{ ( 4 \pi \epsilon )^2 \, \hbar^3 }{ \mu \, e^4 },
	\quad
	\gamma_0 = \frac{ \mu^{3/2} \, e^3 }{ ( 4 \pi \epsilon )^{3/2} \, \hbar^3 },
	\label{eqn:theta_atom_def}	
	\end{equation}
	
	\noindent with $\alpha_0 \approx 5.3 \times 10^{-11}$ m roughly the radius of a free hydrogen atom in its ground state, and $\beta_0 \approx 2.4 \times 10^{-17}$ s the corresponding time unit, given the parameters values from \autoref{tab:SE_nomen_def_exp_val}.
	
\begin{figure}[!h]
	\centering
	\begin{subfigure}[h]{.495\linewidth}
	\centering
	\includegraphics[scale=0.46]{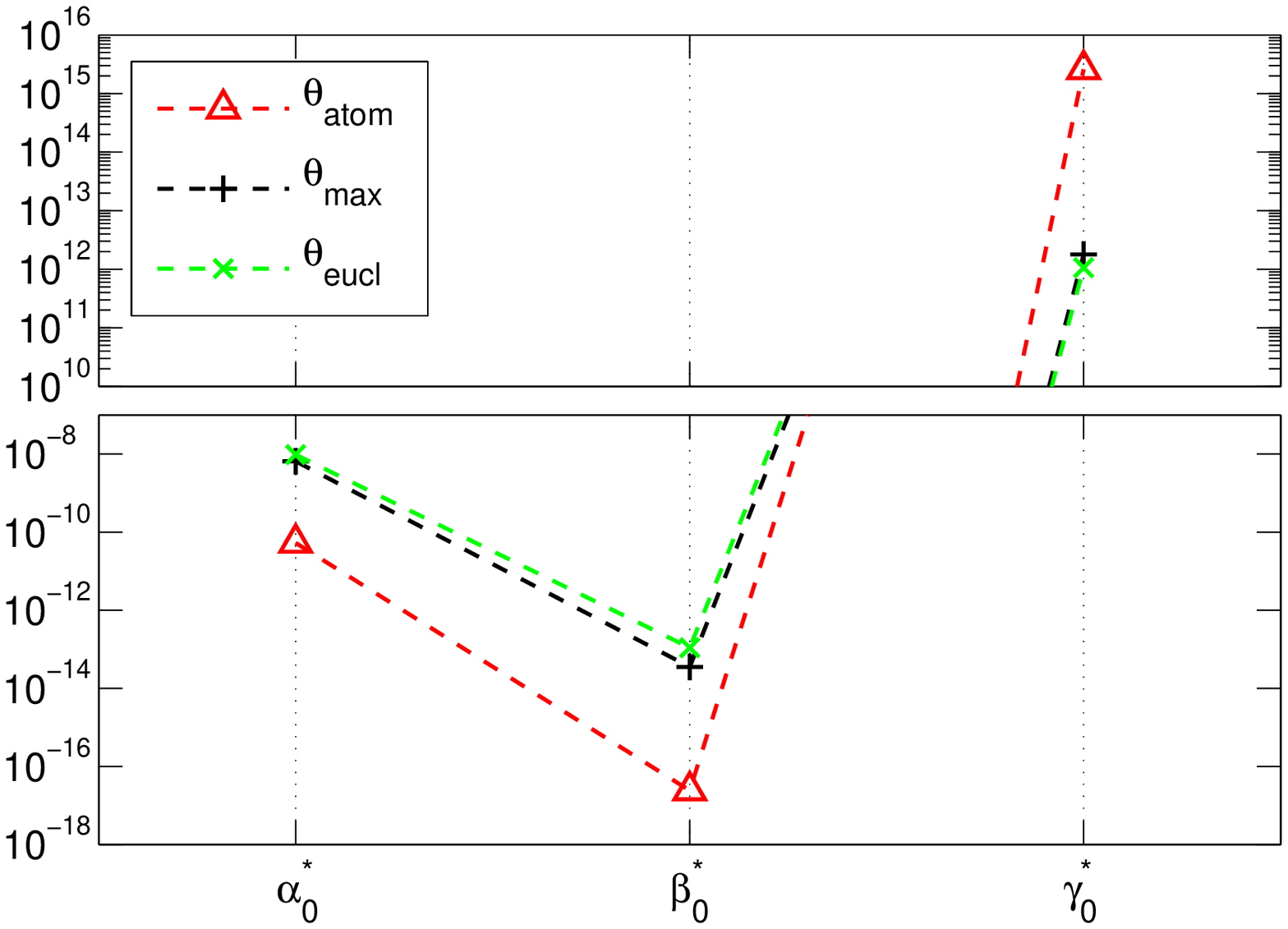}
	\caption{Scaling Factors $\theta^*$.}
	\label{fig:theta_SE}
	\end{subfigure}
	\begin{subfigure}[h]{.495\linewidth}
	\centering
	\includegraphics[scale=0.46]{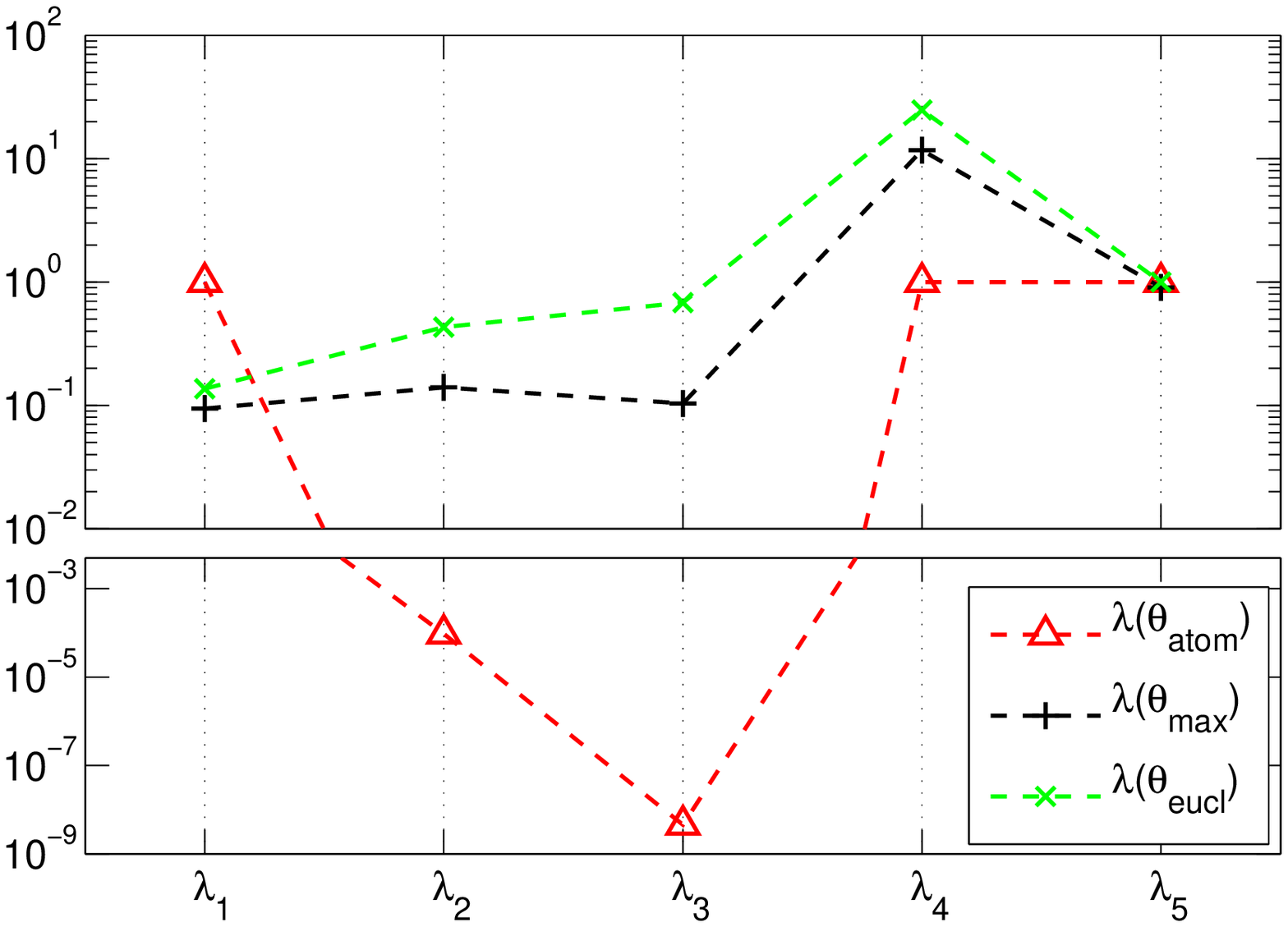}
	\caption{Dimensionless Coefficients $\lambda(\theta^*)$.}
	\label{fig:lambda_SE}
	\end{subfigure}	
	\caption{The Schr\"{o}dinger Equation. (i) Scaling factors $\theta^* \equiv \{ \, \alpha_0^* \equiv \alpha_0 \, / \, c_{\mbox{m}}, \beta_0^* \equiv \beta_0 \, / \, c_{\mbox{s}}, \, \, \gamma_0^* \equiv \gamma_0 \, c_{\mbox{m}}^{3/2} \}$, with $c_{\mbox{m,s}} = 1 \, \mbox{m,s}$, for the SE \eqref{eqn:dimensional_SE}-\eqref{eqn:Dimensional_Norm_Cond_SE} and parameters from \autoref{tab:SE_nomen_def_exp_val}. (ii) Corresponding values of the dimensionless coefficients $\lambda(\theta^*)$ \eqref{eqn:lambdas_def_SE}. The factors $\theta_{\mathrm{atom}}$ [red triangles] are computed as \eqref{eqn:theta_atom_def}, when the \emph{atomic units} are chosen. The factors $\theta^*$ are labelled as $\theta_{\mathrm{eucl}}$ [green crosses] and $\theta_{\max}$ [black crosses], if found by minimisation of the cost function $C_{\mathrm{eucl}}(\theta)$ \eqref{eqn:C_eucl_def} and $C_{\max}(\theta)$ \eqref{eqn:C_max} respectively. The factors $\theta_{\mathrm{eucl}}$ are computed with the analytical formula of \autoref{sec:anal_sol_min_Ceucl}, 
	while the R function \emph{optim} \cite{stats_Rpackage}, with the method SANN \cite{10.2307/3214721}, is applied to minimise $C_{\mathrm{max}}$. The numerical routine is stopped when the maximum number $10^5$ of allowed evaluations of the cost function is exceeded. The simulation requires $\approx 2.7$ sec of computation on a 2.70GHz processor.} 
	\label{fig:scaling_SE}
	\end{figure}

	\autoref{fig:scaling_SE} compares the scaling factors $\theta = \theta_{\mathrm{atom}}$ \eqref{eqn:theta_atom_def}, and corresponding dimensionless coefficients $\lambda(\theta)$ \eqref{eqn:lambdas_def_SE}, with the characteristic constants $\theta = \theta_{\mathrm{eucl}},\theta_{\max}$, found by the Optimal Scaling approach. The numerical values of factors $ \theta_{\mathrm{atom}}, \theta_{\max}$ and $\theta_{\mathrm{eucl}} $ follow a similar trend, as demonstrated in \autoref{fig:theta_SE}. However, the obtained coefficients $\lambda(\theta)$ may lead to different performance of corresponding models. If the atomic factors $\theta=\theta_{\mathrm{atom}}$ \eqref{eqn:theta_atom_def} are chosen, it should be difficult to evaluate the effect of the magnetic field intensity $B$ on the numerical solution $\phi$ of \eqref{eqn:dimensioless_SE}. In particular, variations of $B$, in magnitude comparable to its size, are quenched by the small values of $\lambda_2(\theta_{\mathrm{atom}})$ and $\lambda_3(\theta_{\mathrm{atom}})$, shown in \autoref{fig:lambda_SE}. On the contrary, the Optimal Scaling factors $\theta=\theta_{\mathrm{eucl}},\theta_{\max}$ provide values of $\lambda_{2,3}$ comparable in size with the remaining coefficients $\lambda$, as shown in \autoref{fig:lambda_SE}. As a result, changes in value of $B$ should be better appreciated on the numerical solution $\phi$ of \eqref{eqn:dimensioless_SE}, if the factors $\theta=\theta_{\mathrm{eucl}},\theta_{\max}$ are chosen instead of $\theta=\theta_{\mathrm{atom}}$.
		
\subsection{Case Study III: the Landau-de Gennes Model}
\label{sec:LdGmodel}

	Our next goal is to illustrate the capability of Optimal Scaling (OS) to properly explore limiting behaviours of a given model. In particular, we consider the Landau-de Gennes model, which is commonly used to study properties of nematic liquid crystals \cite{Gartland_2018_liquid_cryst}. As explained in \cite{Gartland_2018_liquid_cryst}, such a model can properly describe the \emph{large-body limit}, achieved when geometric length scales (such as the size of the domain containing the liquid crystal) become large compared to intrinsic length scales (such as the typical size of the liquid crystal molecules). Our objective is to verify if OS is able to reproduce the description of the large-body regime provided in \cite{Gartland_2018_liquid_cryst}.	
	
	The Landau-de Gennes model is expressed in terms of an integral functional of a field $\textbf{Q}$ taking values into the space of symmetric, traceless 3x3 matrices. In its simplest form, the dimensional functional $\mathcal{F}$ can be written as
	
	\begin{equation}
	\mathcal{F}[\textbf{Q}] 
	=
	\int_{\Omega} 
	\left[ 
	\, \frac{L}{2} \, | \nabla \textbf{Q} |^2
	\, + \, 
	\frac{A}{2} \, \mathrm{tr}( \textbf{Q}^2 )
	\, - \, 
	\frac{B}{3} \, \mathrm{tr}( \textbf{Q}^3 )
	\, + \, 
	\frac{C}{4} \, \mathrm{tr}( \textbf{Q}^2 )^2
	\right] dV,
	\label{eqn:dimensional_funct_LdG}
	\end{equation}
	
	\noindent with $L$, $A$, $B$, $C$ the material dependent parameters and $\Omega \subset \mathbb{R}^3$ the domain containing the liquid crystal. Aiming to rewrite in unitless form the functional \eqref{eqn:dimensional_funct_LdG}, one invokes the following change of variables:
	
	\begin{equation}
	\textbf{x} = x_0 \, \bar{\textbf{x}},
	\quad
	\textbf{Q} = Q_0 \, \bar{\textbf{Q}},
	\quad
	\mathcal{F} = \mathcal{F}_0 \, \bar{\mathcal{F}},
	\label{eqn:CoV_LdG}
	\end{equation}	 

	\noindent where $\textbf{x} \in \Omega$ is the (dimensional) coordinate variable, while $\bar{\textbf{x}}$, $\bar{\textbf{Q}}$ and $\bar{\mathcal{F}}$ are the scaled (unitless) counterparts of $\textbf{x}$, $\textbf{Q}$ and $\mathcal{F}$ respectively. In \eqref{eqn:CoV_LdG}, we define $\theta \equiv \{ x_0, Q_0, \mathcal{F}_0 \} \in (0,\infty)^{N_x}$ as the $N_x=3$ characteristic constants to be determined.
	
	\noindent By using \eqref{eqn:CoV_LdG} and $A_{\mathrm{NI}} \equiv B^2 / ( 27 C )$, one obtains the dimensionless formulation of \eqref{eqn:dimensional_funct_LdG}:
	
	\begin{equation}
	\bar{\mathcal{F}}[\bar{\textbf{Q}}] 
	=
	\lambda_1
	\,
	\int_{\bar{\Omega}} 
	\left[ 
	\, \frac{\lambda_2^2}{2} \, | \bar{\nabla} \bar{\textbf{Q}} |^2
	\, + \, 
	\frac{\vartheta}{2} \, \mathrm{tr}( \bar{\textbf{Q}}^2 )
	\, - \, 
	\sqrt{3} \, \lambda_3 \, \mathrm{tr}( \bar{\textbf{Q}}^3 )
	\, + \, 
	\frac{\lambda_4}{4} \, \mathrm{tr}( \bar{\textbf{Q}}^2 )^2
	\right] d\bar{V},
	\label{eqn:dimless_funct_LdG}
	\end{equation}

	\noindent with $\bar{\Omega} = \Omega / x_0$, $\bar{\nabla} = x_0 \nabla$ and $d\bar{V} = dV / x_0^3 $. The unitless parameter $\vartheta \equiv A / A_{\mathrm{NI}}$ does not depend on the scaling factors $\theta = \{ x_0, Q_0, \mathcal{F}_0 \}$. It assumes the values $\vartheta=0,1,9/8$ for the parameters choices suggested in \cite{Gartland_2018_liquid_cryst}. The dimensionless parameters $\lambda_i(\theta)$, with $i = 1, \dots, N_d$ and $N_d=4$, are computed as
	
	\begin{equation}
	\lambda_1(\theta) = \frac{ A_{\mathrm{NI}} \, x_0^3 \, Q_0^2 }
	{ \mathcal{F}_0 },
	\quad
	\lambda_2(\theta) = \frac{ \xi_{\mathrm{NI}} }{ x_0 },
	\quad
	\lambda_3(\theta) = \frac{ B \, Q_0 }{ \sqrt{27} \, A_{\mathrm{NI}} },
	\quad
	\lambda_4(\theta) = \frac{ C \, Q_0^2 }{ A_{\mathrm{NI}} },
	\end{equation}	 
	
	\noindent with $\theta = \{ x_0, Q_0, \mathcal{F}_0 \}$ and $\xi_{\mathrm{NI}} \equiv \sqrt{ L / A_{\mathrm{NI}} }$ a typical length scale of the liquid crystal, called \emph{nematic correlation length}.
	
	In \cite{Gartland_2018_liquid_cryst}, the characteristic constants $\theta = \{ x_0, Q_0, \mathcal{F}_0 \}$ are chosen as
	
	\begin{equation}
	x_0 = R,
	\quad
	Q_0 = \alpha \equiv \frac{ B }{ \sqrt{27} \, C },
	\quad
	\mathcal{F}_0 = \alpha^2 \, A_{\mathrm{NI}} \, R^3,
	\label{eqn:Gartland_factors_LdG}
	\end{equation}

	\noindent so that $N_x=3$ out of $N_d=4$ dimensionless coefficients are equal to 1, i.e. $\lambda_{1,3,4}(\theta) = 1$. As stated above, the \emph{large-body limit} is achieved when geometric length scales (such as the size of the domain $\Omega$) become large compared to intrinsic length scales (for instance, the nematic length $\xi_{\mathrm{NI}}$). Aiming to analyse the large-body limit, the value of $R$ is chosen in \cite{Gartland_2018_liquid_cryst} as the diameter of $\Omega$, typically three orders of magnitude larger than $\xi_{\mathrm{NI}}$:
	
	\begin{equation}
	R = \mathrm{diam}(\Omega) = \xi_{\mathrm{NI}} \, 10^q
	\gg \xi_{\mathrm{NI}},
	\quad
	q=3. 
	\label{eqn:R_large-body_limit_LdG}
	\end{equation}
	
	\noindent As a consequence, the large-body limit leads to the following regime of the functional $\bar{\mathcal{F}}$ parameters:
	
	\begin{equation}
	\lambda_1 = \lambda_3 = \lambda_4 = 1 \gg \lambda_2 = 10^{-q},
	\quad q=3.
	\label{eqn:lambdas_large-body_limit_LdG}
	\end{equation}
	
	\noindent In other words, such a limit assigns a small weight to the term corresponding to $\lambda_2$.
	
	Now, let us verify if the Optimal Scaling procedure is able to recover the physical scaling \eqref{eqn:Gartland_factors_LdG}-\eqref{eqn:R_large-body_limit_LdG} imposed to achieve the parameters regime \eqref{eqn:lambdas_large-body_limit_LdG} of the large-body limit. To do so, we perform the following experiment. 
	
	We consider the cost function $C_{\mathrm{eucl}}(\theta)$ \eqref{eqn:C_eucl_def}, with a choice of $\{ \Theta_i \}_{i=1}^{N_d}$ targeting to drive the orders of magnitude of coefficients $\{ \lambda_i \}_{i=1}^{N_d}$ close to the regime \eqref{eqn:lambdas_large-body_limit_LdG}. In particular, it is possible to select:
	
	\begin{equation}
	\Theta_i = \log_{10}(\lambda_i) = 0,
	\quad i=1,3,4, \quad
	\Theta_2 = \log_{10}(\lambda_2) = -q,
	\quad
	q=3,
	\label{eqn:Theta_LdG}
	\end{equation}	   

	\noindent with $\lambda_1,\dots,\lambda_4$ \eqref{eqn:lambdas_large-body_limit_LdG}. We remark that such a choice of $\Theta_1,\dots,\Theta_4$ differs from what was proposed for the case studies in \autoref{sec:scaling_projectile} and \autoref{sec:scaling_schrodinger}, where $\Theta_i=0$, $\forall i=1,\dots,N_d$. Given \eqref{eqn:Theta_LdG}, the analytical solution derived in \autoref{sec:anal_sol_min_Ceucl} for the argument $\theta_{\mathrm{eucl}} = \{ x_0, Q_0, \mathcal{F}_0 \}$ of minimum of the norm $C_{\mathrm{eucl}}(\theta)$ provides:
	
	\begin{equation}
	x_0 = \xi_{\mathrm{NI}} \, 10^q,
	\quad
	Q_0 = \frac{ B }{ \sqrt{27} \, C},
	\quad
	\mathcal{F}_0 =
	\frac{ B^2 }{ 27 \, C^2}  
	\, A_{\mathrm{NI}} \,
	( \xi_{\mathrm{NI}} \, 10^q )^3.
	\label{eqn:Optimal_Factors_LdG} 
	\end{equation}
	
	\noindent It is easy to check that the \emph{optimal scaling factors} \eqref{eqn:Optimal_Factors_LdG} coincide with the physical scaling \eqref{eqn:Gartland_factors_LdG}-\eqref{eqn:R_large-body_limit_LdG}. Thus, we conclude that Optimal Scaling is capable to explore limiting behaviours of Landau-de Gennes model in extreme ranges of parameters. 
	
	Such a feature of OS may be also helpful in driving dimensionless equations to parameters ranges allowing for asymptotic approximations \cite{Miller_Asympt_Anal}. As shown above, a particular choice of $\{ \Theta_i \}_{i=1}^{N_d}$ in \eqref{eqn:C_eucl_def} can lead to scaling factors giving negligible weights to complex terms. Then, such terms may be discarded, allowing for an asymptotic approximation of the equation to solve.
			 
\section{Modelling Dynamic Development of Latex Particles Morphology}
\label{sec:modelling_DDPM}

	Our next goal is to apply the Optimal Scaling procedure to the Population Balance Equations (PBE) model of the dynamic development of Latex Particles Morphology proposed in \cite{PhDThesis_Rusconi_PMCQS} under the assumptions of \cite{DDPM_2016}. As has been demonstrated in \cite{PhDThesis_Rusconi_PMCQS}, experimental values of parameters of this model may lead to computationally intractable orders of magnitude (e.g. ranging from $-21$ to $36$) of the involved variables.  Thus, this model can be an excellent application for the Optimal Scaling procedure.

\subsection{Dimensional Model}
\label{sec:PBE_model_Latex_Particles}

	The morphology of latex particles can be understood as the pattern of phase-separated domains, 
	describing a given particle.
	The considered pattern is composed by agglomerates of polymers belonging to two separated phases: the so-called \emph{non-equilibrium} and \emph{equilibrium} phases. With the aim to predict the evolution of the morphology of interest, it is possible to focus on the \textbf{distributions} $\tilde{m}(\tilde{v},\tilde{t})$ [L$^{-1}$] and $\tilde{w}(\tilde{v},\tilde{t})$ [L$^{-1}$] of the size $\tilde{v}$ [L] at time $\tilde{t}$ [s] of non-equilibrium and equilibrium polymers clusters, respectively. The volume $\tilde{v}$ is measured in Litres [L] and the elapsed time $\tilde{t}$ in seconds [s]. For any fixed time $\tilde{t}$, the expected number of non-equilibrium and equilibrium clusters with volume $\tilde{v}$ in the interval $[\alpha,\beta]$ is provided by  
	
	\begin{equation}
	\int_\alpha^\beta
	\! \tilde{m}(\tilde{v},\tilde{t}) \, d\tilde{v}
	\quad \mbox{and} \quad
	\int_\alpha^\beta
	\! \tilde{w}(\tilde{v},\tilde{t}) \, d\tilde{v}
	\label{eqn:definiton_tilde_m_w}
	\end{equation}	
	
	\noindent respectively. The distributions $\tilde{m}(\tilde{v},\tilde{t})$ and $\tilde{w}(\tilde{v},\tilde{t})$ are not normalised to 1. Rather, their zero-order moment yields the expected total number of polymers clusters, for any given time $\tilde{t}$.
	
	As detailed in \cite{PhDThesis_Rusconi_PMCQS}, it is possible to show that the distributions $\tilde{m}(\tilde{v},\tilde{t})$ and $\tilde{w}(\tilde{v},\tilde{t})$ satisfy the following system of Population Balance Equations (PBE) \cite{Ramkrishna2000}:
		
	\begin{equation}
	\begin{cases}	
	\frac{ \partial \tilde{m}(\tilde{v},\tilde{t}) }	{ \partial \tilde{t} } 
	& =
	- 
	\frac{ 
	\partial ( \, \tilde{g}(\tilde{v},\tilde{t}) \, \tilde{m}(\tilde{v},\tilde{t}) \, ) }
	{\partial \tilde{v} }
	\, + \, 
	\tilde{n}(\tilde{v},\tilde{t})
	\, - \, 
	\tilde{\mu}(\tilde{v},\tilde{t}) \, \tilde{m}(\tilde{v},\tilde{t}) 
	- 
	\, \tilde{m}(\tilde{v},\tilde{t}) \,
	\int_{0}^{\infty}
	\! \tilde{a}(\tilde{v},\tilde{u},\tilde{t}) 
	\, \tilde{m}(\tilde{u},\tilde{t}) \, d\tilde{u} \\ 
	& \quad 
	+
	\frac{1}{2} \, 
	\int_0^{\tilde{v}}
	\! \tilde{a}(\tilde{v}-\tilde{u},\tilde{u},\tilde{t}) 
	\, \tilde{m}(\tilde{v}-\tilde{u},\tilde{t}) 
	\, \tilde{m}(\tilde{u},\tilde{t}) \, d\tilde{u},
	\quad \quad \quad
	\forall \tilde{v},\tilde{t} \in \mathbb{R}^+, 
	\\	  
	\\  	
	\frac{ \partial \tilde{w}(\tilde{v},\tilde{t}) }	{ \partial \tilde{t} } 
	& =
	- 
	\frac{ 
	\partial ( \, \tilde{g}(\tilde{v},\tilde{t}) \, \tilde{w}(\tilde{v},\tilde{t}) \, ) }
	{\partial \tilde{v} }
	\, + \, 
	\tilde{\mu}(\tilde{v},\tilde{t}) \, \tilde{m}(\tilde{v},\tilde{t}) 
	- 
	\, \tilde{w}(\tilde{v},\tilde{t}) \,
	\int_{0}^{\infty}
	\! \tilde{a}(\tilde{v},\tilde{u},\tilde{t}) 
	\, \tilde{w}(\tilde{u},\tilde{t}) \, d\tilde{u} \\ 
	& \quad 
	+
	\frac{1}{2} \, 
	\int_0^{\tilde{v}}
	\! \tilde{a}(\tilde{v}-\tilde{u},\tilde{u},\tilde{t}) 
	\, \tilde{w}(\tilde{v}-\tilde{u},\tilde{t}) 
	\, \tilde{w}(\tilde{u},\tilde{t}) \, d\tilde{u},
	\quad \quad \quad
	\forall \tilde{v},\tilde{t} \in \mathbb{R}^+, 
	\\
	\\     
    \tilde{m}(\tilde{v},0) & = \tilde{w}(\tilde{v},0) = 0,
	\quad 
	\forall \tilde{v} \in \mathbb{R}^+,
	\quad \quad \quad
	\tilde{m}(0,\tilde{t}) = \tilde{w}(0,\tilde{t}) = 0,
	\quad 
	\forall \tilde{t} \in \mathbb{R}^+.	 
	\end{cases}
	\label{eqn:PBE_latex_part}
	\end{equation}	

	\noindent The \textbf{nucleation rate} $\tilde{n}(\tilde{v},\tilde{t})$ [L$^{-1}$ s$^{-1}$] accounts for the expected number of non-equilibrium clusters formed per unit of time: 
	
	\begin{equation}
    \tilde{n}(\tilde{v},\tilde{t})
    = 
    k_s \, v_c^{-1} 
    \, \tilde{\Phi}(\tilde{t}) 
    \, \tilde{\delta}(\tilde{v}-v_c),
    \label{eqn:DDPM_n}
    \end{equation}
    
    \noindent where $\tilde{\delta}(\tilde{x})$ [L$^{-1}$] is the Dirac delta function.\\
    \noindent The \textbf{growth rate} $\tilde{g}(\tilde{v},\tilde{t})$ [L s$^{-1}$] corresponds to the expected increase of volume per unit of time of a given cluster with size $\tilde{v}$ at time $\tilde{t}$:
    
    \begin{equation}
	\tilde{g}(\tilde{v},\tilde{t})
	=
	\sqrt[3]{36 \pi} \, 	k_d 
	\, \tilde{\Phi}(\tilde{t})
	\, (\tilde{\Psi}(\tilde{t})+1)^{\nicefrac{2}{3}} 
	\, \tilde{v}^{\nicefrac{2}{3}}	 
	+ 
	\frac{ k_p R \bar{V}_{\mathrm{pol2}} }{ \bar{V}_{\mathrm{mon2}} }
	\, \frac{ \tilde{\Psi}(\tilde{t}) }{ \tilde{V}_p(\tilde{t}) }
	\, \tilde{v}.
	\label{eqn:DDPM_g}
	\end{equation}
	
	\noindent The \textbf{migration rate} $\tilde{\mu}(\tilde{v},\tilde{t})$ [s$^{-1}$] is the expected proportion of non-equilibrium clusters moving per unit of time to the equilibrium phase:	
	
	\begin{equation}
	\tilde{\mu}(\tilde{v},\tilde{t}) = k_\mu.
	\label{eqn:DDPM_mu}
	\end{equation} 
	     
    \noindent The \textbf{aggregation rate} $\tilde{a}(\tilde{v},\tilde{u},\tilde{t})$ [s$^{-1}$] corresponds to the expected frequency of coagulations between clusters with volumes $\tilde{v}$ and $\tilde{u}$ at time $\tilde{t}$:
    
    \begin{equation}
    \tilde{a}(\tilde{v},\tilde{u},\tilde{t})
    = 
    k_a \, N_p^{-1} 
    \, (\tilde{\Psi}(\tilde{t})+1)^{\nicefrac{14}{3}}
    \left[ 
    \tilde{v}^{-\nicefrac{1}{3}} 
    +
    \tilde{u}^{-\nicefrac{1}{3}}
    \right].
    \label{eqn:DDPM_a}
    \end{equation}

	As presented in \cite{PhDThesis_Rusconi_PMCQS}, the computation of the rates \eqref{eqn:DDPM_n}-\eqref{eqn:DDPM_a} is coupled with the knowledge of the following non-constant physical quantities.\\
	\noindent \textbf{Expected Total Volume} $\tilde{V}^{\mathrm{mat}}_{\mathrm{pol2}}(\tilde{t})$ [L]:
		
	\begin{equation}
	\begin{cases}
	\frac{d\tilde{V}^{\mathrm{mat}}_{\mathrm{pol2}}(\tilde{t})}{d\tilde{t}}
	& =
	\frac{ k_p R \bar{V}_{\mathrm{pol2}} } { \bar{V}_{\mathrm{mon2}} } 
	\, \frac{ \tilde{\Psi}(\tilde{t}) }{ \tilde{V}_p(\tilde{t}) }
	\, \left[ 
	\tilde{V}^{\mathrm{mat}}_{\mathrm{pol2}}(\tilde{t}) 
	+ 
	V_{\mathrm{pol1}} \right]
	- 
	k_s \, \tilde{\Phi}(\tilde{t})
	- 
	k_d \, \tilde{\Phi}(\tilde{t}) 
	\, \left[ 
	\tilde{\Sigma}_m(\tilde{t})
	+ 
	\tilde{\Sigma}_w(\tilde{t})
	\right], \\ 
	\tilde{V}^{\mathrm{mat}}_{\mathrm{pol2}}(0) & = \, 0.
	\end{cases}
	\label{eqn:DDPM_V_mat_pol2}
	\end{equation}
	
	\noindent \textbf{Expected Total Volume} $\tilde{V}^{\mathrm{c_m}}_{\mathrm{pol2}}(\tilde{t})$ [L]:
	
	\begin{equation}
	\begin{cases}
	\frac{d\tilde{V}^{\mathrm{c_m}}_{\mathrm{pol2}}(\tilde{t})}{d\tilde{t}} 
	& =
	\frac{ k_p R \bar{V}_{\mathrm{pol2}} } { \bar{V}_{\mathrm{mon2}} } 
	\, \frac{ \tilde{\Psi}(\tilde{t}) }{ \tilde{V}_p(\tilde{t}) }
	\, \tilde{V}^{\mathrm{c_m}}_{\mathrm{pol2}}(\tilde{t})
	+ 
	k_s \, \tilde{\Phi}(\tilde{t})
	+ 
	k_d \, \tilde{\Phi}(\tilde{t}) \, \tilde{\Sigma}_m(\tilde{t})
	- 
	k_\mu \tilde{V}^{\mathrm{c_m}}_{\mathrm{pol2}}(\tilde{t}), \\
	\tilde{V}^{\mathrm{c_m}}_{\mathrm{pol2}}(0) & = \, 0.
	\end{cases} 
	\label{eqn:DDPM_V_cm_pol2}
	\end{equation}

	\noindent \textbf{Expected Total Volume} $\tilde{V}^{\mathrm{c_w}}_{\mathrm{pol2}}(\tilde{t})$ [L]:
				
	\begin{equation}
	\begin{cases}
	\frac{d\tilde{V}^{\mathrm{c_w}}_{\mathrm{pol2}}(\tilde{t})}{d\tilde{t}} 
	& = 
	\frac{ k_p R \bar{V}_{\mathrm{pol2}} } { \bar{V}_{\mathrm{mon2}} } 
	\, \frac{ \tilde{\Psi}(\tilde{t}) }{ \tilde{V}_p(\tilde{t}) }
	\, \tilde{V}^{\mathrm{c_w}}_{\mathrm{pol2}}(\tilde{t})
	+ 
	k_d \, \tilde{\Phi}(\tilde{t}) \, \tilde{\Sigma}_w(\tilde{t})
	+ 
	k_\mu \tilde{V}^{\mathrm{c_m}}_{\mathrm{pol2}}(\tilde{t}), \\
	\tilde{V}^{\mathrm{c_w}}_{\mathrm{pol2}}(0) & = \, 0.
	\end{cases} 
	\label{eqn:DDPM_V_cw_pol2}
	\end{equation}
	
	\noindent \textbf{Expected Odds Monomers-Polymers} $\tilde{\Psi}(\tilde{t})$:
	
	\begin{equation}
	\begin{cases}
	\frac{d\tilde{\Psi}(\tilde{t})}{d\tilde{t}} 
	& = 
	- 
	\frac{ k_p R \bar{V}_{\mathrm{pol2}} } { \bar{V}_{\mathrm{mon2}} } 
	\, \frac{ \tilde{\Psi}(\tilde{t}) }{ \tilde{\Psi}(\tilde{t}) + 1 }
	\, \frac
	{ 
	\tilde{\Psi}(\tilde{t}) + 
	\nicefrac{ \bar{V}_{\mathrm{mon2}} }{ \bar{V}_{\mathrm{pol2}} }
	}
	{ \tilde{V}_{\mathrm{pol2}}(\tilde{t}) + V_{\mathrm{pol1}} }, \\
	\tilde{\Psi}(0) & = \, 
	\frac{\bar{M} \, \bar{V}_{\mathrm{mon2}}}{V_{\mathrm{pol1}}}.
	\end{cases} 
	\label{eqn:DDPM_Psi}
	\end{equation}	
	
	\noindent \textbf{Expected Total Volume} $\tilde{V}_{\mathrm{pol2}}(\tilde{t})$ [L]:
			
	\begin{equation}
	\begin{cases}
	\frac{d\tilde{V}_{\mathrm{pol2}}(\tilde{t})}{d\tilde{t}} 
	& = 
	\frac{ k_p R \bar{V}_{\mathrm{pol2}} }{ \bar{V}_{\mathrm{mon2}} } 
	\, 
	\frac{ \tilde{\Psi}(\tilde{t}) }{ \tilde{\Psi}(\tilde{t}) + 1 }, \\
	\tilde{V}_{\mathrm{pol2}}(0) & = \, 0.
	\end{cases} 
	\label{eqn:DDPM_V_pol2}
	\end{equation}
	
	\noindent The following definitions complete the presented model:
	
	\begin{equation}
    \tilde{\Phi}(\tilde{t}) 
    \equiv
    \left[  
    \frac{ \tilde{V}^{\mathrm{mat}}_{\mathrm{pol2}}(\tilde{t}) }
    {
    ( \tilde{\Psi}(\tilde{t}) + 1 )
	( \tilde{V}^{\mathrm{mat}}_{\mathrm{pol2}}(\tilde{t}) + V_{\mathrm{pol1}} )    
    } 
    - 
    \Phi_s
    \right]^+,
    \quad \mbox{with} \quad 
    x^+ \equiv \max{ \{x,0\} },
    \label{eqn:DDPM_Phi}
    \end{equation} 
    
	\begin{equation}
	\tilde{V}_p(\tilde{t})
	\equiv
	\left( \tilde{\Psi}(\tilde{t}) + 1 \right)
	\, \left[
	\tilde{V}^{\mathrm{mat}}_{\mathrm{pol2}}(\tilde{t}) 
	+	
	\tilde{V}^{\mathrm{c_m}}_{\mathrm{pol2}}(\tilde{t}) 
	+	
	\tilde{V}^{\mathrm{c_w}}_{\mathrm{pol2}}(\tilde{t})
	+ 
	V_{\mathrm{pol1}}
	\right]
	\quad \mbox{[L]},
	\label{eqn:DDPM_V_p}
	\end{equation}
		
	\begin{equation}
	\tilde{\Sigma}_{m,w}(\tilde{t}) 
	\equiv
	\sqrt[3]{36 \pi} 
	\, ( \tilde{\Psi}(\tilde{t}) + 1 )^{\nicefrac{2}{3}}
	\int_0^{\infty}
	\! \tilde{v}^{\nicefrac{2}{3}} 
	\, \tilde{m},\tilde{w}(\tilde{v},\tilde{t}) 
	\, d\tilde{v}
	\quad \mbox{[L$^{\nicefrac{2}{3}}$]},
	\label{eqn:DDPM_Sigma_m,w}
	\end{equation}
	
	\noindent where $\tilde{m}$ and $\tilde{w}$ are the solutions of \eqref{eqn:PBE_latex_part}. \autoref{tab:DDPM_nomen_def_exp_val} reports the nomenclature and the experimental values assumed by the physical parameters involved in the computation of the rates \eqref{eqn:DDPM_n}-\eqref{eqn:DDPM_a}.

	\begin{table}[!h]
    \begin{subtable}[t]{0.5\textwidth}
	\centering
	\begin{tabular}[t]{ll}

	\hline
	\textbf{Nomenclature} & \textbf{Experimental Value} \\
	\hline

	\rowcolor[HTML]{EFEFEF}	
	$k_a$ [L$^{\nicefrac{1}{3}}$ s$^{-1}$] & $2 \times 10^{-8}$ L$^{\nicefrac{1}{3}}$ s$^{-1}$ \\	
	
	$k_d$ [L$^{\nicefrac{1}{3}}$ s$^{-1}$] & $5 \times 10^{-8}$ L$^{\nicefrac{1}{3}}$ s$^{-1}$\\  	
	
	\rowcolor[HTML]{EFEFEF}	
	$k_p$ [L mol$^{-1}$ s$^{-1}$] & 850 L mol$^{-1}$ s$^{-1}$ \\ 
	
	$k_s$ [L s$^{-1}$] & $2.5 \times 10^{-5}$ L s$^{-1}$ \\  
	
	\rowcolor[HTML]{EFEFEF}	
	$k_\mu$ [s$^{-1}$] & $10^{-5}$ s$^{-1}$ \\

	$\bar{M}$ [mol] & $2.5$ mol \\
	
	\rowcolor[HTML]{EFEFEF}
	$\bar{V}_{\mathrm{mon2}}$ [L mol$^{-1}$] & 0.1 L mol$^{-1}$ \\  
	
	\hline	
			
	\end{tabular}
	\end{subtable}
	\hspace{\fill}
	\begin{subtable}[t]{0.5\textwidth}
	\centering
	\begin{tabular}[t]{ll}
	
	\hline
	\textbf{Nomenclature} & \textbf{Experimental Value} \\
	\hline

	\rowcolor[HTML]{EFEFEF}	
	$V_{\mathrm{pol1}}$ [L] & 0.25 L \\  

	$v_c$ [L] & $2.5 \times 10^{-22}$ L \\	
	
	\rowcolor[HTML]{EFEFEF}	
	$\bar{V}_{\mathrm{pol2}}$ [L mol$^{-1}$] & 0.095 L mol$^{-1}$ \\   

	$\Phi_s$ & $10^{-3}$ \\
	
	\rowcolor[HTML]{EFEFEF}	
	$N_p$ & $2.8 \times 10^{17}$ \\

	$R$ [mol] & $2.3 \times 10^{-7}$ mol \\

	\rowcolor[HTML]{EFEFEF}	
	& \\
	
	\hline	
    
    \end{tabular}
    \end{subtable}
    	\caption{Latex Particles Morphology Formation. Nomenclature of the parameters involved in the computation of the rates \eqref{eqn:DDPM_n}-\eqref{eqn:DDPM_a}, with the corresponding experimental values. The symbols s, L and mol stand for second, Litre and mole respectively. The data were provided by Prof. J. M. Asua (POLYMAT).}
    	\label{tab:DDPM_nomen_def_exp_val}
    	\end{table}	
	
\subsection{Dimensionless Model}
\label{sec:Dimensionless_PBE_model_Latex_Particles}

	The application of Optimal Scaling (\autoref{sec:optimal_scaling_procedure}) to the dimensional model \eqref{eqn:PBE_latex_part} for Latex Particles Morphology requires the definition of dimensionless counterparts of \eqref{eqn:PBE_latex_part}-\eqref{eqn:DDPM_Sigma_m,w}. We define the scaled counterparts of the variables in the PBE system \eqref{eqn:PBE_latex_part} as:
	
	\begin{equation}
	v \equiv \tilde{v} \, / \, \nu_0,
	\quad	
	t \equiv \tilde{t} \, / \, t_0,
	\quad
	m(v,t) \equiv \tilde{m}(\tilde{v},\tilde{t}) \, / \, m_0,
	\quad
	w(v,t) \equiv \tilde{w}(\tilde{v},\tilde{t}) \, / \, w_0,
	\label{eqn:scaling_PBE_variables}
	\end{equation}	  

	\noindent where $\nu_0$ [L], $t_0$ [s], $m_0$ [L$^{-1}$] and $w_0$ [L$^{-1}$] belong to the set of the strictly positive characteristic constants $\theta$ defined by the change of variables \eqref{eqn:change_of_var}. The variables $\tilde{\Psi}(\tilde{t})$ \eqref{eqn:DDPM_Psi} and $\tilde{\Phi}(\tilde{t})$ \eqref{eqn:DDPM_Phi} are dimensionless and computationally tractable since they are given by the ratios between quantities of the same dimensions and orders of magnitude. Then, we define $\Psi(t)$ and $\Phi(t)$ as the dimensionless counterparts of $\tilde{\Psi}(\tilde{t})$ and $\tilde{\Phi}(\tilde{t})$ respectively:
				 
	\begin{equation}
	\Psi(t) \equiv \tilde{\Psi}(\tilde{t}),
	\quad
	\Phi(t) \equiv \tilde{\Phi}(\tilde{t}).
	\end{equation}	  
	
	\noindent The quantities $\tilde{V}^{\mathrm{mat}}_{\mathrm{pol2}}(\tilde{t})$ \eqref{eqn:DDPM_V_mat_pol2}, $\tilde{V}_{\mathrm{pol2}}(\tilde{t})$ \eqref{eqn:DDPM_V_pol2} and $\tilde{V}_p(\tilde{t})$ \eqref{eqn:DDPM_V_p} are scaled by the factors $M_0$ [L], $P_0$ [L] and $\Pi_0$ [L] respectively:
	
	\begin{equation}
	V^{\mathrm{mat}}_{\mathrm{pol2}}(t) \equiv 
	\tilde{V}^{\mathrm{mat}}_{\mathrm{pol2}}(\tilde{t}) \, / \, M_0,
	\quad
	V_{\mathrm{pol2}}(t) \equiv 
	\tilde{V}_{\mathrm{pol2}}(\tilde{t}) \, / \, P_0,
	\quad
	V_p(t) \equiv 
	\tilde{V}_p(\tilde{t}) \, / \, \Pi_0.
	\end{equation}
		
	\noindent The Dirac delta function $\tilde{\delta}(\tilde{v}-v_c)$ [L$^{-1}$] is scaled by the factor $\delta_0$ [L$^{-1}$]:
	
	\begin{equation}
	\delta(v-\lambda_c) \equiv \tilde{\delta}(\tilde{v}-v_c) \, / \, \delta_0,
	\label{eqn:scaling_delta}
	\end{equation}	 	
	
	\noindent with the dimensionless coefficient $\lambda_c$ computed as

	\begin{equation}
	\lambda_c \equiv 
	\int_0^{\infty} \! v \, \delta(v-\lambda_c) \, dv =
	\frac{v_c}{\nu_0^2 \, \delta_0}.
	\label{eqn:lambda_c}
	\end{equation}	

	\noindent The definitions \eqref{eqn:DDPM_Sigma_m,w} of the variables $\tilde{\Sigma}_m(\tilde{t})$ and $\tilde{\Sigma}_w(\tilde{t})$ suggest the proper scaling factors:
	
	\begin{equation}
	\Sigma_m(t) \equiv 
	\frac{\tilde{\Sigma}_m(\tilde{t})}
	{\sqrt[3]{36 \pi} \, \nu_0^{\nicefrac{5}{3}} \, m_0},
	\quad
	\Sigma_w(t) \equiv 
	\frac{\tilde{\Sigma}_w(\tilde{t})}
	{\sqrt[3]{36 \pi} \, \nu_0^{\nicefrac{5}{3}} \, w_0}.
	\end{equation}

	\noindent By integrating \eqref{eqn:PBE_latex_part}, multiplied by $\tilde{v}$, with respect to $\tilde{v}$ over $\mathbb{R}^+$, it is possible to show that the quantities $\tilde{V}^{\mathrm{c_m}}_{\mathrm{pol2}}(\tilde{t})$ \eqref{eqn:DDPM_V_cm_pol2} and $\tilde{V}^{\mathrm{c_w}}_{\mathrm{pol2}}(\tilde{t})$ \eqref{eqn:DDPM_V_cw_pol2} correspond to the first-order moments of $\tilde{m}(\tilde{v},\tilde{t})$ and $\tilde{w}(\tilde{v},\tilde{t})$: 
		
	\begin{equation}
	\tilde{V}^{\mathrm{c_m}}_{\mathrm{pol2}}(\tilde{t})
	=
	\int_0^{\infty} \! \tilde{v} \, \tilde{m}(\tilde{v},\tilde{t}) \, d\tilde{v},
	\quad
	\tilde{V}^{\mathrm{c_w}}_{\mathrm{pol2}}(\tilde{t})
	=
	\int_0^{\infty} \! \tilde{v} \, \tilde{w}(\tilde{v},\tilde{t}) \, d\tilde{v},
	\quad
	\forall \tilde{t} \in \mathbb{R}^+.
	\label{eqn:first_order_mom_dimensional}
	\end{equation}
	
	\noindent Given \eqref{eqn:first_order_mom_dimensional}, the dimensionless counterparts $V^{\mathrm{c_{m,w}}}_{\mathrm{pol2}}(t)$ of the variables $\tilde{V}^{\mathrm{c_{m,w}}}_{\mathrm{pol2}}(\tilde{t})$ are computed as:
	
	\begin{equation}
	V^{\mathrm{c_m}}_{\mathrm{pol2}}(t)
	\equiv
	\frac{ \tilde{V}^{\mathrm{c_m}}_{\mathrm{pol2}}(\tilde{t}) }
	{ \nu_0^2 \, m_0},
	\quad
	V^{\mathrm{c_w}}_{\mathrm{pol2}}(t)
	\equiv
	\frac{ \tilde{V}^{\mathrm{c_w}}_{\mathrm{pol2}}(\tilde{t}) }
	{ \nu_0^2 \, w_0}.	 
	\label{eqn:scaling_first_ord_mom}
	\end{equation}
	
	\noindent In agreement with \eqref{eqn:first_order_mom_dimensional}, the proposed scaling \eqref{eqn:scaling_PBE_variables}-\eqref{eqn:scaling_first_ord_mom} ensures the dimensionless variables $V^{\mathrm{c_m}}_{\mathrm{pol2}}(t)$ and $V^{\mathrm{c_w}}_{\mathrm{pol2}}(t)$ to be the first-order moments of the dimensionless distributions $m(v,t)$ and $w(v,t)$.
	
	Given the characteristic constants $\theta \equiv \{ \nu_0, t_0, m_0, w_0, M_0, P_0, \Pi_0, \delta_0 \}$ and the corresponding scaled quantities, it is possible to rewrite the PBE system \eqref{eqn:PBE_latex_part} and the rate functions \eqref{eqn:DDPM_n}-\eqref{eqn:DDPM_a} in terms of the defined dimensionless variables. The equations \eqref{eqn:dimless_PBE_DDPM}-\eqref{eqn:dimless_Sigma_m,w} summarise the arising dimensionless model, where the parameters are specified in \autoref{tab:DDPM_param_scaling}.
	
	\begin{equation}
	\begin{cases}	
	\frac{ \partial m(v,t) }	{ \partial t } 
	& =
	- 
	\frac{ 
	\partial ( \, g(v,t) \, m(v,t) \, ) }
	{\partial v }
	\, +
	\, n(v,t)
	\, - 
	\, \mu_m(v,t) \, m(v,t) 
	- 
	\, m(v,t) \,
	\int_{0}^{\infty}
	\! a_m(v,u,t) \, m(u,t) \, du \\ 
	& \quad 
	+
	\frac{1}{2} \, 
	\int_0^v
	\! a_m(v-u,u,t) \, m(v-u,t) \, m(u,t) \, du,
	\quad \quad \quad
	\forall v,t \in \mathbb{R}^+, 
	\\	  
	\\  	
	\frac{ \partial w(v,t) }	{ \partial t } 
	& =
	- 
	\frac{ 
	\partial ( \, g(v,t) \, w(v,t) \, ) }
	{\partial v }
	\, + 
	\, \mu_w(v,t) \, m(v,t) 
	- 
	\, w(v,t) \,
	\int_{0}^{\infty}
	\! a_w(v,u,t) \, w(u,t) \, du \\ 
	& \quad 
	+
	\frac{1}{2} \, 
	\int_0^v
	\! a_w(v-u,u,t) \, w(v-u,t) \, w(u,t) \, du,
	\quad \quad \quad
	\forall v,t \in \mathbb{R}^+, 	   
	\\
	\\     
    m(v,0) & = w(v,0) = 0,
	\quad 
	\forall v \in \mathbb{R}^+,
	\quad \quad \quad
	m(0,t) = w(0,t) = 0,
	\quad 
	\forall t \in \mathbb{R}^+,		 
	\end{cases}
	\label{eqn:dimless_PBE_DDPM}
	\end{equation}
	
	\begin{equation}
    a_{m,w}(v,u,t) \equiv
    \, \lambda^{m,w}_a
    \, (\Psi(t)+1)^{\nicefrac{14}{3}}
    \left[ 
    v^{-\nicefrac{1}{3}} + u^{-\nicefrac{1}{3}}
    \right],
    \label{eqn:dimless_aggr_rate}
    \end{equation}

    \begin{equation}
	g(v,t) \equiv
	\, \lambda_d 
	\, \Phi(t) 
	\, (\Psi(t)+1)^{\nicefrac{2}{3}} 
	\, v^{\nicefrac{2}{3}}	 
	+ 
	\, \lambda_p
	\, \Psi(t) 
	\, V_p(t)^{-1}
	\, v,
	\label{eqn:dimless_growth_rate}
	\end{equation}
	
	\begin{equation}
    n(v,t) \equiv
	\, \lambda_n \, \Phi(t) \, \delta(v-\lambda_c),
	\quad \mbox{with } \delta(x) \mbox{ the Dirac delta},
	\label{eqn:dimless_nucl_rate}
    \end{equation}
   
	\begin{equation}
	\mu_{m,w}(v,t) \equiv \lambda^{m,w}_\mu,
	\label{eqn:dimless_phase_tr_rate}
	\end{equation} 
    
	\begin{equation}
	\begin{cases}
	\frac{dV^{\mathrm{mat}}_{\mathrm{pol2}}(t)}{dt}
	& =
	\, \lambda_p
	\, \frac{ \Psi(t) }{ V_p(t) }
	\, \left[ 
	\, V^{\mathrm{mat}}_{\mathrm{pol2}}(t) 
	\, + \, 
	\lambda^{\mathrm{mat}}_{\mathrm{pol1}} \, 
	\right]
	- 
	\, \Phi(t) 
	\, \left[
	\, \lambda^{\mathrm{mat}}_s
	+
	\, \lambda^{\mathrm{mat}}_{\mathrm{d_m}} \, \Sigma_m(t) + 
	\, \lambda^{\mathrm{mat}}_{\mathrm{d_w}} \, \Sigma_w(t)
	\right], \\
	V^{\mathrm{mat}}_{\mathrm{pol2}}(0) & = \, 0,
	\end{cases}
	\label{eqn:dimless_V_mat_pol2}
	\end{equation}    
    	
	\begin{equation}
	\begin{cases}
	\frac{dV^{\mathrm{c_m}}_{\mathrm{pol2}}(t)}{dt}
	& =
	\, \lambda_p
	\, \frac{ \Psi(t) }{ V_p(t) }
	\, V^{\mathrm{c_m}}_{\mathrm{pol2}}(t)
	+ 
	\, \Phi(t) 
	\, \left[
	\, \lambda^m_s
	+
	\, \lambda_d \, \Sigma_m(t)  
	\, \right]
	-
	\, \lambda^m_\mu \, V^{\mathrm{c_m}}_{\mathrm{pol2}}(t), \\
	V^{\mathrm{c_m}}_{\mathrm{pol2}}(0) & = \, 0,
	\end{cases}
	\label{eqn:dimless_V_cm_pol2}
	\end{equation}
	
	\begin{equation}
	\begin{cases}
	\frac{dV^{\mathrm{c_w}}_{\mathrm{pol2}}(t)}{dt}
	& =
	\, \lambda_p
	\, \frac{ \Psi(t) }{ V_p(t) }
	\, V^{\mathrm{c_w}}_{\mathrm{pol2}}(t)
	+ 
	\, \lambda_d \, \Phi(t) 
	\, \Sigma_w(t) 
	+
	\, \lambda^w_\mu \, V^{\mathrm{c_m}}_{\mathrm{pol2}}(t), \\
	V^{\mathrm{c_w}}_{\mathrm{pol2}}(0) & = \, 0,
	\end{cases}
	\label{eqn:dimless_V_cw_pol2}
	\end{equation}
		
	\begin{equation}
	\begin{cases}
	\frac{d\Psi(t)}{dt}
	& = 
	- \lambda^{\mathrm{pol2}}_p 
	\, \frac{ \Psi(t) }{ \Psi(t) + 1 }
	\, \frac{ \Psi(t) + \Psi_r }
	{ V_{\mathrm{pol2}}(t) + \lambda^{\mathrm{pol2}}_{\mathrm{pol1}} }, \\
	\Psi(0) & = \, \bar{\Psi},
	\end{cases} 
	\label{eqn:dimless_Psi}
	\end{equation}	
		
	\begin{equation}
	\begin{cases}
	\frac{dV_{\mathrm{pol2}}(t)}{dt} 
	& = 
	\, \lambda^{\mathrm{pol2}}_p 
	\, \frac{ \Psi(t) }{ \Psi(t) + 1 }, \\
	V_{\mathrm{pol2}}(0) & = \, 0,
	\end{cases} 
	\label{eqn:dimless_V_pol2}
	\end{equation}
		
	\begin{equation}
    \Phi(t) \equiv
    \left[  
    \frac{ V^{\mathrm{mat}}_{\mathrm{pol2}}(t) }
    { ( \Psi(t) + 1 ) 
    ( V^{\mathrm{mat}}_{\mathrm{pol2}}(t) 
    + 
    \lambda^{\mathrm{mat}}_{\mathrm{pol1}} ) } 
    - 
    \Phi_s
    \right]^+,
    \quad \mbox{with} \quad 
    x^+ \equiv \max{ \{x,0\} },
    \end{equation} 
    
	\begin{equation}
	V_p(t) \equiv
	\, \left( \Psi(t)+1 \right)
	\, \left[
	\lambda^{\mathrm{mat}}_p \, V^{\mathrm{mat}}_{\mathrm{pol2}}(t)
	+	
	\lambda^m_p \, V^{\mathrm{c_m}}_{\mathrm{pol2}}(t)
	+	
	\lambda^w_p \, V^{\mathrm{c_w}}_{\mathrm{pol2}}(t)
	+ 
	\lambda^{\mathrm{pol1}}_p
	\right],
	\end{equation}
		
	\begin{equation}
	\Sigma_{m,w}(t) \equiv
	\, (\Psi(t)+1)^{\nicefrac{2}{3}} 
	\int_0^{\infty}
	\! v^{\nicefrac{2}{3}} \, m,w(v,t) \, dv.
	\label{eqn:dimless_Sigma_m,w}
	\end{equation}
			
	\begin{table}[!h]
    	\begin{subtable}[t]{0.5\textwidth}
	\centering
	\begin{tabular}[t]{ll}

	\hline
	\textbf{Parameters $\lambda$} & \textbf{$\lambda(\theta)$} \\
	\hline
	
	\rowcolor[HTML]{EFEFEF}
	$\lambda^m_a$ & $ k_a \, m_0 \, \nu_0^{\nicefrac{2}{3}} \, t_0  \, / \, N_p$ \\

	$\lambda_d$ & $ \sqrt[3]{36 \pi} \, k_d \, t_0  \, / \, \sqrt[3]{\nu_0} $ \\

	\rowcolor[HTML]{EFEFEF}
	$\lambda_n$ & $ k_s \, \delta_0 \, t_0 \, / \, ( m_0 \, v_c ) $ \\

	$\lambda^m_\mu$ & $k_\mu \, t_0 $ \\

	\rowcolor[HTML]{EFEFEF}
	$\lambda^{\mathrm{mat}}_{\mathrm{d_m}}$ &
	$ \sqrt[3]{36 \pi} \, k_d \, m_0 \, \nu_0^{\nicefrac{5}{3}} \, t_0 \, / \, M_0 $ \\
	
	$\lambda^m_s$ & $ k_s \, t_0 \, / \, ( \nu_0^2 \, m_0 )$\\
		
	\rowcolor[HTML]{EFEFEF}	
	$\lambda^{\mathrm{pol2}}_{\mathrm{pol1}}$ & 
	$ V_{\mathrm{pol1}} \, / \, P_0 $ \\
	
	$\lambda^m_p$ & $ \nu_0^2 \, m_0 \, / \, \Pi_0 $\\

	\rowcolor[HTML]{EFEFEF}	
	$\lambda^{\mathrm{pol2}}_p$ & $ k_p \, R \, \bar{V}_{\mathrm{pol2}} \, t_0 \, / \, ( \bar{V}_{\mathrm{mon2}} \, P_0 )$ \\
	
	$\lambda^{\mathrm{mat}}_p$ & $ M_0 \, / \, \Pi_0 $\\

	\hline
	\textbf{Other Parameters} & \textbf{Definition} \\ 
	\hline

	\rowcolor[HTML]{EFEFEF}		
 	$\bar{\Psi}$ & $ \bar{M} \, \bar{V}_{\mathrm{mon2}} \, / \, V_{\mathrm{pol1}} $ \\
	\hline	
	
	\end{tabular}
	\end{subtable}
	\hspace{\fill}
	\begin{subtable}[t]{0.5\textwidth}
	\centering
	\begin{tabular}[t]{ll}	
		
	\hline
	\textbf{Parameters $\lambda$} & \textbf{$\lambda(\theta)$} \\
	\hline  
	
	\rowcolor[HTML]{EFEFEF}
	$\lambda^w_a$ & $ k_a \, w_0 \, \nu_0^{\nicefrac{2}{3}} \, t_0  \, / \, N_p$ \\

	$\lambda_p$ & $ k_p \, R \, \bar{V}_{\mathrm{pol2}} \, t_0 \, / \, ( \bar{V}_{\mathrm{mon2}} \, \Pi_0 )$ \\

	\rowcolor[HTML]{EFEFEF}
	$\lambda_c$ & $ v_c \, / \, ( \nu_0^2 \, \delta_0 )$\\

	$\lambda^w_\mu$ & $k_\mu \, m_0 \, t_0  \, / \, w_0$ \\

	\rowcolor[HTML]{EFEFEF}
	$\lambda^{\mathrm{mat}}_{\mathrm{d_w}}$ & $ \sqrt[3]{36 \pi} \, k_d \, w_0 \, \nu_0^{\nicefrac{5}{3}} \, t_0 \, / \, M_0 $ \\
	
	$\lambda^{\mathrm{mat}}_s$ & $ k_s \, t_0 \, / \, M_0 $\\
		
	\rowcolor[HTML]{EFEFEF}	
	$\lambda^{\mathrm{mat}}_{\mathrm{pol1}}$ & $ V_{\mathrm{pol1}} \, / \, M_0 $ \\[0.1cm]
	
	$\lambda^w_p$ &  $ \nu_0^2 \, w_0 \, / \, \Pi_0 $ \\

	\rowcolor[HTML]{EFEFEF}	
	$\lambda^{\mathrm{pol1}}_p$ & $ V_{\mathrm{pol1}} \, / \, \Pi_0 $ \\
	
	& \\

	\hline
	\textbf{Other Parameters} & \textbf{Definition} \\ 
	\hline

	\rowcolor[HTML]{EFEFEF}		
 	$\Psi_r$ & $ \bar{V}_{\mathrm{mon2}}  \, / \, \bar{V}_{\mathrm{pol2}} $ \\
	\hline	
	
    \end{tabular}
    \end{subtable}
    	\caption{Latex Particles Morphology Formation. Dimensionless parameters $\lambda(\theta)$, with $\theta \equiv \{ \nu_0, t_0, m_0, w_0, M_0, P_0, \Pi_0, \delta_0 \}$, of the equations \eqref{eqn:dimless_PBE_DDPM}-\eqref{eqn:dimless_Sigma_m,w}, arising from the scaling of the PBE model \eqref{eqn:PBE_latex_part}, with rates \eqref{eqn:DDPM_n}-\eqref{eqn:DDPM_a}. \autoref{tab:DDPM_nomen_def_exp_val} provides experimental values of the physical parameters.}
 	\label{tab:DDPM_param_scaling}
	\end{table}	

\subsection{Optimal Scaling of Dynamic Development of Latex Particles Morphology}
\label{sec:scaling_DDPM}

	The examination of the model for dynamic development of Latex Particles Morphology, introduced in \autoref{sec:PBE_model_Latex_Particles}, reveals that the values of the physical parameters given in \autoref{tab:DDPM_nomen_def_exp_val} lead to computationally intractable orders of magnitude of the involved variables, e.g. $\tilde{v} \propto \nu_0 \approx 10^{-17}$ L, $\tilde{m} \propto m_0 \approx 10^{32}$ L$^{-1}$ and $\tilde{w} \propto w_0 \approx 10^{32}$ L$^{-1}$. The scaling procedure described in \autoref{algo:optimal_scaling} can address this problem by providing the feasible values for the dimensionless coefficients $\lambda$ and the corresponding variables. 
	
\begin{figure}[!h]
	\centering
	\begin{subfigure}[h]{.455\linewidth}
	\centering
	\includegraphics[scale=0.43]{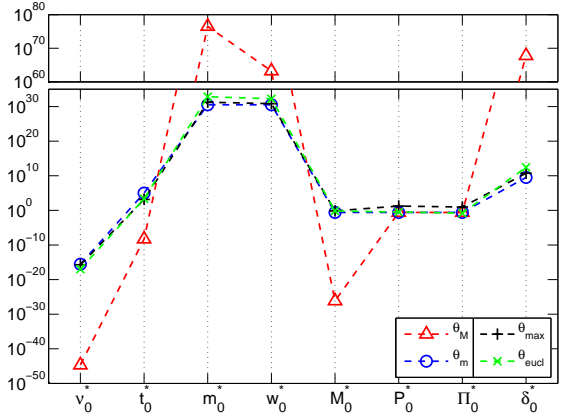}
	\caption{Scaling Factors $\theta^*$.}
	\label{fig:theta_DDPM}
	\end{subfigure}
	\begin{subfigure}[h]{.535\linewidth}
	\centering
	\includegraphics[scale=0.52]{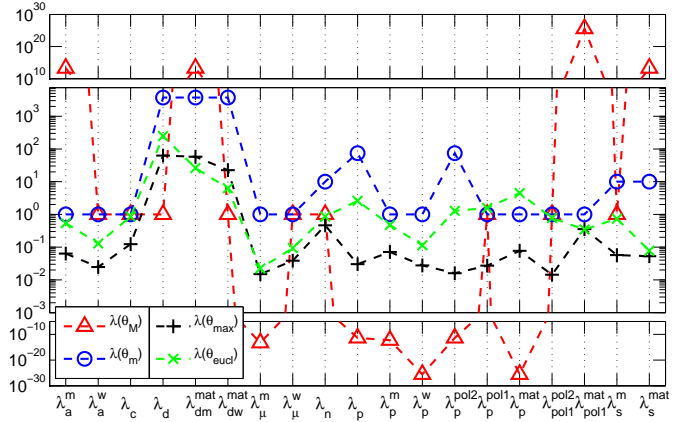}
	\caption{Dimensionless Coefficients $\lambda(\theta^*)$.}
	\label{fig:lambda_DDPM}
	\end{subfigure}	
	\caption{Latex Particles Morphology Formation. (i) Scaling factors $\theta^* \equiv \{ \, \nu_0^* \equiv \nu_0 \, / \, c_{\mbox{L}}, t_0^* \equiv t_0 \, / \, c_{\mbox{s}}, \, \, m_0^* \equiv m_0 \, c_{\mbox{L}}, \, w_0^* \equiv w_0 \, c_{\mbox{L}}, \, M_0^* \equiv M_0 \, / \, c_{\mbox{L}}, \, P_0^* \equiv P_0 \, / \, c_{\mbox{L}}, \, \Pi_0^* \equiv \Pi_0 \, / \, c_{\mbox{L}}, \, \delta_0^* \equiv \delta_0 \, c_{\mbox{L}} \}$, with $c_{\mbox{s,L}} = 1 \, \mbox{s,L}$, obtained with various scaling methods for the model described in \autoref{sec:PBE_model_Latex_Particles}. (ii) Corresponding values of the dimensionless coefficients $\lambda(\theta^*)$, defined in \autoref{tab:DDPM_param_scaling} (\autoref{sec:Dimensionless_PBE_model_Latex_Particles}). The factors $\theta^*$ are labelled as $\theta_{\mathrm{eucl}}$ (green crosses) if found as solutions of the minimisation problem \eqref{eqn:theta_opt_def} by means of the analytical formula of \autoref{sec:anal_sol_min_Ceucl} for $C(\theta) = C_{\mathrm{eucl}}(\theta)$ \eqref{eqn:C_eucl_def}. The label $\theta_{\max}$ (black crosses) refers to the numerical solution of \eqref{eqn:theta_opt_def} obtained using the Simulated Annealing Algorithm 
\cite{10.2307/3214721} for $C(\theta) = C_{\max}(\theta)$ \eqref{eqn:C_max}. The factors $\theta_M$ (red triangles) and $\theta_m$ (blue circles) are defined as in \eqref{eqn:theta_M_m_definition}. They provide the solutions of the traditional scaling method \cite{Holmes2009_ND}, reported in \autoref{sec:sol_trad_scaling}, when the ratio $r$ \eqref{eqn:def_ratio_r} assumes its maximum and minimum values respectively.} 
	\label{fig:scaling_DDPM}
	\end{figure}

	In \autoref{sec:Dimensionless_PBE_model_Latex_Particles} we derived the dimensionless counterpart of the considered model along with the resulting equations \eqref{eqn:dimless_PBE_DDPM}-\eqref{eqn:dimless_Sigma_m,w} and dimensionless coefficients $\lambda(\theta)$ defined in \autoref{tab:DDPM_param_scaling}. The optimal scaling factors $\theta_{\mathrm{opt}}$ \eqref{eqn:theta_opt_def} then can be computed using the analytical solution provided in \autoref{sec:anal_sol_min_Ceucl} for $C(\theta) = C_{\mathrm{eucl}}(\theta)$ \eqref{eqn:C_eucl_def}. The resulting factors $\theta_{\mathrm{opt}}$ are labelled as $\theta_{\mathrm{eucl}}$, and shown in \autoref{fig:scaling_DDPM} (green crosses), with the corresponding coefficients $\lambda(\theta_{\mathrm{eucl}})$.
	
	For an arbitrary choice of the cost function $C(\theta)$ in \eqref{eqn:theta_opt_def}, numerical routines, such as the Simulated Annealing Algorithm 
\cite{10.2307/3214721}, can be used to perform the optimisation. For example, for $C(\theta) = C_{\max}(\theta)$ \eqref{eqn:C_max} we implemented \autoref{algo:optimal_scaling} in 
R and chose the function 
	\emph{optim} \cite{stats_Rpackage}, with the method SANN, to perform Simulated Annealing. We label as $\theta_{\max}$ 
	the numerical values of $\theta_{\mathrm{opt}}$, obtained in this manner,  and present them in \autoref{fig:scaling_DDPM} (black crosses). The minimisation routine has been stopped when the maximum number $10^5$ of allowed evaluations of the cost function was exceeded. The simulation required 
$\approx 4.8$ sec of computation on a 2.70GHz processor.
	
	\autoref{fig:scaling_DDPM} suggests that the cost functions $C_{\mathrm{eucl}}(\theta)$ and $C_{\max}(\theta)$ have close arguments of minimum. In addition, the ratio $r$ between the maximum and the minimum coefficients $\lambda$ is $\approx 10^4$, i.e. $r = \lambda_d/\lambda^m_\mu \approx 10^4$. Such a difference looks computationally tractable, especially if compared with the original dimensional equation, where $ m_0 / \nu_0 \approx 10^{49} $ L$^{-2}$. Clearly, the proposed scaling procedure is able to reduce the computationally intractable orders of magnitude, caused by the experimental values of the physical parameters.
	
	Aimed to compare the presented results of Optimal Scaling with other scaling methods, we use the traditional scaling procedure \cite{Holmes2009_ND} to find the factors $\theta \equiv \{ \nu_0, t_0, m_0, w_0, M_0, P_0, \Pi_0, \delta_0 \} \in (0,\infty)^{N_x}$, with $N_x=8$. As reported in 	
\autoref{sec:sol_trad_scaling}, we must select $N_x=8$ out of the $N_d=19$ coefficients $\lambda$, defined in \autoref{tab:DDPM_param_scaling} in \autoref{sec:Dimensionless_PBE_model_Latex_Particles}. Among all possible selections we only consider the choices for which the linear system \eqref{eqn:system_to_solve_traditional_scaling} is numerically solvable, i.e. the absolute value of the determinant of the matrix to invert is bigger than $\varepsilon=10^{-12}$. For all chosen combinations of coefficients $\lambda$,  we numerically solve the system \eqref{eqn:system_to_solve_traditional_scaling} to find the corresponding factors $\theta = \theta_k$, with the subscript $k \in \mathbb{N}$ indexing the chosen combination of coefficients $\lambda$.

	As mentioned before, the computational tractability of the arising coefficients $\lambda(\theta_k)$ and corresponding variables can be quantified by the ratio $r(\theta_k)$:
	
	\begin{equation}
	r(\theta_k) \equiv 
    \frac
    { \max_{i=1, \dots, N_d} \lambda_i(\theta_k) }
    { \min_{i=1, \dots, N_d} \lambda_i(\theta_k) }.
    \label{eqn:def_ratio_r}
    	\end{equation}
    	
\begin{figure}[!h]
	\centering
	\includegraphics[scale=0.4]{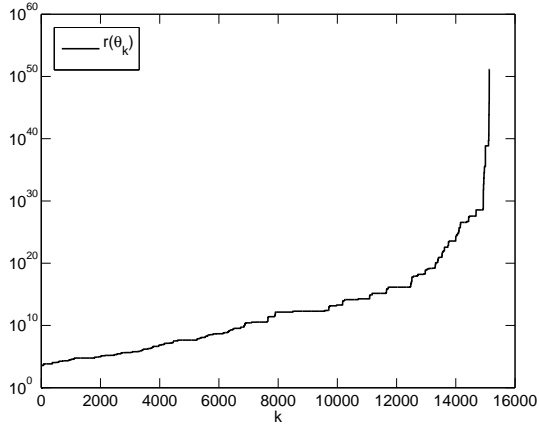}
	\caption{Latex Particles Morphology Formation. Ratio $r(\theta_k)$ \eqref{eqn:def_ratio_r}. The subscript $k \in \mathbb{N}$ enumerates all possible choices of $N_x=8$ out of $N_d=19$ coefficients $\lambda$, defined in \autoref{tab:DDPM_param_scaling} (\autoref{sec:Dimensionless_PBE_model_Latex_Particles}), such that the linear system \eqref{eqn:system_to_solve_traditional_scaling} is numerically solvable, i.e. the absolute value of the determinant of the matrix to invert is bigger than $\varepsilon=10^{-12}$. For the sake of readability, the combinations of coefficients $\lambda$ have been sorted 
	according 
	to the value of $r(\theta_k)$.} 
	\label{fig:max_min_lambda_trad_scaling_DDPM}
	\end{figure}

	\noindent To provide a complete description of the behaviour of the ratio $r$ \eqref{eqn:def_ratio_r}, we show in \autoref{fig:max_min_lambda_trad_scaling_DDPM} the values of $r(\theta_k)$ for all possible choices of $N_x=8$ out of $N_d=19$ coefficients $\lambda$, such that the linear system \eqref{eqn:system_to_solve_traditional_scaling} is numerically solvable. The figure demonstrates that a half the combinations of coefficients $\lambda$ gives $r > 10^{10}$. In other words, the traditional scaling is likely to lead to computationally intractable orders of magnitude for the dimensionless variables, unless the coefficients $\lambda$ are properly chosen.	
		
	In order to compare the coefficients $\lambda$ resulting from various considered scaling approaches, we define as $\theta_M$ and $\theta_m$ those scaling factors $\theta_k$ for which the ratio $r(\theta_k)$ \eqref{eqn:def_ratio_r} achieves its maximum and minimum value respectively: 

	\begin{equation}
	\theta_M \equiv \mathop{\mathrm{argmax}}_{k} r(\theta_k),
	\quad
	\theta_m \equiv \mathop{\mathrm{argmin}}_{k} r(\theta_k),
    	\label{eqn:theta_M_m_definition}
	\end{equation}

	\noindent where the index $k \in \mathbb{N}$ enumerates all choices of $N_x=8$ out of $N_d=19$ coefficients $\lambda$, such that the linear system \eqref{eqn:system_to_solve_traditional_scaling} is numerically solvable. \autoref{fig:scaling_DDPM} shows the values of $\theta_M$ (red triangles) and $\theta_m$ (blue circles), with corresponding coefficients $\lambda(\theta_M)$ and $\lambda(\theta_m)$.
	
	\autoref{fig:theta_DDPM} suggests that the factors $\theta_m$ are close to the arguments $\theta_{\mathrm{eucl}}$ and $\theta_{\max}$ of minimum of the cost functions $C_{\mathrm{eucl}}(\theta)$ and $C_{\max}(\theta)$ respectively. In addition, the factors $\theta_m$ ensure computationally tractable values for the coefficients $\lambda(\theta_m)$ with $r(\theta_m) \approx 10^4$, as shown in \autoref{fig:lambda_DDPM}. The proximity of $\theta_m$ to $\theta_{\mathrm{eucl}}$ and $\theta_{\max}$ is in agreement with the conclusions of \autoref{sec:scaling_projectile}, namely, that the scaling factors found by Optimal Scaling (\autoref{tab:scaling_projectile}, Methods (d)-(e)) are close to the traditional factors with the smallest ratio $r = \max_i \lambda_i / \min_i \lambda_i$ (\autoref{tab:scaling_projectile}, Method (a)).
	
	The factors $\theta_M$ (red triangles) and the corresponding coefficients $\lambda(\theta_M)$ shown in \autoref{fig:scaling_DDPM} illustrate the worst choice of factors $\theta$ for the traditional scaling method reported in \autoref{sec:sol_trad_scaling}. In this case, $\theta_M$ (red triangles) are far from $\theta_m \approx \theta_{\mathrm{eucl}} \approx \theta_{\max} $ (crosses and circles) (\autoref{fig:theta_DDPM}), the coefficients $\lambda(\theta_M)$ are computationally intractable (\autoref{fig:lambda_DDPM}) and $r(\theta_M) \approx 10^{50}$.
	
	The provided results demonstrate that the traditional scaling approach described in \autoref{sec:sol_trad_scaling} can compete in performance with the Optimal Scaling procedure only when it gives rise to a scheme which satisfies the following requirements:  $(i)$ the linear system \eqref{eqn:system_to_solve_traditional_scaling} to be solvable and $(ii)$ the ratio $r$ \eqref{eqn:def_ratio_r} is minimal. Obviously, the significant efforts required for finding such a scheme are not comparable with the well-defined and straightforward solution offered by the Optimal Scaling algorithm.
		
\section{Numerical Treatment of Dimensionless Model for Latex Particles Morphology}
\label{sec:num_treat_DDPM}
	
	Solving the PBE model \eqref{eqn:dimless_PBE_DDPM} derived in \autoref{sec:Dimensionless_PBE_model_Latex_Particles} and \autoref{sec:scaling_DDPM} is far from being a trivial task, because of its numerical and modelling complexities. The numerical difficulties include $(i)$ potential inaccuracies in a computed solution for highly aggregating processes, $(ii)$ numerical instabilities for growth-dominated systems, $(iii)$ increased stiffness for processes involving rapid particles nucleation, $(iv)$ numerical implementation of singular nucleation rates in the presence of aggregation and growth mechanisms, $(v)$ domain errors for high-order aggregation kernels. Several integration methods have been developed for solving similar models (see, for example, \cite{Alexopoulos2004, Alexopoulos2005, Roussos2005, Meimaroglou2006}). However, the majority of such numerical methods is able to deal only with a limited range of rate functions. Here we propose a general discretization method for the numerical treatment of the model \eqref{eqn:dimless_PBE_DDPM} and investigate the effect of applied scaling procedures on performance of such a method. We call it \emph{Generalised Method Of Characteristics} (GMOC). As the name suggests, the Method Of Characteristics (MOC) \cite{Scott2003} is a special case of GMOC, as will be shown below.
	
\subsection{Generalised Method Of Characteristics}
\label{sec:GMOC}

	Since the equations for $m(v,t)$ and $w(v,t)$ in \eqref{eqn:dimless_PBE_DDPM} can be solved analogously, we limit our discussion to the solution of the following Population Balance Equations (PBE) system \cite{Ramkrishna2000}:

	\begin{equation}
	\begin{cases}
	\frac{ \partial m(v,t) } {\partial t}
	& = 
	- 
	\frac{ \partial ( \, g(v,t) \, m(v,t) \, )}
	{\partial v}
	\, + \,
	n(v,t) 	 
	\, - \,
	\mu(v,t) \, m(v,t) 
	- 
	\, m(v,t) \,
	\int_{0}^{\infty}
	\! a(v,u,t) \, m(u,t) \, du	
	\\ 
	& \quad
	+
	\frac{1}{2} \, 
	\int_0^v
	\! a(v-u,u,t) \, m(v-u,t) \, m(u,t) \, du,
	\quad \forall v,t \in \mathbb{R}^+,
	\\ \\ 
    	m(v,0) & = \omega_0(v) \ge 0,
	\quad \forall v \in \mathbb{R}^+,
	\quad \quad \quad
	m(0,t) = 0,
	\quad 
	\forall t \in \mathbb{R}^+.
	\end{cases}
	\label{eqn:PBE}
	\end{equation}
	
	\noindent We propose to evaluate the solution $m(v,t)$ of \eqref{eqn:PBE} along $N$ curves, belonging to 
	
	$ P \equiv \left\{ (v,t) \in \mathbb{R}^2: v,t \ge 0 \right\}$, with the prescribed form $ v = \varphi_k(t)$, $\forall k=1, \dots, N$. Given $ m_k(t) \equiv \left. m(v,t) \right|_{v=\varphi_k(t)}$, the chain rule provides the time derivative of $m_k(t)$. As a result, one obtains the ODE system arising from \eqref{eqn:PBE} by evaluating the solution $m(v,t)$ on the curves $v=\varphi_k(t)$, $\forall k=1, \dots, N$, $\forall t \in \mathbb{R}^+$:
	
	\begin{align}
	\frac{dm_k(t)}{dt} 
	& =
	\left[
	\frac{ d\varphi_k(t) } { dt }
	- 
	g(\varphi_k(t),t)
	\right]
	\left.
	\frac{ \partial m(v,t) } {\partial v} 
	\right|_{v=\varphi_k(t)} 
	-	
	\, \left[
	\rho(\varphi_k(t),t) 
	+\mu(\varphi_k(t),t)
	\right]
	\, m_k(t) 
	\, +
	\nonumber \\
	&
	\quad
	+ \, n(\varphi_k(t),t)
	+ \, A^+(\varphi_k(t),t;m(\cdot,t))
	- \, A^-(\varphi_k(t),t;m(\cdot,t)),	
	\label{eqn:systemODE_GMOC}
	\end{align}

	\noindent where $ \rho(v,t) \equiv \partial g(v,t) / \partial v$ and
	
	\begin{align}
	A^+(v,t;m(\cdot,t)) 
	& \equiv 
	\frac{1}{2} \, \int_0^v \! a(v-u,u,t) \, m(v-u,t) \, m(u,t) \, du, 
	\label{eqn:A+_GMOC} \\
	A^-(v,t;m(\cdot,t)) 
	& \equiv 
	m(v,t) \, \int_0^{\infty} \! a(v,u,t) \, m(u,t) \, du.
	\label{eqn:A-_GMOC}
	\end{align}
	
	\noindent The particular choice of the curves $v=\varphi_k(t)$, for $k=1, \dots, N$, as the solutions of the Cauchy problems 
	
	\begin{equation}
	\varphi_k'(t) = g(\varphi_k(t),t),
	\quad \forall t \in \mathbb{R}^+,
	\quad \varphi_k(0) = \bar{\varphi}_k > 0,	
	\quad \forall k=1, \dots, N,
	\label{eqn:characteristics_MOC}
	\end{equation}
		
	\noindent allows cancelling the partial derivatives with respect to $v$ in \eqref{eqn:systemODE_GMOC}. The curves $v = \varphi_k(t)$ in \eqref{eqn:characteristics_MOC} correspond to the \emph{characteristics} defined by MOC, showing that GMOC embeds MOC. Moreover, the Method of Lines (ML) \cite{Schiesser1991} can be also recovered from GMOC by imposing:	
	
	\begin{equation}
	\varphi_k(t) = \varphi_k(0),
	\quad \forall t \in \mathbb{R}^+,
	\quad \forall k=1, \dots, N.	
	\label{eqn:ML_curves}
	\end{equation}
	
	\noindent The definition \eqref{eqn:characteristics_MOC}, i.e. MOC, avoids the numerical errors caused by the discretization of the partial derivatives with respect to $v$ in \eqref{eqn:systemODE_GMOC}. However, the characteristics grid \eqref{eqn:characteristics_MOC} may lead to the inefficient integration of the PBE system \eqref{eqn:PBE}, if other dynamical mechanisms make the solution $m(v,t)$ behave differently from what is prescribed by the function $g(v,t)$ \cite{Kumar1997-III}. In addition, MOC can only work if the characteristic curves are not crossing each other, with the consequent formation of a shock \cite{Scott2003}. The characteristics may also fail to cover some parts of the domain $P$, if the given curves are diverging in the considered region \cite{Scott2003}.
	
	For these reasons, the advantage of GMOC is in the flexible choice of the curves $v=\varphi_k(t)$. This can be especially beneficial in the case when the solution $m(v,t)$ has a complex shape potentially leading to numerical instabilities. However, the appropriate choice of the curves $v=\varphi_k(t)$ is a non-trivial and problem-specific task. The numerical experiments shown in \autoref{sec:num_res} will rely on the constant curves \eqref{eqn:ML_curves}, i.e. ML.
	
	\autoref{sec:GMOC_impl} concludes the implementation of GMOC, explaining how to impose the initial and boundary conditions of \eqref{eqn:PBE} in the ODE \eqref{eqn:systemODE_GMOC}. \autoref{sec:GMOC_impl} also provides the numerical schemes for approximation of the partial derivatives with respect to $v$, the integral terms and the time evolution in \eqref{eqn:systemODE_GMOC}.
	
\subsection{Numerical Experiments}
\label{sec:num_res}

	In \autoref{sec:scaling_DDPM} we considered various scaling procedures, including the newly derived Optimal Scaling, for nondimensionalization of the PBE model \eqref{eqn:PBE_latex_part} and found that the range of magnitude of $\lambda$ coefficients in resulting dimensionless equations \eqref{eqn:dimless_PBE_DDPM} may vary from $10^{4}$ to $10^{50}$ depending on a choice of a scaling procedure. Our objective is to investigate how the variation of $\lambda$ coefficients affects the computational performance and accuracy of the resolution method (GMOC in this case) applied to \eqref{eqn:dimless_PBE_DDPM}.
	
\begin{figure}[!h]
	\centering
	\begin{subfigure}[h]{.5\linewidth}
	\centering
	\includegraphics[scale=0.47]{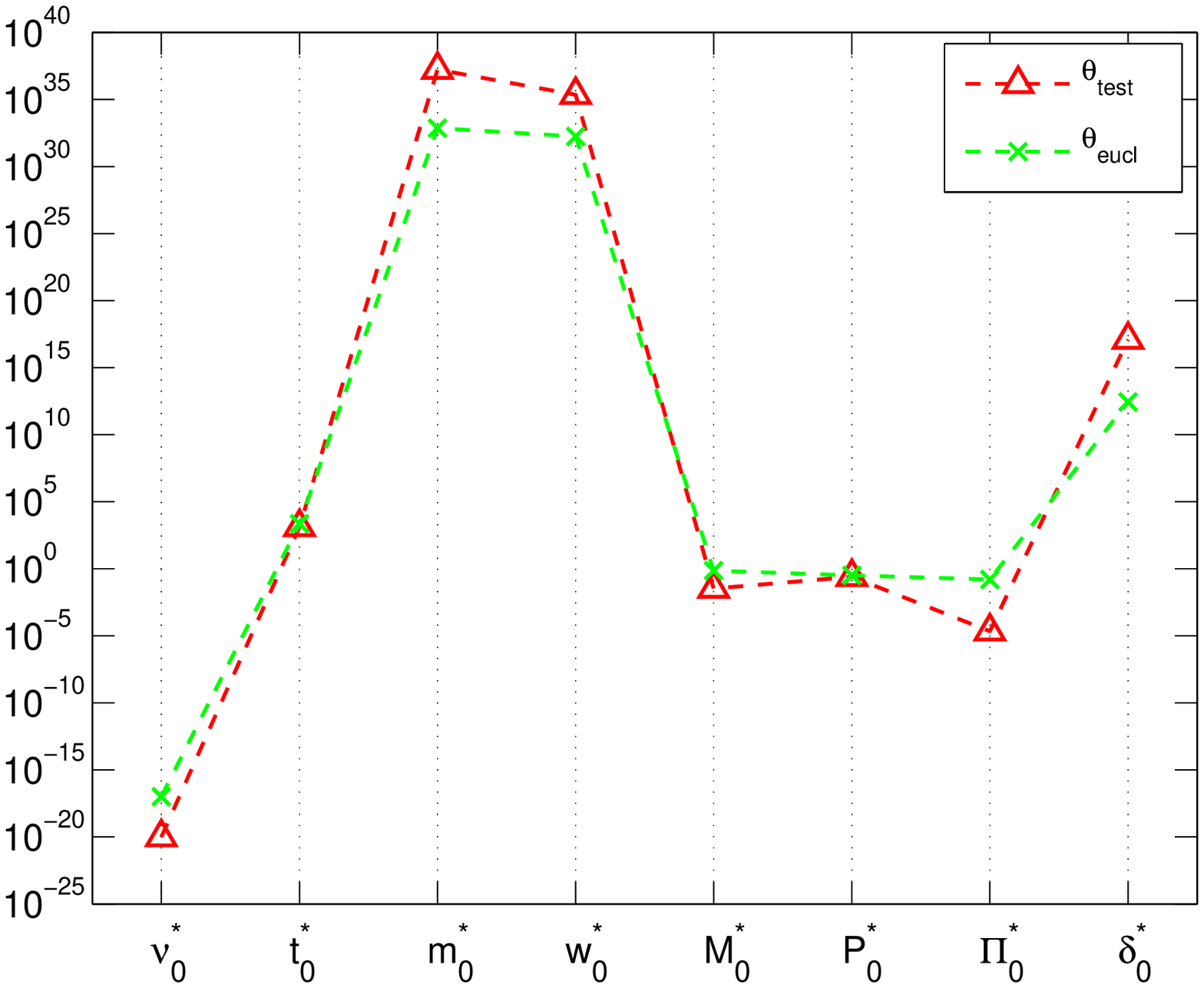}
	\caption{Scaling Factors $\theta^*$.}
	\label{fig:theta_DDPM_test}
	\end{subfigure}
	\begin{subfigure}[h]{.49\linewidth}
	\centering
	\includegraphics[scale=0.47]{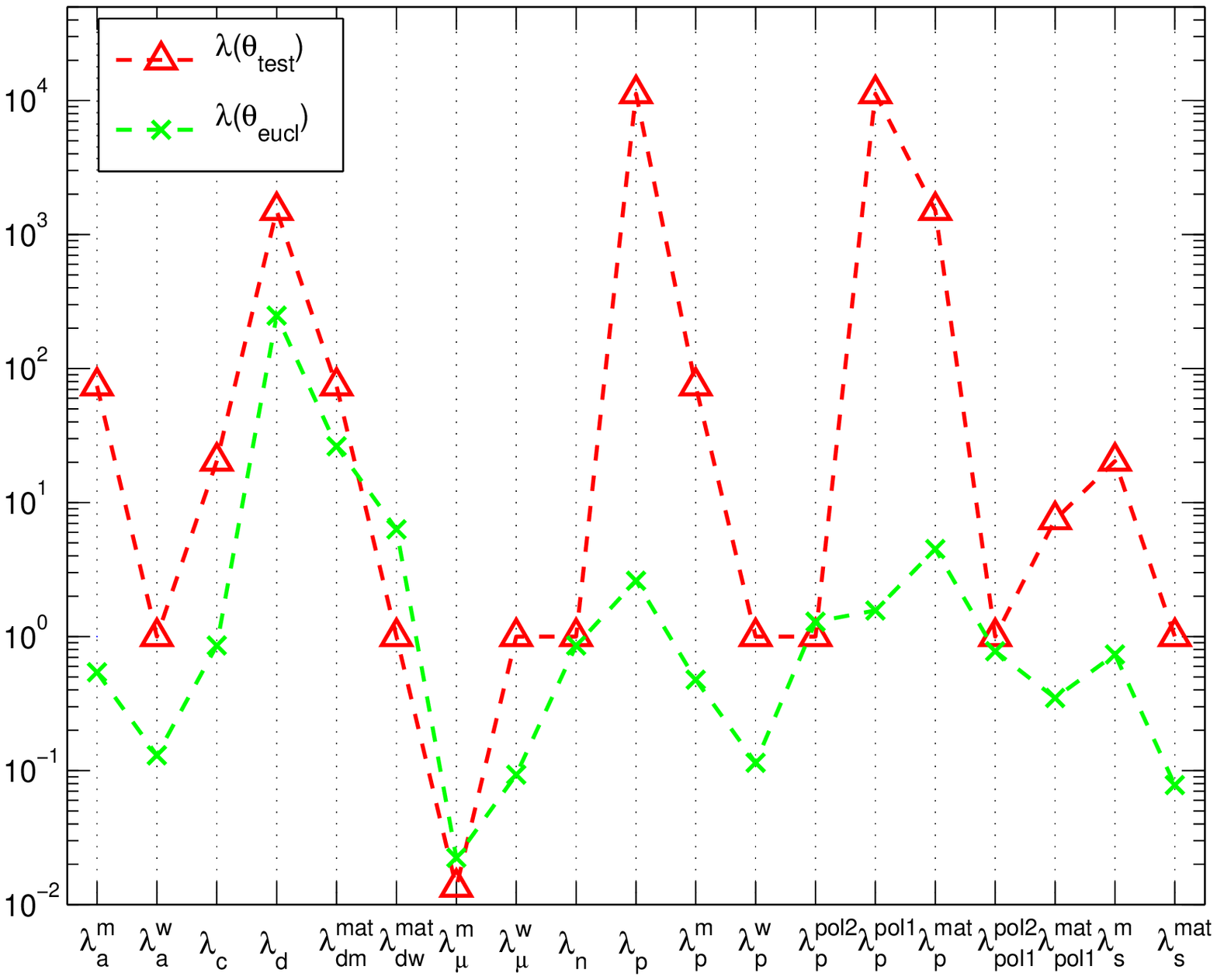}
	\caption{Dimensionless Coefficients $\lambda(\theta^*)$.}
	\label{fig:lambda_DDPM_test}
	\end{subfigure}	
	\caption{Latex Particles Morphology Formation. (i) Scaling factors $\theta^* \equiv \{ \, \nu_0^* \equiv \nu_0 \, / \, c_{\mbox{L}}, t_0^* \equiv t_0 \, / \, c_{\mbox{s}}, \, \, m_0^* \equiv m_0 \, c_{\mbox{L}}, \, w_0^* \equiv w_0 \, c_{\mbox{L}}, \, M_0^* \equiv M_0 \, / \, c_{\mbox{L}}, \, P_0^* \equiv P_0 \, / \, c_{\mbox{L}}, \, \Pi_0^* \equiv \Pi_0 \, / \, c_{\mbox{L}}, \, \delta_0^* \equiv \delta_0 \, c_{\mbox{L}} \}$, with $c_{\mbox{s,L}} = 1 \, \mbox{s,L}$, for the model described in \autoref{sec:PBE_model_Latex_Particles}. (ii) Corresponding values of the dimensionless coefficients $\lambda(\theta^*)$, defined in \autoref{tab:DDPM_param_scaling} (\autoref{sec:Dimensionless_PBE_model_Latex_Particles}). The factors $\theta^*$ are labelled as $\theta_{\mathrm{eucl}}$ (green crosses), if found by the Optimal Scaling approach using the analytical formula of \autoref{sec:anal_sol_min_Ceucl} for $C(\theta) = C_{\mathrm{eucl}}(\theta)$ \eqref{eqn:C_eucl_def}. The constants $\theta_{\mathrm{test}}$ (red triangles) are obtained with the traditional scaling approach described in \autoref{sec:sol_trad_scaling}.} 
	\label{fig:scaling_DDPM_test}
	\end{figure}

	As discussed in \autoref{sec:Dimensionless_PBE_model_Latex_Particles}, the dimensionless counterpart \eqref{eqn:dimless_PBE_DDPM} of \eqref{eqn:PBE_latex_part} is obtained by means of the scaling factors $\theta$, prescribed by a particular scaling methods. We consider the factors $\theta = \theta_{\mathrm{eucl}}$, obtained by the Optimal Scaling procedure for $C(\theta) = C_{\mathrm{eucl}}(\theta)$ \eqref{eqn:C_eucl_def} and $\theta = \theta_{\mathrm{test}}$ found by the traditional approach described in \autoref{sec:sol_trad_scaling}. To provide an illustrative comparison, we choose the factors $\theta_{\mathrm{test}}$ such that $r(\theta_{\mathrm{test}}) \gg r(\theta_{\mathrm{eucl}})$, where $r(\theta)$ is defined in \eqref{eqn:def_ratio_r}. The factors $\theta_{\mathrm{eucl}}$, $\theta_{\mathrm{test}}$ and the corresponding  parameters $\lambda$ are compared in \autoref{fig:theta_DDPM_test} and \autoref{fig:lambda_DDPM_test} respectively. In particular, \autoref{fig:lambda_DDPM_test} confirms that
	
	\begin{equation}
	r(\theta_{\mathrm{test}}) \approx 10^6 
	\gg 
	r(\theta_{\mathrm{eucl}}) \approx 10^4.
	\end{equation}
	
	The effect of a choice of $\theta$ on the efficiency of the numerical treatment of the dimensionless equations \eqref{eqn:dimless_PBE_DDPM} is evaluated by the following experiment. First, we set the integration domains for the dimensional model \eqref{eqn:PBE_latex_part} as: 
	
	\begin{equation}
	\tilde{v} \in [0,\tilde{v}_{\max}]
	\quad \mbox{and} \quad
	\tilde{t} \in [0,\tilde{T}_{\max}],
	\quad \mbox{with} \quad
	\tilde{v}_{\max} = 10^{-16} \mbox{ L}
	\quad \mbox{and} \quad
	\tilde{T}_{\max} = 450 \mbox{ s}.	
	\label{eqn:dimensional_interval_of_integration}
	\end{equation}
	
	\noindent As motivated in \autoref{sec:Dimensionless_PBE_model_Latex_Particles}, the dimensionless counterpart \eqref{eqn:dimless_PBE_DDPM} of \eqref{eqn:PBE_latex_part} should then be integrated within the properly scaled domains: 
	
	\begin{equation}
	v \in [0,v_{\max}]
	\quad \mbox{and} \quad
	t \in [0,T_{\max}],
	\quad \mbox{with} \quad
	v_{\max} \equiv \tilde{v}_{\max} \, / \, \nu_0,
	\quad
	T_{\max} \equiv \tilde{T}_{\max} \, / \, t_0,
	\label{eqn:dimensionless_interval_of_integration}
	\end{equation}
	
	\noindent where $\tilde{v}_{\max}$ and $\tilde{T}_{\max}$ are chosen as in \eqref{eqn:dimensional_interval_of_integration}, while $\nu_0$ and $t_0$ depend on a particular choice of $\theta \equiv \{ \nu_0, t_0, m_0, w_0, M_0, P_0, \Pi_0, \delta_0 \}$, i.e. $\theta = \theta_{\mathrm{eucl}}$ or $\theta = \theta_{\mathrm{test}}$. We choose GMOC (\autoref{sec:GMOC}) as an integration technique and use the same numbers $N$ and $M$ of grids points (see \autoref{tab:settings_DDPM}) for each experiment in order to equalise the computational effort required for evaluation of solutions $m(v,t)$ and $w(v,t)$ of \eqref{eqn:dimless_PBE_DDPM} when $\theta = \theta_{\mathrm{eucl}}$ or $\theta = \theta_{\mathrm{test}}$. The accuracy of the solutions $m(v,t)$ and $w(v,t)$ of \eqref{eqn:dimless_PBE_DDPM} can be quantified by the error functions
	
	\begin{equation}
	\varepsilon_{m,w}(t) \equiv
	\frac{ | V^{\mathrm{c_{m,w}}}_{\mathrm{pol2}}(t) - F_{m,w}(t) | }
	{ \left[ 
	\int_0^{T_{\max}} 
	\! V^{\mathrm{c_{m,w}}}_{\mathrm{pol2}}(t)^2 
	\, dt \right]^{1/2} },
	\quad \mbox{with} \quad	
	F_{m,w}(t) \equiv \int_0^{\infty} \! v \, m,w(v,t) \, dv,
	\quad
	\forall t \in [0,T_{\max}],
	\label{eqn:error_fun_def}
	\end{equation}

	\noindent since it is possible to show that the auxiliary variables $V^{\mathrm{c_{m,w}}}_{\mathrm{pol2}}(t)$ \eqref{eqn:dimless_V_cm_pol2}-\eqref{eqn:dimless_V_cw_pol2} must be equal to the first-order moments $F_{m,w}(t)$, as discussed in \autoref{sec:Dimensionless_PBE_model_Latex_Particles}.
	
	The integration of \eqref{eqn:dimless_PBE_DDPM} by GMOC requires the point-wise evaluation \eqref{eqn:systemODE_GMOC} of the function $n(v,t)$ \eqref{eqn:dimless_nucl_rate}. The numerical evaluation of the Dirac delta $\delta(v-\lambda_c)$ in \eqref{eqn:dimless_nucl_rate} is performed by assuming that the source term $n(v,t)$ spreads over a specified size range, as suggested in \cite{Alexopoulos2005}. In particular, we propose the following approximation:
	
	\begin{equation}
	\delta(v-\lambda_c) \approx \mathcal{N}(v;\lambda_c,\sigma_c), 
	\quad \forall v \in \mathbb{R}^+,
	\quad \lambda_c \gg \sigma_c > 0,
	\label{eqn:approx_GMOC_Dirac_delta}
	\end{equation}	 
	
	\noindent where $\mathcal{N}(v;\mu,\sigma)$ denotes the probability density function of the Gaussian with mean $\mu$ and standard deviation $\sigma$. In order to make the Gaussian very narrow, the standard deviation $\sigma_c$ should be chosen much smaller than $\lambda_c$.
	
\begin{figure}[!h]
	\centering
	\begin{subfigure}[h]{.495\linewidth}
	\centering
	\includegraphics[scale=0.475]{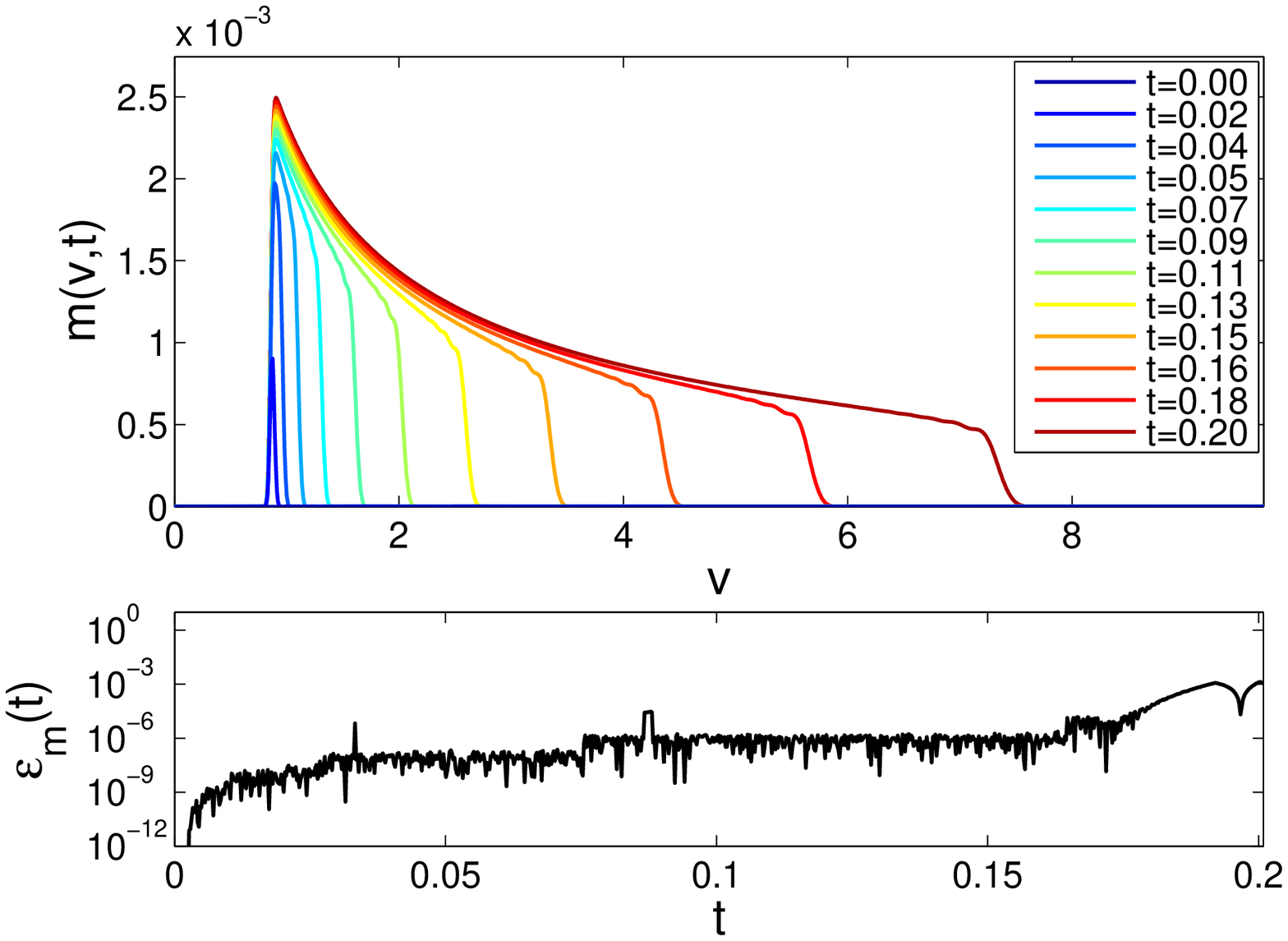}
	\captionsetup{width=0.9\linewidth}
	\caption{Solution $m(v,t)$ of \eqref{eqn:dimless_PBE_DDPM} and error $\varepsilon_m(t)$ \eqref{eqn:error_fun_def}, for $\lambda(\theta)$ with $\theta=\theta_{\mathrm{eucl}}$.}
	\label{fig:m(v,t)_dCeucl0}
	\end{subfigure}
	\begin{subfigure}[h]{.495\linewidth}
	\centering
	\includegraphics[scale=0.475]{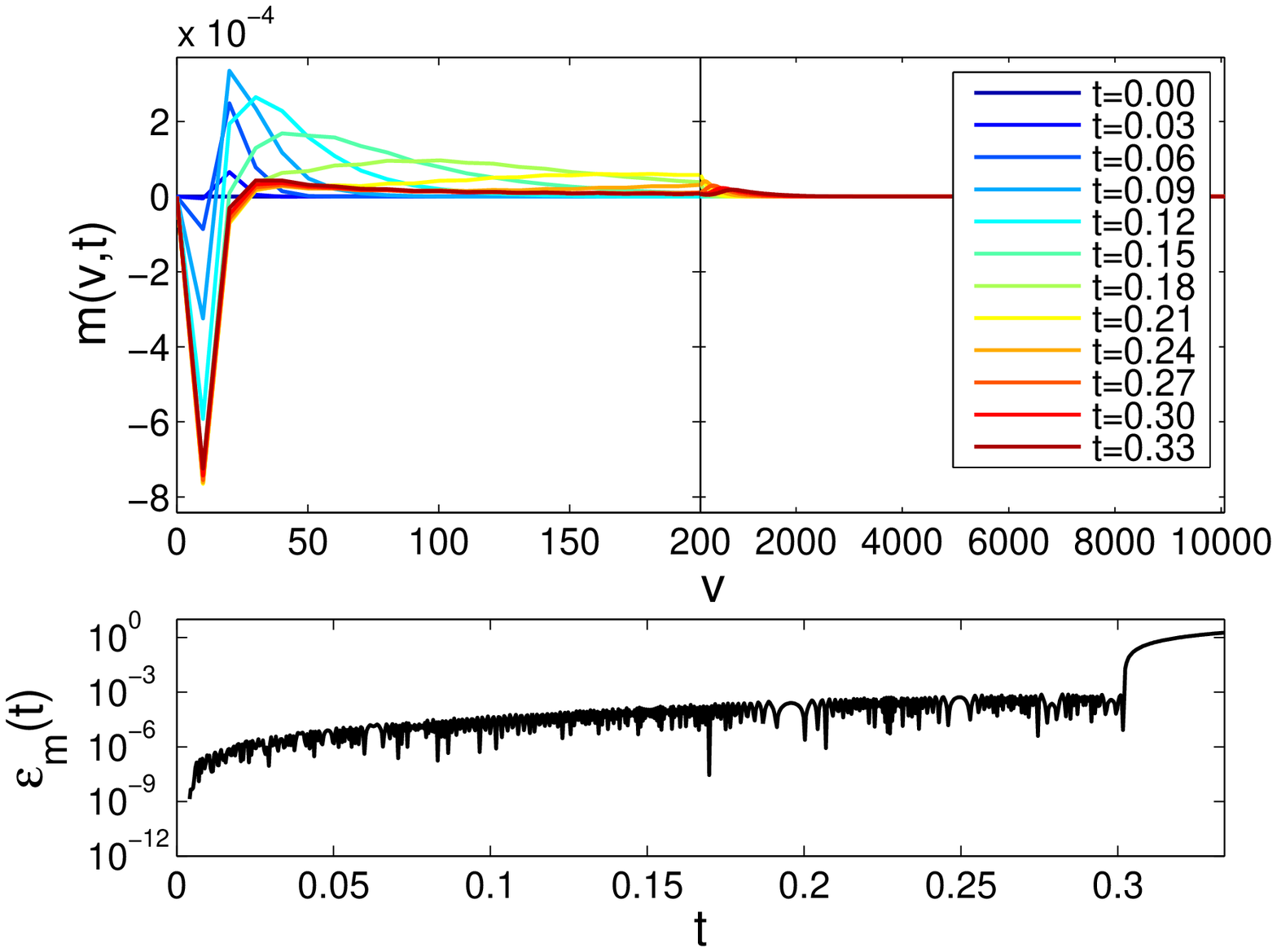}
	\captionsetup{width=0.9\linewidth}	
	\caption{Solution $m(v,t)$ of \eqref{eqn:dimless_PBE_DDPM} and error $\varepsilon_m(t)$ \eqref{eqn:error_fun_def}, for $\lambda(\theta)$ with $\theta=\theta_{\mathrm{test}}$.}
	\label{fig:m(v,t)_BadScaling}
	\end{subfigure}	
	\begin{subfigure}[h]{.495\linewidth}
	\centering
	\includegraphics[scale=0.475]{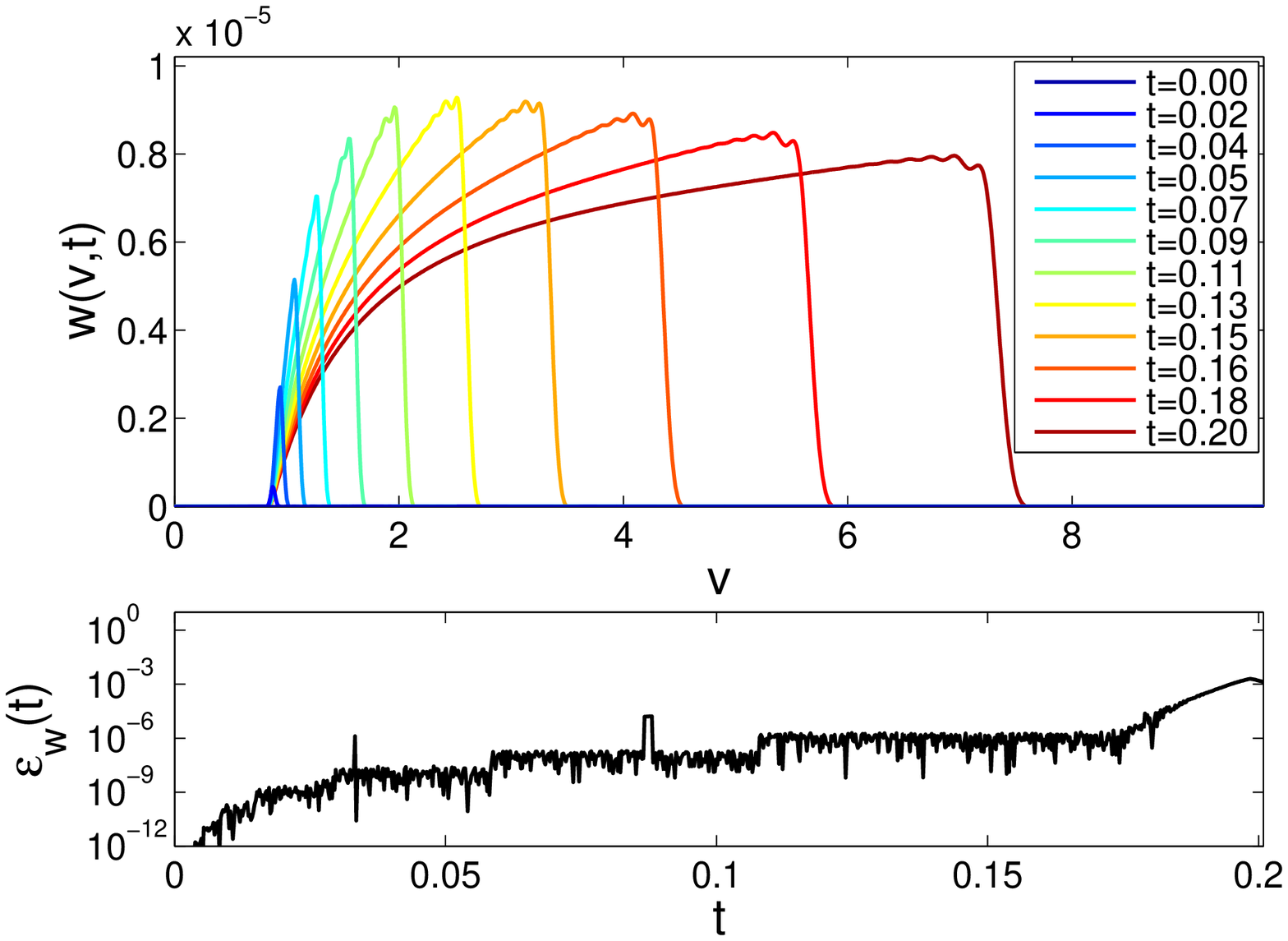}	\captionsetup{width=0.9\linewidth}	
	\caption{Solution $w(v,t)$ of \eqref{eqn:dimless_PBE_DDPM} and error $\varepsilon_w(t)$ \eqref{eqn:error_fun_def}, for $\lambda(\theta)$ with $\theta=\theta_{\mathrm{eucl}}$.}
	\label{fig:w(v,t)_dCeucl0}
	\end{subfigure}
	\begin{subfigure}[h]{.495\linewidth}
	\centering
	\includegraphics[scale=0.475]{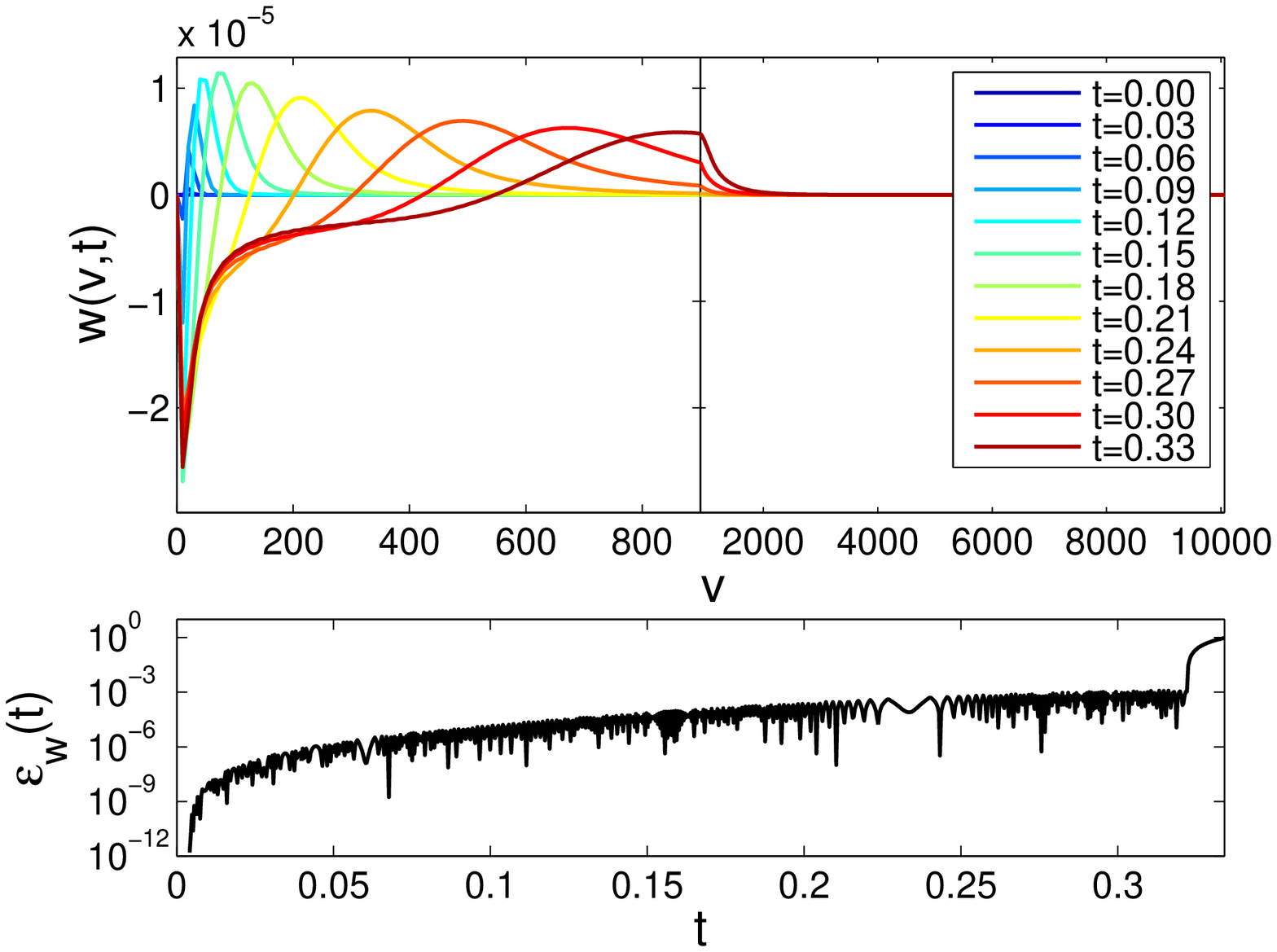}
	\captionsetup{width=0.9\linewidth}
	\caption{Solution $w(v,t)$ of \eqref{eqn:dimless_PBE_DDPM} and error $\varepsilon_w(t)$ \eqref{eqn:error_fun_def}, for $\lambda(\theta)$ with $\theta=\theta_{\mathrm{test}}$.}
	\label{fig:w(v,t)_BadScaling}
	\end{subfigure}	
	\caption{Latex Particles Morphology Formation. Numerical solutions $m(v,t)$ and $w(v,t)$ of \eqref{eqn:dimless_PBE_DDPM}, from \autoref{sec:Dimensionless_PBE_model_Latex_Particles}, with corresponding errors $\varepsilon_{m,w}(t)$ \eqref{eqn:error_fun_def}. The parameters $\lambda(\theta)$, defined in \autoref{tab:DDPM_param_scaling}, are evaluated for (i),(iii) $\theta=\theta_{\mathrm{eucl}}$ and (ii),(iv) $\theta=\theta_{\mathrm{test}}$. Settings and computational times are specified in \autoref{tab:settings_DDPM}.} 
	\label{fig:solution_PBE_test_scaling}
	\end{figure}

	\autoref{tab:settings_DDPM} summarises settings and computational times of performed simulations, corresponding to $\theta = \theta_{\mathrm{eucl}}$ and $\theta = \theta_{\mathrm{test}}$. The numerical solutions $m(v,t)$ and $w(v,t)$ of \eqref{eqn:dimless_PBE_DDPM} are shown in \autoref{fig:solution_PBE_test_scaling}, with corresponding errors $\varepsilon_{m,w}(t)$ \eqref{eqn:error_fun_def}.
	
	\begin{table}[!h]
	\footnotesize
	\centering
	\begin{tabular}{p{1.5cm} p{2.15cm} p{2.85cm} p{2.7cm} p{2.5cm} p{1.4cm} p{1.5cm}}
    
	\hline
    
	\textbf{Scaling \newline Factors} & 
	\textbf{Technical \newline Details} & 
	\textbf{Space Grid $\textbf{v}$ \newline Grid Size $N$} & 
    \textbf{Time Grid $\textbf{t}$ \newline Grid Size $M$} & 
    	\textbf{Software \& \newline Hardware} & 
    	\textbf{Error} &
	\textbf{CPU Time} \\
       
    	\hline
    	
    	\rowcolor[HTML]{EFEFEF}
   		
   	$\theta=\theta_{\mathrm{eucl}}$ 
   	\newline \autoref{fig:theta_DDPM_test} 
   	\newline (green crosses) & 
   	Approximation \eqref{eqn:approx_GMOC_Dirac_delta} 
   	with $\sigma_c = \lambda_c / 50$ &
   	$\textbf{v} = \left\{ \varphi_k = k h \right\}_{k=0}^N$
   	\newline $h = v_{\max}/N$
   	\newline $v_{\max} \approx 9.7$
   	\newline $N = 10^3$ &
   	$\textbf{t} = \left\{ t_k = k \tau \right\}_{k=0}^M$
   	\newline $\tau = T_{\max}/M$
	\newline $T_{\max} \approx 0.2$  
   	\newline $M = 9 \times 10^4$ &
   	C++ \newline \textbf{BCAM code} 
   	\newline 64-bit Linux OS \newline 2.40GHz proc. &
   	$\varepsilon_{m,w}(t)$ \eqref{eqn:error_fun_def} in 
   	\newline Fig. \ref{fig:m(v,t)_dCeucl0} and \ref{fig:w(v,t)_dCeucl0} & 
   	$1.1 \times 10^4$ sec \\   	   	
   	
   	$\theta=\theta_{\mathrm{test}}$ 
   	\newline \autoref{fig:theta_DDPM_test} 
   	\newline (red triangles) & 
   	Approximation \eqref{eqn:approx_GMOC_Dirac_delta}
   	with $\sigma_c = \lambda_c / 50$ &
   	$\textbf{v} = \left\{ \varphi_k = k h \right\}_{k=0}^N$
   	\newline $h = v_{\max}/N$
	\newline $v_{\max} \approx 10^4$   	
   	\newline $N = 10^3$ &
   	$\textbf{t} = \left\{ t_k = k \tau \right\}_{k=0}^M$
	\newline $\tau = T_{\max}/M$
	\newline $T_{\max} \approx 0.33$
   	\newline $M = 9 \times 10^4$ &
   	C++ \newline \textbf{BCAM code} 
   	\newline 64-bit Linux OS \newline 2.40GHz proc. &
   	$\varepsilon_{m,w}(t)$ \eqref{eqn:error_fun_def} in 
   	\newline Fig. \ref{fig:m(v,t)_BadScaling} and \ref{fig:w(v,t)_BadScaling} & 
   	$1.1 \times 10^4$ sec \\      	
   	
	\hline
   	   	
	\end{tabular}
	\caption{Latex Particles Morphology Formation. Settings and computational times for integrating the dimensionless model \eqref{eqn:dimless_PBE_DDPM} (\autoref{sec:Dimensionless_PBE_model_Latex_Particles}), using the GMOC method (\autoref{sec:GMOC}). The numerical solutions of \eqref{eqn:dimless_PBE_DDPM} are shown in \autoref{fig:solution_PBE_test_scaling} within the integration domains \eqref{eqn:dimensionless_interval_of_integration}. \textbf{BCAM code} refers to the in-house package.}	
	\label{tab:settings_DDPM}
    \end{table}	 	   
    
	In the light of the data shown in \autoref{fig:solution_PBE_test_scaling}, we can conclude that the choice of the scaling factors $\theta$ has a significant effect on the quality of the performed simulation. Given the same computational effort, i.e. the same numbers $N$ and $M$ of grids points (\autoref{tab:settings_DDPM}), the integration performed with the scaling factors $\theta_{\mathrm{eucl}}$ ensures errors $\varepsilon_{m,w}(t)$ up to three orders of magnitude smaller than the errors observed with the use of the factors $\theta_{\mathrm{test}}$.
	
	In addition, the simulation performed with the factors $\theta_{\mathrm{test}}$ is affected by strong oscillations to negative values of the solutions $m(v,t)$ and $w(v,t)$, as shown in \autoref{fig:m(v,t)_BadScaling} and \autoref{fig:w(v,t)_BadScaling}, whereas no such oscillations are observed in  \autoref{fig:m(v,t)_dCeucl0} and \autoref{fig:w(v,t)_dCeucl0} corresponding to the factors $\theta_{\mathrm{eucl}}$. The oscillations to negative values of the solutions $m(v,t)$ and $w(v,t)$ are not physical, since the solutions of \eqref{eqn:dimless_PBE_DDPM} must be always non-negative, as proven in \autoref{sec:non_negative}.
	
	The strong oscillations in \autoref{fig:m(v,t)_BadScaling} and \autoref{fig:w(v,t)_BadScaling} can be explained by the inability of the space grid $\textbf{v}$ to deal with the narrow distribution $\mathcal{N}(v;\lambda_c,\sigma_c)$ in \eqref{eqn:approx_GMOC_Dirac_delta}. If $\theta=\theta_{\mathrm{test}}$, the data from \autoref{tab:settings_DDPM} show that the grid size $h \approx 10$ is bigger than the width $\sigma_c \approx 0.4$. As a consequence, the grid $\textbf{v}$ is not fine enough to deal with the narrow density $\mathcal{N}(v;\lambda_c,\sigma_c)$. On the contrary, the scaling $\theta=\theta_{\mathrm{eucl}}$ allows for $h \approx 9.7 \times 10^{-3} < \sigma_c \approx 1.7 \times 10^{-2}$, leading to the reduction of numerical oscillations in \autoref{fig:m(v,t)_dCeucl0} and \autoref{fig:w(v,t)_dCeucl0}. The values of $h$ and $\sigma_c$ are strictly related to the choice of factors $\theta = \{ \nu_0, t_0, m_0, w_0, M_0, P_0, \Pi_0, \delta_0 \}$, since they can be computed as
		 
	 \begin{equation}
	 h = \frac{ \tilde{v}_{\max} }{ \nu_0 \, N }
	 \quad \mbox{and} \quad
	 \sigma_c = \frac{ v_c }{ 50 \, \nu_0^2 \, \delta_0 },	 
	 \end{equation}
	 
	\noindent where $\tilde{v}_{\max}$, $N$ and $v_c$ have been fixed.
	
	In conclusion, the scaling factors $\theta$ have a strong impact on performance and accuracy of the numerical treatment of the resulting equations, and the Optimal Scaling procedure, as the name suggests, provides an optimal choice of such factors.  	
	
\section{Conclusions \& Discussion}
\label{sec:concl_discuss}

	We propose a novel scaling procedure for the dimensionless reformulation of an arbitrary equation expressed in physical units. As any standard scaling method, the new approach potentially simplifies the analysis of the underlying equation, and reduces its numerical complexity.	
	 At the same time, by its very design,
	 our method assures optimal and computationally tractable orders of magnitude for the terms involved in the resulting dimensionless equation.
	  Defined in this way, the Optimal Scaling technique
	   relies on the optimisation of the characteristic constants 
	    which determine the
	   magnitudes of physical quantities, i.e. scaling factors.
	
	The method is independent of a particular resolution methodology, applied to a given equation, since it can be performed once, without interfering with the chosen solver. In this sense, the scaling procedure should never require a significant computational effort. Moreover, we suggest a specific choice of a cost function to be minimised which, without a performance loss, allows for an analytical computation of scaling factors. This analytical solution is general and holds for a wide range of equations. We validated the Optimal Scaling method  by applying it to three popular problems - the simple Projectile Problem, the Schr\"{o}dinger Equation for a hydrogen atom in a magnetic field and the Landau-de Gennes model for liquid crystals.
	
	One of the important properties of the efficiency of a scaling routine is its ability to produce scaled parameters of similar orders of magnitude in order to avoid severe round-off errors, and improve conditioning of the considered problem. We introduce a metric for quantification of the similarity among scaled (dimensionless) quantities, which is a ratio of maximal and minimal values of scaled parameters. If compared with other traditional scaling approaches \cite{Holmes2009_ND, Scaling_Langtangen} applied to two tested problems, Optimal Scaling decreases the ratio value from $10^5$ to $10^2$ in the case of Projectile Problem and from $10^8$ to $10^2$ for Schr\"{o}dinger Equation.
	
	In addition, the reported data demonstrate that Optimal Scaling ensures a strong reduction of the range of orders of magnitude assumed by physical quantities. In the example of Schr\"{o}dinger Equation (SE), our methodology results in decreasing a ratio between maximal and minimal parameters values from $10^{25}$ (dimensional) to $10^2$ (dimensionless), whereas for the Projectile Problem these values are respectively $10^3$ and $10^2$. Despite such a strong transformation, Optimal Scaling allows for an accurate estimation of characteristic features of a given system. As an example, we show that the factors, provided by Optimal Scaling, follow the same trend as the often used atomic units for SE. Also, we demonstrate that the Optimal Scaling procedure preserves the qualitative behaviour of the flow in the phase space of position and momentum of the Projectile Problem. Finally, we prove that Optimal Scaling correctly describes the large-body limit of Landau-de Gennes model for liquid crystals.
	
	The outlined features of Optimal Scaling make it an excellent candidate for application to the recently proposed PBE model for Latex Particles Morphology formation \cite{DDPM_2016,PhDThesis_Rusconi_PMCQS}. The Optimal Scaling method, applied to this model, gives rise to a new dimensionless computationally tractable PBE model with the parameters differing by 4 instead of 49 orders of magnitude as appears in the original dimensional model. We demonstrate that for this system, the traditional scaling approach \cite{Holmes2009_ND, Scaling_Langtangen} is comparable in performance with the Optimal Scaling procedure only when the former generates a scheme which satisfies the following requirements: $(i)$ the underlying linear system is solvable and $(ii)$ the ratio between maximal and minimal parameters is minimal. Obviously, the significant efforts required for finding such a scheme are not comparable with the well-defined and straightforward solution offered by the Optimal Scaling algorithm, which automatically satisfies those conditions. Moreover, Optimal Scaling can be viewed as an intelligent and efficient searching algorithm for such a scheme.
	
	To solve the resulting dimensionless PBE model for Latex Particles Morphology formation, we introduce a novel flexible numerical approach called Generalised Method Of Characteristics (GMOC). GMOC integrates the equations of interest along properly defined characteristic curves and can be beneficial to the problems where the solutions have complex shapes, potentially leading to numerical instabilities. 
	
	We apply GMOC with the constant characteristic curves to the dimensionless PBE equations, obtained using different scaling approaches, in order to evaluate the effect of those approaches on the overall computational performance of the solver. Our numerical experiments reveal that a choice of scaling factors has a significant effect on performance of a PBE solver. The integration carried out with the optimally scaled factors yields the errors up to three orders of magnitude smaller than the those obtained with alternative scaling techniques. Finally, the use of Optimal Scaling helps avoid  unphysical numerical oscillations present in the numerical solutions obtained with the traditional scaling methods. In summary, a particular choice of scaling factors has a strong impact on performance of numerical solver, and the proposed Optimal Scaling procedure, as the name suggests, allows for the optimal choice of such factors.

\section*{Acknowledgments}
\addcontentsline{toc}{section}{Acknowledgments}

	This research is supported by the Spanish Ministry of Science, Innovation and Universities: MTM2016-76329-R (AEI/FEDER, EU), MTM2017-82184-R (DESFLU) and BCAM Severo Ochoa accreditation SEV-2017-0718. The Basque Government is acknowledged for support through ELKARTEK Programme (grant KK-2018/00054) and BERC 2018-2021 program. The authors thank J.M. Asua and S. Hamzehlou (POLYMAT, Spain) for valuable discussions.

\appendix
\counterwithin{equation}{section}
\renewcommand{\theequation}{\thesection.\arabic{equation}}

\renewcommand{\thesection}{A}
\section{Solution for Traditional Scaling Factors $\theta$}
\label{sec:sol_trad_scaling}

	Here we explain how to find the scaling factors $\theta$ defined in \eqref{eqn:change_of_var}, following the traditional scaling procedure explained in \cite{Holmes2009_ND}. As discussed in \autoref{sec:optimal_scaling_procedure}, the factors $\theta \in (0,\infty)^{N_x}$ are found by imposing $ N = \min \{ N_x, N_d \} $ coefficients $\lambda(\theta) \in (0,\infty)^{N_d}$ \eqref{eqn:dimensionless_equation} equal to 1. If $N_d > N_x$, one must select $N_x$ out of the $N_d$ coefficients $\lambda(\theta)$ and the factors $\theta$ are found as the solution of 
			
	\begin{equation}
	\lambda_c(\theta) = 1,
	\quad \forall c=1, \dots, N_x,
	\label{eqn:system_to_solve_trad_factors}
	\end{equation}	   
	
	\noindent where the label $c=1, \dots, N_x$ enumerates the chosen $N_x$ out of $N_d$ coefficients $\lambda_i(\theta)$, $i=1, \dots, N_d$. As mentioned in \autoref{sec:anal_sol_min_Ceucl}, the Buckingham $\Pi$-theorem \cite{Barenblatt2003_book} ensures the shape  \eqref{eqn:HP_shape_lambda} for coefficients $\lambda_c(\theta)$ in \eqref{eqn:system_to_solve_trad_factors}. Discarding the units of measure, the change of variables $\theta_j = 10^{\rho_j}$, $\forall j =1, \dots, N_x$, leads to
	
	\begin{equation}
	\sum_{j=1}^{N_x}
	\alpha^c_j \, \rho_j 
	=
	- \log_{10}(\kappa_c),
	\quad
	\forall c = 1, \dots, N_x,	
	\label{eqn:system_to_solve_traditional_scaling}
	\end{equation}	 
	
	\noindent where $c$ enumerates the chosen $N_x$ out of $N_d$ coefficients $\lambda$ to be imposed to 1. The solution of \eqref{eqn:system_to_solve_traditional_scaling} provides $\theta = \{ 10^{\rho_j} \}_{j=1}^{N_x}$.

\renewcommand{\thesection}{B}
\section{Non-Negative Solutions for Latex Particles Morphology}
\label{sec:non_negative}
	
	Our objective is to show that the solutions $m(v,t)$ and $w(v,t)$ of the dimensionless system \eqref{eqn:dimless_PBE_DDPM}, derived in \autoref{sec:Dimensionless_PBE_model_Latex_Particles}, must be always non-negative. As before, it is enough to prove this statement for the solution $m(v,t)$ only, as the identical arguments can be applied to $w(v,t)$. In other words, we want to demonstrate that the solution $m(v,t)$ of \eqref{eqn:PBE} is always non-negative:	
	
	\begin{equation}
	m(v,t) \ge 0, 
	\quad \forall v,t \in \mathbb{R}^+.
	\label{eqn:non_negative_m}
	\end{equation}	
	
	Let us define $t^* > 0$ as the minimal time at which there exists such a volume $v=v^*>0$ that $m(v^*,t^*)=0$. Assuming the continuous evolution of $m(v,t)$ from a positive initial data of \eqref{eqn:PBE}, we have 
	
	\begin{equation}
	m(v,t^*) > 0,
	\quad 
	\forall v \in (0,v^*) \cup (v^*,\infty), 
	\quad
	m(v^*,t^*) = 0.
	\label{eqn:v_t_star_definition}
	\end{equation}
	
	\noindent Equation \eqref{eqn:v_t_star_definition} indicates $v^*$ as a point of minimum for the function $m(v,t^*)$ of volume $v$. Under regularity assumptions, \eqref{eqn:v_t_star_definition} implies $ \partial m(v,t) / \partial v |_{\substack{v=v^* \\ t=t^*}} = 0 $ and the transport term in \eqref{eqn:PBE} vanishes for $v=v^*$ and $t=t^*$. Given the null transport term and the non-negative functions $a$ and $n$, the time derivative of $m(v,t)$ is non-negative at $v=v^*$ and $t=t^*$: 
	
	\begin{equation}
	\left. \frac{ \partial m(v,t) } {\partial t}
	\right|_{\substack{v=v^* \\ t=t^*}}
	= 
	\, n(v^*,t^*)
	+ \frac{1}{2} \, 
	\int_0^{v^*}
	\! a(v^*-u,u,t^*) \, m(v^*-u,t^*) \, m(u,t^*) \, du \ge 0.
	\label{eqm:time_der_m(v_t_star)}
	\end{equation}	 
	
	\noindent As soon as $m(v,t)$ touches the zero level, its time derivative \eqref{eqm:time_der_m(v_t_star)} becomes non-negative, repulsing $m$ to positive values. In other words, \eqref{eqn:non_negative_m} follows.
	
	The statement \eqref{eqn:non_negative_m} can be generalised to the cases $(i)$ the growth rate $g$ explicitly depends on the solution $m$, i.e. $g=g(m,v,t)$, and $(ii)$ the extra term $\partial^2 m(v,t) / \partial v^2$ is considered in \eqref{eqn:PBE}. The previous argument follows because the transport term is still null, for $v=v^*$ and $t=t^*$, and $\partial^2 m(v,t) / \partial v^2$ only adds a non-negative term in \eqref{eqm:time_der_m(v_t_star)}, with $v^*$ being a point of minimum value.	

\renewcommand{\thesection}{C}
\section{Implementation of Generalised Method Of Characteristics}
\label{sec:GMOC_impl}

	In what follows, we provide details of the implementation of the Generalised Method Of Characteristics or GMOC (\autoref{sec:GMOC}) for solving the PBE \eqref{eqn:PBE}. We explain how to impose the initial and boundary conditions of \eqref{eqn:PBE} in the ODE \eqref{eqn:systemODE_GMOC}, as well as how to numerically treat the partial derivatives with respect to $v$, the integral terms and the time evolution in \eqref{eqn:systemODE_GMOC}.
	
	\textbf{Initial and Boundary Conditions}. The initial condition of \eqref{eqn:PBE} imposes:
	
	\begin{equation}
	m_k(0) = \omega_0(\varphi_k(0)),
	\quad \forall k=1, \dots, N.
	\end{equation}
	
	\noindent The curve $\varphi_0(t) = 0$, $\forall t \in \mathbb{R}^+$, is defined to prescribe the boundary condition of \eqref{eqn:PBE}:
	
	\begin{equation}
	\left. m(v,t) \right|_{v=\varphi_0(t)} 
	\equiv	
	m_0(t) = 0, 
	\quad \forall t \in \mathbb{R}^+.
	\end{equation}
	
	\noindent The information carried by $\varphi_0(t)$ and $m_0(t)$ is used to approximate the partial derivatives and the integral terms, as explained below. As a result, the boundary condition of \eqref{eqn:PBE} is imposed in the ODE system \eqref{eqn:systemODE_GMOC}.
	
	\textbf{Partial Derivatives}. Given $\left\{ \varphi_k(t), m_k(t) \right\}_{k=0}^N$ for all $t \in \mathbb{R}^+$, finite difference schemes can approximate the partial derivatives $\left. \partial m(v,t) / \partial v \right|_{v=\varphi_k(t)}$, $\forall k=1, \dots, N$. As stated in \cite{Mesbah2009}, such schemes may suffer from numerical diffusion and a common solution consists in using high-order approximation schemes. Motivated by the statement, we select a fourth-order accurate scheme.

	Given the hyperbolic Partial Differential Equation \eqref{eqn:PBE}, it is possible to define the domain of dependence $D(v,t)$ as the subset of $ P \equiv \left\{ (v,t) \in \mathbb{R}^2: v,t \ge 0 \right\}$, such that $m(v,t)$ only depends on the values of the solution inside $D(v,t)$. The negative integral term in \eqref{eqn:PBE} makes the domain $D(v,t)$ be the set $\mathbb{R}^+ \times [0,t)$. Both forward and backwards points have an influence on the solution at $(v,t)$, motivating the choice of a central scheme for approximating the partial derivatives with respect to $v$. Assuming the condition
	
	\begin{equation}
	\varphi_k(t) - \varphi_{k-1}(t) = h(t) > 0,
	\quad \forall k=1, \dots, N,
	\, \forall t \in \mathbb{R}^+,
	\label{eqn:hp_uniform_grid}
	\end{equation}	  
	
	\noindent we propose the following fourth-order accurate scheme \cite{Fornberg1988}:
	
	\begin{equation}
	\left.
	\frac{ \partial m(v,t) } {\partial v} 
	\right|_{v=\varphi_k(t)}
	\approx
	\begin{cases}
	\frac{-m_{k+2}(t)+8m_{k+1}(t)-8m_{k-1}(t)+m_{k-2}(t)}
	{12h(t)},
	& \forall k=2, \dots, N-2,
	\\
	\frac{25m_N(t)-48m_{N-1}(t)+36m_{N-2}(t)-16m_{N-3}(t)+3m_{N-4}(t)}
	{12h(t)},
	& \mbox{if } k=N,
	\\
	\frac{3m_N(t)+10m_{N-1}(t)-18m_{N-2}(t)+6m_{N-3}(t)-m_{N-4}(t)}
	{12h(t)},
	& \mbox{if } k=N-1,
	\\
	\frac{m_4(t)-6m_3(t)+18m_2(t)-10m_1(t)-3m_0(t)}
	{12h(t)},
	& \mbox{if } k=1.
	\end{cases}
	\label{eqn:finite_diff_scheme}
	\end{equation}
		
	\noindent The lack of either backwards or forward grid points does not allow the central scheme for $k=1,N-1,N$, being replaced by asymmetric formulas.
	
	\textbf{Integral Terms}. Given $\left\{ \varphi_k(t), m_k(t) \right\}_{k=0}^N$ for all $t \in \mathbb{R}^+$, we approximate the integral terms $A^-$ and $A^+$ in \eqref{eqn:systemODE_GMOC} by the quadrature rules	
	
	\begin{equation}
	\int_0^{\infty}
	\! l(u,t) \, m(u,t) \, du
	\approx
	\sum_{j=0}^N
	\, p_j(t) 
	\, l(\varphi_j(t),t)
	\, m_j(t),
	\quad \forall t \in \mathbb{R}^+,
	\label{eqn:QR_GMOC}
	\end{equation}
	
	\noindent where the function $l(u,t)$ is defined as
		
	\begin{equation}
	l(u,t) \equiv 
	\mathbbm{1}_{(0,\infty)}(u)
	\, a(\varphi_k(t),u,t),
	\quad \forall k=1, \dots, N,
	\end{equation}
	
	\noindent for $A^-$ \eqref{eqn:A-_GMOC}, while
		
	\begin{equation}
	l(u,t) \equiv 
	\mathbbm{1}_{(0,\varphi_k(t))}(u) 
	\, a(\varphi_k(t)-u,u,t)
    \, m(\varphi_k(t)-u,t),
	\quad \forall k=1, \dots, N,
	\label{eqn:convol_kernel}
	\end{equation}	
			
	\noindent for $A^+$ \eqref{eqn:A+_GMOC}, with $\mathbbm{1}_X(x)=1$ if $ x \in X $, zero otherwise. The weights $p_j(t)$, $j=0, \dots, N$, in \eqref{eqn:QR_GMOC} specify the quadrature rule to be used. Assumed \eqref{eqn:hp_uniform_grid}, the fourth-order accurate composite Simpson's rule \cite{Integration_CompRules} is chosen in agreement with the accuracy of the numerical scheme \eqref{eqn:finite_diff_scheme}. Taking advantage of \eqref{eqn:hp_uniform_grid}, it is possible to evaluate $m(\varphi_k(t)-u,t)$ for all $u=\varphi_j(t)$, such that $\varphi_j(t) \in (0,\varphi_k(t))$:
	
	\begin{equation}
	m(\varphi_k(t)-\varphi_j(t),t)
	=
	m(\varphi_{k-j}(t),t)
	=
	m_{k-j}(t).
	\label{eqn:eval_convol_kernel}
	\end{equation}
	
	\noindent In the case the hypothesis \eqref{eqn:hp_uniform_grid} does not hold, the evaluation of $m(\varphi_k(t)-\varphi_j(t),t)$ requires the interpolation of the data $\left\{ \varphi_k(t), m_k(t) \right\}_{k=0}^N$.
	
	\textbf{Time Evolution}. The state vector $ y(t) \equiv \left\{ m_1(t), \dots, m_N(t) \right\}$ corresponds to the solution of the Cauchy problem:

	\begin{equation}    
    y' = F(y), \quad y(t_0) = y_0,
    \label{eqn:Chaucy_Problem_GMOC}
    \end{equation}
    
    \noindent where the function $F: [0,\infty)^N \to \mathbb{R}^N$ is defined by the Right-Hand Side of \eqref{eqn:systemODE_GMOC}, provided the approximation schemes \eqref{eqn:finite_diff_scheme} and \eqref{eqn:QR_GMOC}. In agreement with the accuracy of the considered schemes, the fourth-order accurate Runge-Kutta method (RK4) \cite{Delin2012} is chosen to integrate \eqref{eqn:Chaucy_Problem_GMOC}. For the sake of simplicity, the time step $\tau>0$ is assumed to be constant for all time $t$. 

\section*{References}
\addcontentsline{toc}{section}{References}

\bibliography{refs.bib}

\newcommand{\etalchar}[1]{$^{#1}$}
\begin{thebibliography}{MKHVdH09}

\bibitem[AK05]{Alexopoulos2005}
A.~H. Alexopoulos and C.~Kiparissides.
\newblock {\em Chem. Eng. Sci.}, 60:4157--4169, 2005.

\bibitem[ARK04]{Alexopoulos2004}
A.~H. Alexopoulos, A.~I. Roussos, and C.~Kiparissides.
\newblock {\em Chem. Eng. Sci.}, 59:5751--5769, 2004.

\bibitem[B\'92]{10.2307/3214721}
C.~J.~P. B\'elisle.
\newblock {\em J. Appl. Prob.}, 29:885--895, 1992.

\bibitem[Bar96]{Barenblatt1996_book}
G.~I. Barenblatt.
\newblock {\em Scaling, self-similarity, and intermediate asymptotics}.
\newblock Cambridge University Press, Cambridge, 1996.

\bibitem[Bar03]{Barenblatt2003_book}
G.~I. Barenblatt.
\newblock {\em Scaling}.
\newblock Cambridge University Press, Cambridge, 2003.

\bibitem[BRS93]{10.2307/3690278}
C.~J.~P. B\'elisle, H.~E. Romeijn, and R.~L. Smith.
\newblock {\em Math. Oper. Res.}, 18:255--266, 1993.

\bibitem[BSOT88]{0953-4075-21-20-001}
S.~Bivona, W.~Schweizer, P.~F. O'Mahony, and K.~T. Taylor.
\newblock {\em J. Phys. B: At., Mol. Opt. Phys.}, 21:L617, 1988.

\bibitem[Dem97]{doi:10.1137/1.9781611971446}
J.~Demmel.
\newblock {\em {Applied Numerical Linear Algebra}}.
\newblock Society for Industrial and Applied Mathematics, 1997.

\bibitem[For88]{Fornberg1988}
B.~Fornberg.
\newblock {\em Math. Comp.}, 51:699--706, 1988.

\bibitem[FS96]{Fassbinder_1996_Hmagn}
P.~Fassbinder and W.~Schweizer.
\newblock {\em Phys. Rev. A}, 53:2135--2139, 1996.

\bibitem[GJ18]{Gartland_2018_liquid_cryst}
E.~C. Gartland~Jr.
\newblock {\em Math. Model. Anal.}, 23:414--432, 2018.

\bibitem[HLA16]{DDPM_2016}
S.~Hamzehlou, J.~R. Leiza, and J.~M. Asua.
\newblock {\em Chem. Eng. J.}, 304:655--666, 2016.

\bibitem[Hol09]{Holmes2009_ND}
M.~H. Holmes.
\newblock {\em Introduction to the foundations of applied mathematics}.
\newblock Springer, New York, 2009.

\bibitem[KGV83]{Kirk1983}
S.~Kirkpatrick, C.~D. Gelatt, and M.~P. Vecchi.
\newblock {\em Science}, 220:671--680, 1983.

\bibitem[KP79]{Kahan:1979:PFS:1057520.1057522}
W.~Kahan and J.~Palmer.
\newblock {\em SIGNUM Newsl.}, 14:13--21, 1979.

\bibitem[KR97]{Kumar1997-III}
S.~Kumar and D.~Ramkrishna.
\newblock {\em Chem. Eng. Sci.}, 52:4659--4679, 1997.

\bibitem[LP16]{Scaling_Langtangen}
H.~P. Langtangen and G.~K. Pedersen.
\newblock {\em Scaling of Differential Equations}.
\newblock Springer, first edition, 2016.

\bibitem[Mil06]{Miller_Asympt_Anal}
P.~D. Miller.
\newblock {\em {Applied Asymptotic Analysis}}.
\newblock American Mathematical Soc., 2006.

\bibitem[MKHVdH09]{Mesbah2009}
A.~Mesbah, H.~J.~M. Kramer, A.~E.~M. Huesman, and P.~M.~J. Van~den Hof.
\newblock {\em Chem. Eng. Sci.}, 64:4262--4277, 2009.

\bibitem[MRK06]{Meimaroglou2006}
D.~Meimaroglou, A.~I. Roussos, and C.~Kiparissides.
\newblock {\em Chem. Eng. Sci.}, 61:5620--5635, 2006.

\bibitem[Nay00]{Nayfeh_PM2000}
A.~Nayfeh.
\newblock {\em Perturbation Methods}.
\newblock Wiley-VCH, 2000.

\bibitem[Per73]{0022-3700-6-9-002}
I.~C. Percival.
\newblock {\em J. Phys. B: At., Mol. Phys.}, 6:L229, 1973.

\bibitem[{R C}19]{stats_Rpackage}
{R Core Team and contributors worldwide}.
\newblock {The R Stats Package}, 2019.

\bibitem[RAK05]{Roussos2005}
A.~I. Roussos, A.~H. Alexopoulos, and C.~Kiparissides.
\newblock {\em Chem. Eng. Sci.}, 60:6998--7010, 2005.

\bibitem[Ram00]{Ramkrishna2000}
D.~Ramkrishna.
\newblock {\em Population Balances: Theory and Applications to Particulate
  Systems in Engineering}.
\newblock Academic Press, San Diego, 2000.

\bibitem[Rus18]{PhDThesis_Rusconi_PMCQS}
S.~Rusconi.
\newblock {\em Probabilistic Modelling of Classical and Quantum Systems}.
\newblock PhD thesis, UPV/EHU - University of the Basque Country, 2018.

\bibitem[Sar03]{Scott2003}
S.~Sarra.
\newblock {\em JOMA}, 3, 2003.

\bibitem[Sch91]{Schiesser1991}
W.~E. Schiesser.
\newblock {\em The Numerical Method of Lines}.
\newblock Academic Press, first edition, 1991.

\bibitem[SM03]{Integration_CompRules}
E.~S\"uli and D.~Mayers.
\newblock {\em An Introduction to Numerical Analysis}.
\newblock Cambridge University Press, Cambridge, 2003.

\bibitem[SR97]{ode45_Matlab}
L.~F. Shampine and M.~W. Reichelt.
\newblock {\em SIAM J. Sci. Comput.}, 18:1--22, 1997.

\bibitem[TC12]{Delin2012}
D.~Tan and Z.~Chen.
\newblock {\em MSME}, 7:1--10, 2012.

\bibitem[\v{C}85]{Cerny1985}
V.~\v{C}ern\'y.
\newblock {\em J. Optim. Theory Appl.}, 45:41--51, 1985.

\bibitem[YL91]{YaoLi1991}
X.~Yao and G.~Li.
\newblock {\em J. Comput. Sci. Tech.}, 6:329--338, 1991.

\bibitem[ZSM{\etalchar{+}}93]{Zabinsky1993}
Z.~B. Zabinsky, R.~L. Smith, J.~F. McDonald, H.~E. Romeijn, and D.~E. Kaufman.
\newblock {\em J. Global Optim.}, 3:171--192, 1993.

\end{thebibliography}

\end{document}